**Measured distribution of cloud chamber tracks from radioactive decay: a new empirical approach to investigating the quantum measurement problem**

Jonathan F. Schonfeld
Center for Astrophysics | Harvard and Smithsonian
60 Garden St., Cambridge, Massachusetts 02138 USA
jschonfeld@cfa.harvard.edu
ORCID ID# 0000-0002-8909-2401

**Abstract** Using publicly available video of a diffusion cloud chamber with a very small radioactive source, I measure the spatial distribution of where tracks start, and consider possible implications. This is directly relevant to the quantum measurement problem and its possible resolution, and appears never to have been done before. The raw data are relatively uncontrolled, leading to caveats that should guide future, more tailored experiments. Aspects of the results may suggest a modification to Born's rule at very small wavefunction, with possibly profound implications for the detection of extremely rare events such as proton decay, but other explanations are not ruled out. Speculatively, I introduce two candidate small-wavefunction Born rule modifications, a hard cutoff, and an offset model with a stronger underlying physical rationale. Track distributions from decays in cloud chambers represent a previously unappreciated way to probe the foundations of quantum mechanics, and a novel case of wavefunctions with macroscopic signatures.

**1. Introduction**

Measurement occupies a privileged role in the conventional formulation of quantum theory. Between measurements, a wave function evolves smoothly according to Schroedinger's equation; but measurement itself is widely supposed to entail moments when the wavefunction changes discontinuously by application of random projections. This juxtaposition of smooth and discontinuous is the quantum measurement problem [1]. A solution to the problem could take the form of a demonstration that instantaneous projections are really idealized "effective representations" of more complicated processes governed entirely by smooth Schroedinger evolution. Such a solution would expose limits to projection-based formulations. Such limits could be consequential for quantum computing, in which quantum measurement is carried out on industrial scale; and for detection of events, such as proton decay, whose extreme rarity could be an intrinsic challenge for canonical measurement phenomenology.

Quantum computing and proton-decay searches employ very sophisticated measurement technology. Since 2020 I've focused my own quantum-measurement research on the lower-tech cloud chamber, because its underlying physics is much simpler. In a recent paper [2], I explored the processes by which a cloud chamber detects single charged particles emitted by the simplest radioactive decays. I identified a mechanism to explain the origination of cloud chamber tracks without appeal to random projections, and derived an emergent position-space Born rule that describes the distribution in space and time of tracks' starting points. Deviations from this Born rule would presumably result when droplet formation in supersaturated vapors deviates from conventional idealizations. The Born distribution in Reference [2] depends on position in a very simple way, and should be easier to verify experimentally than the familiar double-slit interference pattern, but I have been unable to find applicable experiments on the statistics of cloud-chamber track locations. So I have conducted a search on the internet for videos of cloud-chamber tracks induced by radioactive decay, and have measured a track distribution myself in one video to compare with Reference [2]. The purpose of this paper is to report my methodology and findings.

As will be apparent, it may be premature to draw firm conclusions from comparison between theory and measurement, but aspects of the results may suggest a modification of the position-space Born rule at extremely small wavefunction, although other, more mundane candidate mechanisms cannot be ruled out. Such a Born rule modification could have profound implications for the detection of extremely rare processes such as proton decay. Caveats for any interpretation of our measurements arise from many uncontrolled aspects of the underlying data. Issues include: uncontrolled temperature and vapor density in the cloud chamber; uncalibrated placement of the video camera; suboptimal and varying placement of the illumination source; and indeterminate thermal characteristics of the sample's mounting fixture. These caveats are all good justifications for building an apparatus tailored to the purpose at hand, instrumented to reveal and diagnose deviations between measurement and theory. Judged against this standard, the present work is a proof of principle, a necessary first step before committing time and expense to a more rigorous experiment.

In the next section, I review the microscopic theory of cloud chamber detection behavior developed in Reference [2], together with the conventional formulation of measurement in quantum mechanics. In Section 3, I recast the theory of Reference [2] into a form better matched to the data analyzed here. In Section 4, I survey cloud chamber videos available online, explain why I chose the particular video analyzed here, and list the video's relevant technical specifications and limitations. In section 5, I explain how I extracted track coordinates, and I graph the distribution of coordinates for direct comparison with the mathematics in Section 3. In Section 6, I compare data with modified predictions derived from two speculative Born rule refinements. Section 7 contains some concluding remarks.

## 2. Microscopic theory of cloud chamber behavior

Reference [2] addresses cloud chamber detection of slow decays in which one heavy particle transforms into another by emitting a single, spinless nonrelativistic light charged particle. We paraphrase the gist of Reference [2]:

A cloud chamber is an enclosure containing air supersaturated with a condensable vapor, which can be water but is more typically ethyl alcohol. The chamber is cooled from the bottom, so that supersaturation, and therefore favorability for charged-particle track formation, is greatest in a "sensitive layer" at the bottom. Conventional wisdom has it that when a charged particle passes through the chamber, it ionizes air molecules, and the ions nucleate visible vapor droplets (drive them supercritical). However, this assumes the charged particle wavefunction is very collimated (so the particle can be treated as a point at one location moving in one direction), while the actual wavefunction of an emitted charged particle near the initial source is not collimated in any meaningful sense. Indeed, the wavefunction near the source at three-dimensional position **x** and time $t$ is given by

$$\psi(\pmb{x}, t) \sim -i Y(\Omega) \sqrt{\frac{\gamma}{v}} \frac{1}{\|\pmb{x}\|} exp\left\{\left(\frac{\|\pmb{x}\|}{v} - t\right)\left(\frac{\gamma}{2} - i\frac{mv^2}{\hbar}\right)\right\}, \qquad (2.1)$$

where Y is a spherical harmonic, $\Omega$ is solid angle, we adopt the convention that **x** is defined relative to the location of the initial decaying particle, $\|\mathbf{x}\|$ is distance from the initial heavy particle, $m$ is emitted particle mass, $v$ is emitted particle speed, $\gamma$ is the conventional decay e-folding rate [i.e. ln(2) divided by half-life], and we ignore an irrelevant overall phase factor exp(i[total energy]$t/\hbar$). (In what follows, we treat Y as constant because decays in the particular data we analyze below are s-wave.) It should be noted that Equation (2.1) holds outside the decay source's interaction radius, which we can treat as zero because we're interested in length scales characteristic of the cloud chamber, while the interaction radius is on a nuclear scale. Equation (2.1) ignores the possibility that enclosure-induced boundary conditions could modify the wavefunction. This requires a more careful analysis, beyond the scope of this paper; but see a related remark in Section 5. Equation (2.1) also ignores the impact of multiple small-cross-section interactions with the atoms of the gas that makes up the cloud chamber medium; we will return to this point at the end of this section, and again in Section 5.

By contrast with the conventional picture, the wavefunction of Equation (2.1) interacts with an already existing vapor droplet (generated randomly due to thermal fluctuations in the most common case of diffusion cloud chambers) that is just barely sub-critical. A barely sub-critical droplet turns out to have a very large amplitude of interaction with the wavefunction of Equation (2.1), so that even a very weak wavefunction can provoke the subcritical droplet to grow quickly in a supercritical fashion and become visible, and provoke the wavefunction to collimate. This is so because, in the presence of a droplet that's already formed, single-molecule ionization can proceed with very small energy loss, since ion-induced potential energy due to droplet polarization can nearly balance electron excitation energy. This near-degeneracy drives the cross section of this quantum Coulomb interaction to singularity, and leads to a collimated free emitted-particle wavefunction in the final state.

Within this picture, I argued in Reference [2] that the probability per unit time and unit volume to find an emitted-particle track originating at three-dimensional position **x** and time $t$ is approximately

$$P(\mathbf{x},t) \sim \rho A v \tau |\psi(\mathbf{x},t)|^2, \qquad (2.2)$$

where $\rho$ and $\tau$ are constants characteristic of the cloud chamber medium, and $A$ is a constant characteristic of the ionization process. Substituting Equation (2.1) for $\psi$, and assuming small $\gamma$ (slow decays), this reduces to

$$P(\mathbf{x},t) \propto \frac{1}{\|x\|^2}. \qquad (2.3)$$

Equation (2.2) is a Born rule, in the sense that it asserts a proportionality between a measurement probability and the squared absolute value of a wavefunction in a particular coordinate system. The general, textbook formulation of Born's rule holds that

1. Any observable quantity corresponds to a Hermitian operator **M** acting on the wavefunction of the object to which measurement is applied.
2. Any measurement of **M** can result only in some eigenvalue $\mu$ of **M**.
3. Which specific eigenvalue is observed is intrinsically random, with probability $|<\psi|\psi_\mu>|^2$, where $\psi_\mu$ is the eigenvector of **M** corresponding to $\mu$.
4. The measurement drives the wavefunction to transform discontinuously into the projection

$$\frac{\langle\psi|\psi_\mu\rangle}{|\langle\psi|\psi_\mu\rangle|}\psi_\mu \qquad (2.4)$$

The cloud chamber provides a particularly interesting commentary on these provisions.

- Commentary on provision 1: For a cloud chamber, **M** is clearly the operator whose eigenstates are defined by position **x**, although it's not obvious that detection of an emitted particle is equivalent to position measurement in any canonical sense.
- Commentary on provision 2: The eigenvalue rule is a tautology, because **x** is the only basis for which the cloud chamber Born rule is derived. (A more formal way to say this is that this Born rule is *contextual*, and consequently the rigorous conclusion of Gleason's theorem [29], that the Born rule cannot be violated, doesn't apply.)
- Commentary on provision 3: This probability rule is clearly true for a cloud chamber if Equation (2.2) is correct, except that the randomness is not intrinsic. Instead, randomness in this case reflects the random nature of thermal fluctuations in the underlying detecting medium.
- Commentary on provision 4: This projection rule can't be true as stated for a cloud chamber, since any emitted particle, first detected at **x**, immediately and rapidly flees away. Reference [2] explains why this projection rule doesn't apply in some other, more generalized sense either (again vitiating Gleason's theorem).

The cloud chamber case suggests that the full axiomatization #1-4 goes too far, but that the narrower $|\psi|^2$ proportionality in Equation (2.2) still applies, at least approximately. This paper tests whether – and how much – this position-space Born rule is actually *true*.

Before proceeding to quantitative analysis, let us return to the earlier remark that Equation (2.1) ignores multiple small-cross-section interactions between the emitted particle wavefunction and the gaseous cloud chamber medium. Clearly, Equation (2.1) represents a steady flow of wavefunction square norm outward from the center at rate $\gamma\exp(-\gamma t)$. Each gas atom with a nonzero interaction cross section is like an obstacle in a stream, in that it gives rise to a thin, low-square-norm wake of emitted-particle wavefunction on its downstream side (possibly combined with the atom promoted to an excited state). The wake is a very thin cone – opening angle $\lambda/(4\sigma/\pi)^{1/2}$ for emitted particle wavelength $\lambda$ and cross section $\sigma$ – and doesn't reduce the overall square norm that flows away from the center (there's no backflow). [Note this is *not* the strong collimation that marks the start of a visible cloud chamber track [2]; that channels nearly all the system's square norm.] Under these circumstances, the emitted particle wavefunction still should look locally like the plane wave $\exp(im\mathbf{v}\cdot\mathbf{x}/\hbar)$ (**v** has magnitude $v$ and points from source to the local area in question) regardless of the internal state of the gas. This is all that the reasoning in Reference [2], Section 4 really relies on, so Equation (2.2), with Equation (2.1) substituted for $\psi$, should survive. Things get more complicated when wakes beget more wakes by encountering more obstacles: opening angles may then widen successively in a stepwise Gaussian random process. If the typical number of random steps is large enough, one can imagine the overall square-norm flow becoming disordered, departing from purely radial and threatening the viability of Equations (2.1) and (2.2). This requires further study, beyond the scope of this paper, but we will attempt to coarsely quantify this possibility in Section 5. In any case, this underscores the importance of examining real data.

## 3. Cumulative radial distribution seen in a two-dimensional image

In the scenario highlighted in the next section, the cloud chamber is a thin flat circular enclosure (Petri dish) viewed from above, with a sensitive layer of depth $a$ beginning on the dish floor. The radioactive source is propped at height $b$ above the floor of the chamber, so the two-dimensional density corresponding to Equation (2.3) actually recorded by the camera is

$$D(x,y) \propto \int_{-b}^{a-b} \left(\frac{1}{R^2 + z^2}\right) dz, \qquad (3.1)$$

where $R=(x^2+y^2)^{1/2}$ is two-dimensional distance from the decay source. As a practical matter, we won't have enough statistics to do a good job measuring $D$ as a function of $x$ and $y$, so we'll default to the cumulative radial distribution, proportional to the following:

$$\int_0^R 2\pi r dr \int_{-b}^{a-b} \left(\frac{1}{r^2 + z^2}\right) dz = \int_0^R 2\pi r dr \left\{ \int_0^b + \int_0^{a-b} \right\} \left(\frac{1}{r^2 + z^2}\right) dz$$
$$= C(R,b) + C(R, a-b) \qquad (3.2)$$

where

$$C(R,b) \equiv \pi b \ln\left[1 + \left(\frac{R}{b}\right)^2\right] + 2\pi R \cdot Arctan\left(\frac{b}{R}\right). \qquad (3.3)$$

## 4. Data

A Google search with terms "cloud chamber video" produces at least the twenty-three distinct examples in References [3-25] (all but one referring to diffusion cloud chambers). Many of these don't apply here. References [3-12] show cloud chamber activity only from background radiation in the ambient environment. References [13, 14] provide clips of only a few seconds each. References [15, 16] show tracks from thorium rods, not point sources, and with uncontrolled viewing geometries. References [17-20] show tracks from lumps of material, making it impossible to look at small $R$. Also, by virtue of their nontrivial masses, they produce so many tracks that it's difficult to separate one track from another, even running the videos in slow motion. In Reference [21] the lump is quite small, but the cloud chamber track footage is very brief. References [22-25] all use small samples mounted at the ends of needles, i.e. the right physical geometry and nicely separable tracks. But Reference [22] shows only 8 sec of track activity, with an enclosure of indeterminate size. Reference [23] also involves an enclosure of indeterminate size, as well as a variety of camera angles that complicate geometrical viewing analysis. Reference [24] has the best image quality, but uses three needles at the same time, so it's difficult to determine which track comes from which needle. Reference [25] seems to be the closest to "just right:" the enclosure is a Petri dish, whose dimensions are standardized, and the camera is located directly above (more or less) and fixed throughout the observation, which lasts a full 1 min 30 sec.

Detailed experimental specifications and caveats for Reference [25] are as follows, and a sample video frame is shown in Figure 1:
- Chamber dimensions (presumed standard Petri dish): diameter 100mm, depth 15mm.
- Radioactive source: labeled $^{210}$Pb, but, because of decay chain, admixed with $^{210}$Bi and $^{210}$Po. $^{210}$Pb and $^{210}$Bi are beta emitters (half-lives 22.3 years and 5.0 days, respectively), and $^{210}$Po is an s-wave alpha emitter (138 days) [26]. Nominal source activity is 0.01 μCi (https://www.spectrumtechniques.com/products/sources/needle-sources/).
- Radioactive source fixture: eye of a needle, whose other end is stuck in a cork. (The radioactive region is about 4mm long [R. Stevens, Spectrum Techniques Inc., private communication], treated mathematically in this paper as a point. Some discussion of the impact of this point idealization can be found at the end of Section 5.) The entire assembly, including cork, is enclosed in the Petri dish.
- Location of radioactive source: xy coordinate of radioactive source not quite centered in Petri dish; vertical placement (in $z$) is unspecified, but Figure 1 shows that the cork has radius 13mm at its base, narrowing to 10mm at its far end resting on the dish floor at a distance of 20mm; and the needle point is another 20mm beyond that. So elementary trigonometry says the needle point is ~3.5mm above the floor of the Petri dish, i.e. $b$=3.5mm.
- Alcohol concentration: value and horizontal-plane homogeneity uncontrolled.
- Chamber temperature: value and horizontal-plane homogeneity uncontrolled; cooled with dry ice underneath.
- Depth of supersaturated sensitive layer: The theory in Reference [27] suggests that sensitive-layer thickness is roughly a fixed percentage of dish depth for given boundary temperatures. That reference's Figure 2.9 corresponds to the temperature boundary conditions in this paper and indicates that the depth of the sensitive layer should be 10%-20% of the dish depth, i.e. 1.5mm-3mm here. This range would put the needle point above the top of the sensitive layer. As we shall see later, the video of Reference [25] seems to show that the needle point must actually lie *inside* the sensitive layer, so, for data analysis purposes, I expand the range of possible sensitive-layer depths for consideration to 1mm-4mm.
- Source and fixture thermal characteristics: Indeterminate.
- Illumination: flashlight, handheld from the side, angle and brightness variable. Shadow of needle against dish floor is clearly visible, and complicates track measurement in its vicinity.
- Video camera location: Fixed, directly over top of Petri dish, pointing approximately straight down.

- Frame rate: 30fps; when downloaded to Microsoft Video Editor, frames are labeled minute:second:hundredth, and the hundredth is always of the form multiple-of-10 plus either 0, 3 or 6. Presumably 3 and 6 are rounded from 3.33 and 6.66.
- Average rate of new track formation (see below): ~2.5/sec.
- First frame with cloud chamber tracks: 3:15:03, frame #5852.
- Last frame with cloud chamber tracks: 4:45:46, frame #8565.
- Notable anomalies: Tracks once formed seem to drift in a clockwise rotation.

## 5. Measurement

I downloaded the video of Reference [25] as an mp4 file onto a Dell laptop with screen resolution 1366x768, and manually stepped through frames 5852-8565 using Microsoft Video Editor in full-screen mode. Knowing the actual Petri dish dimensions, I was able to calibrate 0.16mm/pixel. Every time I encountered a new track that appeared to point away from the source, I put the cursor over what I thought was the track's likely starting point, left-clicked, and read out the click pixel coordinates using the application "MacroRecorder," [https://www.macrorecorder.com/] (and note that in the y direction, pixels count from the top). I stored the coordinates in the first tab of a spreadsheet (see Supplemental Material at [URL to be furnished] for spreadsheet contents). For every new track, the spreadsheet shows its video time, frame number, starting-point pixel coordinates, starting-point xy coordinates in mm relative to the radioactive source, and two-dimensional distance $R$ from the source. Each track is labeled short/long and diffuse/sharp based on eyeball judgements, as an aid to independent auditing. All tracks visible in this video are attributable to alpha particles (R. Schumacher, Carnegie Mellon University, private communication).

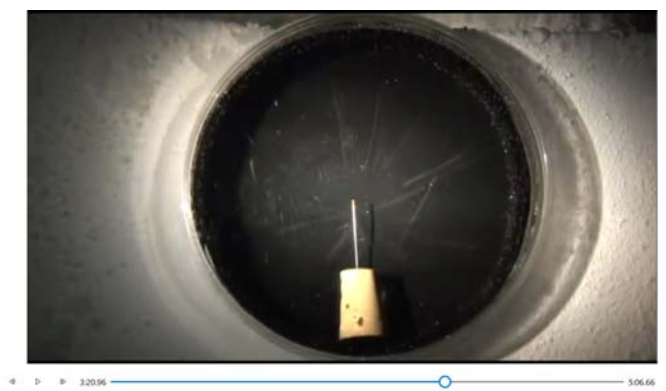

Figure 1: Frame #6030 from Reference [25].

Readers are invited to check track identification directly against the actual video. It will be seen that the first appearance of a track is often difficult to discern, because tracks start faint and then "develop" over time as their constituent droplets expand. When stepping through the video, it may be necessary to look over a short time period past the frame at which a new track is indicated, to see that the faint smudge at that location in that frame will in fact develop into a track. The reader will also see that track-origin identification is often made difficult and even ambiguous by the non-ideal placement of camera and illumination source.

In principle, one must distinguish between alpha and beta tracks because the theory in Section 2 applies to the alpha emitter $^{210}$Po but not to the beta emitters $^{210}$Pb and $^{210}$Bi. Since beta decay also involves a neutrino, it is emission of two particles with an energy spectrum for the outgoing charged particle, rather than emission of a single charged particle with a unique outgoing energy. However, beta tracks are too tenuous to be seen in this video (R. Schumacher, Carnegie Mellon University, private communication).

To compare with the results in Section 3, I first put together a cumulative distribution. I sort the spreadsheet on the values of $R$ for alpha tracks, then number the values starting at the smallest, and finally graph the pairs [$R$, sort index] on a scatter plot. In Figures 2a and 2b, I compare this (blue) directly with vertically scaled graphs (red) of Equation (3.2) with $b$=3.5mm, and $a$=2.25mm or 4mm. For $a$=2.25mm (when the radioactive source lies above the sensitive layer) it also seems prudent to at least consider the possibility that all tracks actually originate at the radioactive source, but become visible only when they cross the roof of the sensitive layer. For this reason, Figure 2a also includes a vertically scaled cumulative radial distribution (purple) of points at which rays that exit the source cross the roof of the sensitive layer; the match to raw data is poor. I choose the overall scale of the red curve to give the best match over the broadest range of R values, and the vertical scale of the purple curve to best approximate the apparent asymptotic behavior of the raw data.

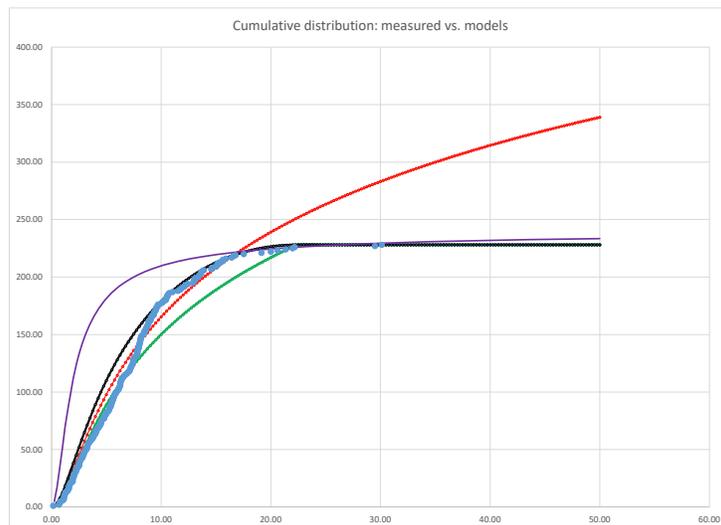

Figure 2a: Measured (blue) vs. non-cutoff (red), cutoff (green and black), and sensitive-layer-roof-crossing (purple) theoretical cumulative radial distributions (number of counts), as functions of $R$ (in mm). The model curves assume $b$=2.25mm. Vertical scales of green and black points are set so that at the largest $R$ (=50mm, edge of the Petri dish) they match the blue scatter.

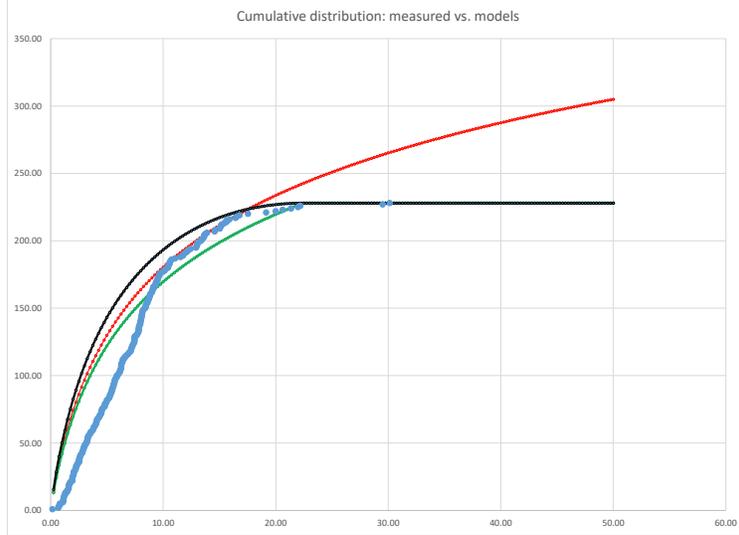

Figure 2b: Measured (blue) vs. non-cutoff (red) and cutoff (green and black) theoretical cumulative radial distributions (number of counts), as functions of $R$ (in mm). The model curves assume $b$=4mm. Vertical scales of green and black points are set so that at the largest $R$ (=50mm, edge of the Petri dish) they match the blue scatter. The roof-crossing curve is absent because in this case the needlepoint is inside the sensitive layer.

The blue and red data sets line up, to varying degrees, until the measured curve appears to turn over and prematurely flatten between about $R$=17mm and 22.5mm. This means a shortfall in observed detections that originate at – rather than just pass through – large $R$. (According to Reference [2], track origination requires the presence of a vapor droplet that is sub-critical but just barely so; track continuation does not). This could mean various things: (i) The theory in Reference [2] could be wrong. (ii) The dearth of track originations for $R$>22.5mm could be a temperature periphery effect: Maybe the Petri dish is warmer at its extremities, making it harder for subcritical vapor droplets to form spontaneously by thermal fluctuation. The volume outside $R$=22.5mm is about 80% of the entire Petri dish, straining the concept of "extremity," but a more careful analysis, beyond the scope of this paper, seems warranted. (iii) The dearth of track originations for $R$>22.5mm could be a corollary of the simple fact that 22.5mm is similar in order of magnitude to the stopping distance for a 5.3 MeV alpha particle in air [28]. But coincidence seems a much more likely explanation for this similarity, because track origination and termination seem like such radically different phenomena. (iv) The dearth of track originations for $R$>22.5mm could be related to the fact the 22.5mm and the Petri dish depth, 15mm, are of similar orders of magnitude; but I can't come up with a physical mechanism that plays on that similarity. (v) The dearth of track originations for $R$>22.5 could be related to enclosure-induced boundary conditions whose neglect we mentioned in Section 2, but it's hard to see how that could reduce the wavefunction by even as much as a single order of magnitude (this is the remark referred in the paragraph containing Equation (2.1)). (vi) Perhaps Equations (2.1) and (2.2) break down for large $R$ because of the cumulative impact of multiple small-cross-section interactions, as first discussed at the end of Section 2. To begin to quantify this possibility, imagine a circular atomic cross section of diameter 1Å and use the atomic (not molecular) density of air $2 \times 10^{27} \text{m}^{-3}$. Then in 22.5mm, a cone formed in the wake of an alpha-atom interaction encounters (22.5mm)x($\pi/4$ Å$^2$)x($2 \times 10^{27}\text{m}^{-3}$)~$3 \times 10^5$ such obstacles. So, referring back to the end of Section 2, the overall opening angle expands by random walk to $(\lambda/(1\text{Å}))\times(3\times10^5)^{1/2}$. Using $\lambda=6\times10^{-5}\text{Å}$ for a 5.3 MeV alpha particle, this is ~0.03 radian ~ 2°. This does not seem like a large number, but further study could show otherwise.

The video itself seems at odds with the idea that the needle point lies above the sensitive layer. Otherwise, the apparent length of tracks would be bounded above by some small multiplier times the distance from source to track origin (3.5/1.25~3 if $b$=3.5mm and $a$=2.25mm), but this isn't seen. So I'm inclined toward $a$=4mm as closer to the truth.

Perhaps the extreme shortfall of track originations beyond $R$=22.5mm is actually a breakdown of the Born rule within the framework of Reference [2]: Maybe, beyond $R$=22.5mm, the alpha wavefunction in Equation (2.1) is so attenuated that the chamber simply runs out of subcritical droplets close enough to critical for such a weak wavefunction to push into visibility. One way to test this would be to use a different alpha emitter with roughly 1/5 the half-life of $^{210}$Po. Then the factor $\gamma$ in Equation (2.1) would guarantee that the wavefunction wouldn't get weak enough to fail to start tracks until $R$ increased by a factor of ~2.2, from 22.5mm to beyond the full 50mm radius of the dish, and then track originations would be seen throughout the chamber. (This assumes that $t$ in Equation (2.1) is small enough that the factor $e^{-\gamma t}$ is O(1) for both the $^{210}$Po of Reference [25] and the hypothetical comparison alpha emitter.) Alternatively, one could use the inexpensive alpha emitter $^{241}$Am with roughly 1,000 times the half-life of $^{210}$Po. Then no track origination might be observed beyond $22.5/(1,000)^{1/2}$ ~ 2/3mm.

Violation of Born's rule here wouldn't necessarily clash with prior supporting evidence in its favor, because the wavefunction in this case is so tiny. According to Equation (2.1) (ignoring the exponential factor and using $Y=1/(4\pi)^{1/2}$), the value of |wavefunction|$^2$ at $R$=22.5mm is approximately $10^{-13}\text{m}^{-3} \sim (2\times10^4\text{m})^{-3}$. The distance $2\times10^4$m is vast compared to the length scales presumably characteristic of the wavefunction in any laboratory double-slit experiment I'm aware of. Nevertheless, even though this |wavefunction|$^2$ is very small numerically, its phenomenological impact here – between 80 and 110 track starts missing beyond 22.5mm – is much too large to dismiss as observational noise. In any case, the phenomenology in this paper and reference [2] is very particular to cloud chambers, so it remains to be seen how or even if it relates to prior searches for Born rule violations in other physical situations. These include, for example, Reference [30], which places an experimental limit on Born rule violation arising from hypothetical cubic interference terms in qutrit measurement.

One might have expected a track-origination deficit at *small R* because the radioactive source generates heat that can suppress vapor supersaturation. That might be seen in Figure 2b; but then again maybe not, because the source may be too small to generate consequential heat. Indeed, numerical calculations (unpublished) suggest that an apparent deficit at small radius is more likely an artifact that comes about because the radioactive atoms extend a few mm away from the end of the needle.

## 6. Cutoff models

The rollover in the blue data points suggests one modify the Born rule so that probability cuts off below some value of $|\psi|^2$ (that may depend on details of the measuring apparatus). This would be like scaling quantum efficiency by a wavefunction-dependent factor that is unity when wavefunction is sufficiently large, and zero when sufficiently small. I illustrate two speculative modifications in Figure 3. The green trace corresponds to a naïve hard cutoff. The black trace corresponds to an offset Born rule, i.e. probability proportional to max($|\psi|^2$-cutoff,0).

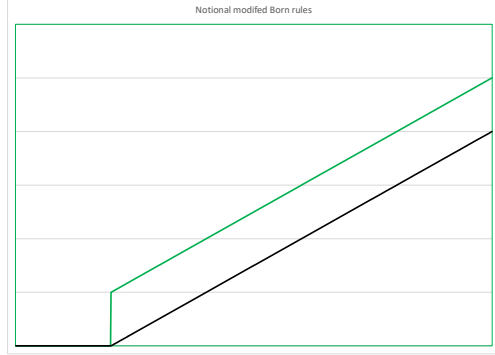

Figure 3: Notional modifications of Born rule. Horizontal axis represents |wavefunction|$^2$ and vertical axis represents probability density. Units are arbitrary. The wavefunction value at the cutoff would depend on particulars of the measurement system.

The hard cutoff is an arbitrary ansatz, but the offset model has a physical motivation arising from the underlying discreteness of vapor droplets made of finitely many distinct molecules. To see this, start from the development in Reference [2]: I argued that a cloud chamber track originates near a subcritical vapor droplet of radius $R_d$ when the following inequality holds ($R_c$ is critical radius)

$$Av\tau|\psi|^2 > R_c - R_d. \qquad (6.1)$$

I introduced $\rho$, the probability of droplet formation per unit time, unit volume and unit interval in $R_d$, and then obtained Equation (2.2) for the probability that Inequality (6.1) happens per unit time and unit volume. Now suppose that, because of molecular discreteness, $R_d$ can't actually get closer to $R_c$ than some minimum $\delta$. Then Inequality (6.1) is refined to

$$Av\tau|\psi|^2 > R_c - R_d > \delta, \qquad (6.2)$$

and the probability per unit time and unit volume must become

$$P(\mathbf{x},t) \sim \rho Av\tau |\psi(\mathbf{x},t)|^2 - \rho\delta = \rho Av\tau \left(|\psi(\mathbf{x},t)|^2 - \frac{\delta}{Av\tau}\right) \qquad (6.3)$$

when the quantity in parentheses is greater than zero, and zero otherwise, i.e. the offset model with cutoff $\delta/Av\tau$.

It is easy to derive modified cumulative distributions for the models in Figure 3. Suppose that, for either model, the cutoff is the value of $|\psi|^2$ corresponding to Equation (2.1) at $\|\mathbf{x}\|=K$ for some distance $K$ (and assume, in Equation (2.1), that we can ignore any spread in the factor exp(-$\gamma t$/2) among the $^{210}$Po nuclei that are generated by decay of Reference [25]'s initial $^{210}$Pb sample). Then, for $K>b$ and $>(a-b)$, a hard cutoff turns Equation (3.2) into

$$\int_{-b}^{a-b} dz \int_0^{\min(R,\sqrt{K^2-z^2})} \left(\frac{1}{r^2+z^2}\right) 2\pi r\, dr = C_1(R,b,K) + C_1(R,a-b,K), \qquad (6.4)$$

where

$$C_1(R,b,K) = C(R,b) \text{ for } R < \sqrt{K^2-b^2},$$

$$= 2\pi b\left[\ln\left|\frac{K}{b}\right|+1\right] \text{ for } R > K,$$

$$=2\pi\left\{R\cdot Arctan\left(\frac{\sqrt{K^2-R^2}}{R}\right) + b\left[\ln\left|\frac{K}{b}\right|+1\right] - \sqrt{K^2-R^2}\right\} \text{ otherwise} \qquad (6.5)$$

An offset turns Equation (3.2) into

$$\int_{-b}^{a-b} dz \int_0^{\min(R,\sqrt{K^2-z^2})} \left[\frac{1}{r^2+z^2} - \frac{1}{K^2}\right] 2\pi r\, dr = C_2(R,b,K) + C_2(R,a-b,K), \qquad (6.6)$$

where

$$C_1(R,b,K) - C_2(R,b,K) = \frac{\pi b R^2}{K^2} \text{ for } R < \sqrt{K^2-b^2},$$

$$= \pi b\left(1 - \frac{b^2}{3K^2}\right) \text{ for } R > K,$$

$$= \pi\left\{b\left(1 - \frac{b^2}{3K^2}\right) - \frac{2}{3}\frac{(K^2-R^2)^{\frac{3}{2}}}{K^2}sgn(b)\right\} \text{ otherwise.} \qquad (6.7)$$

The green and black curves in Figure 2 are the same as Equations (6.4) and (6.6), respectively, for $K=22.5$mm, scaled so that they match the measured data at the largest value of $R$. It is notable that, for $b=2.25$mm, the offset model appears to provide a more faithful match to the data, at least beyond 8mm. For $b=4$mm, the data doesn't seem to favor one model over the other.

If cutoff models like these apply to detectors beyond cloud chambers, it could have profound implications for detection of extremely rare processes such as proton decay. True probabilities of detection could be orders of magnitude *smaller* than expected from naïve Born rule arguments, and therefore today's accepted bounds on the proton lifetime could be mistakenly *long* by orders of magnitude.

## 7. Concluding remarks

Using opportunistic data, I have carried out a proof of principle for a previously unappreciated, low-cost probe of quantum mechanics fundamentals. It is also a novel tabletop example of a wavefunction with a signature visible to the naked eye without cryogenic equipment (e.g., superconductivity) or micron-scale detectors and finely controlled coherence (e.g., double-slit interference). Aspects of the data in this paper may already show violation of a position-space Born rule, although other explanations are not ruled out.

The ideal experiment tailored to this science would address the caveats identified in Section 4. It would also include the following improvements:
- Several interchangeable alpha emitters with half-lives spread over a few orders of magnitude.
- Automated track detection and classification, to eliminate reliance on eyeball judgement.
- Track detection in 3D, and sufficient statistics to measure three-dimensional distributions.
- Sufficient control and instrumentation to see deviations from Born rule (if we have not seen them here already for $R>22.5$mm).
- Sufficient instrumentation to probe dynamics of individual droplet formation.

**Acknowledgements**

I am grateful to Steve Gagnon (Jefferson Laboratory), Reinhard Schumacher (Carnegie-Mellon University), Roger Stevens (Spectrum Techniques) and Frank Taylor (MIT) for helpful correspondence, and to the reviewers for helpful comments.

**Author contributions**

The single author of this paper is solely responsible for its content.

**Competing interests**

The author declares no competing interests.

| TAB #1: VIDEO DATA | | | | | | | | | | | | |
|---|---|---|---|---|---|---|---|---|---|---|---|---|
| Comment | Min | Sec | 1/100 | Frame # | Track length | Track thickness | X (pixels) | Y (pixels) | X source offset (mm) | Y source offset (mm) | R | |
| CALIBRATION | | | | | | | | | | | | |
| Pinpoint (source) | 3 | 28 | 3 | 6242 | N/A | N/A | 703 | 345 | | | | |
| Boundary below point | 3 | 28 | 3 | 6242 | N/A | N/A | 705 | 631 | | | | |
| Boundary above point | 3 | 28 | 3 | 6242 | N/A | N/A | 699 | 6 | | | Diameter = | 625 | pixels |
| Boundary left of point | 3 | 28 | 3 | 6242 | N/A | N/A | 405 | 340 | | | Pixel = | 0.16 | mm |
| Boundary right of point | 3 | 28 | 3 | 6242 | N/A | N/A | 1049 | 342 | | | Height (a) = | 2.25 | mm |
| Last data frame | 4 | 45 | 46 | 8565 | N/A | N/A | N/A | N/A | | | Height (b) = | 3.5 | mm |
| First data frame | 3 | 15 | 3 | 5852 | N/A | N/A | N/A | N/A | | | | |
| | | | | | | | | | | | | |
| TRACK STARTS | | | | | | | | | | | | |
| | 3 | 15 | 20 | 5857 | Long | Sharp | 731 | 393 | 4.48 | -7.68 | 8.89 | |
| | 3 | 15 | 20 | 5857 | Short | Diffuse | 735 | 424 | 5.12 | -12.64 | 13.64 | |
| | 3 | 15 | 56 | 5868 | Short | Diffuse | 757 | 285 | 8.64 | 9.6 | 12.92 | |
| | 3 | 15 | 70 | 5872 | Short | Diffuse | 762 | 286 | 9.44 | 9.44 | 13.35 | |
| | 3 | 15 | 93 | 5879 | Long | Diffuse | 827 | 380 | 19.84 | -5.6 | 20.62 | |
| | 3 | 16 | 23 | 5888 | Short | Sharp | 619 | 404 | -13.44 | -9.44 | 16.42 | |
| | 3 | 16 | 30 | 5890 | Long | Sharp | 712 | 382 | 1.44 | -5.92 | 6.09 | |
| | 3 | 16 | 33 | 5891 | Short | Diffuse | 729 | 370 | 4.16 | -4 | 5.77 | |
| | 3 | 16 | 43 | 5894 | Long | Diffuse | 788 | 334 | 13.6 | 1.76 | 13.71 | |
| | 3 | 16 | 53 | 5897 | Short | Diffuse | 608 | 304 | -15.2 | 6.56 | 16.56 | |
| | 3 | 17 | 20 | 5917 | Long | Diffuse | 758 | 285 | 8.8 | 9.6 | 13.02 | |
| | 3 | 17 | 26 | 5919 | Short | Diffuse | 653 | 276 | -8 | 11.04 | 13.63 | |
| | 3 | 17 | 46 | 5925 | Short | Diffuse | 735 | 390 | 5.12 | -7.2 | 8.83 | |
| | 3 | 17 | 56 | 5928 | Short | Diffuse | 726 | 371 | 3.68 | -4.16 | 5.55 | |
| | 3 | 17 | 56 | 5928 | Short | Sharp | 727 | 376 | 3.84 | -4.96 | 6.27 | |
| | 3 | 17 | 86 | 5937 | Short | Sharp | 789 | 376 | 13.76 | -4.96 | 14.63 | |
| | 3 | 17 | 90 | 5938 | Long | Sharp | 700 | 278 | -0.48 | 10.72 | 10.73 | |
| | 3 | 18 | 26 | 5949 | Long | Sharp | 769 | 292 | 10.56 | 8.48 | 13.54 | |
| | 3 | 18 | 43 | 5954 | Long | Sharp | 759 | 352 | 8.96 | -1.12 | 9.03 | |
| | 3 | 18 | 66 | 5961 | Long | Sharp | 818 | 312 | 18.4 | 5.28 | 19.14 | |
| | 3 | 18 | 93 | 5969 | Short | Sharp | 729 | 295 | 4.16 | 8 | 9.02 | |
| | 3 | 19 | 26 | 5979 | Short | Diffuse | 674 | 384 | -4.64 | -6.24 | 7.78 | |

| | | | | | | | | | | | |
|---|---|---|---|---|---|---|---|---|---|---|---|
| | 3 | 19 | 60 | 5989 | Short | Diffuse | 769 | 397 | 10.56 | -8.32 | 13.44 | |
| | 3 | 19 | 93 | 5999 | Long | Sharp | 720 | 289 | 2.72 | 8.96 | 9.36 | |
| | 3 | 20 | 43 | 6014 | Long | Sharp | 686 | 287 | -2.72 | 9.28 | 9.67 | |
| | 3 | 21 | 3 | 6032 | Short | Diffuse | 717 | 381 | 2.24 | -5.76 | 6.18 | |
| | 3 | 21 | 23 | 6038 | Long | Diffuse | 800 | 294 | 15.52 | 8.16 | 17.53 | |
| | 3 | 21 | 96 | 6060 | Short | Diffuse | 720 | 282 | 2.72 | 10.08 | 10.44 | |
| | 3 | 22 | 53 | 6077 | Short | Diffuse | 735 | 356 | 5.12 | -1.76 | 5.41 | |
| | 3 | 22 | 86 | 6087 | Long | Sharp | 715 | 280 | 1.92 | 10.4 | 10.58 | |
| | 3 | 23 | 23 | 6098 | Long | Sharp | 743 | 256 | 6.4 | 14.24 | 15.61 | |
| | 3 | 23 | 70 | 6112 | Short | Diffuse | 515 | 335 | -30.08 | 1.6 | 30.12 | |
| | 3 | 23 | 70 | 6112 | Long | Diffuse | 625 | 291 | -12.48 | 8.64 | 15.18 | |
| | 3 | 23 | 73 | 6113 | Long | Sharp | 653 | 375 | -8 | -4.8 | 9.33 | |
| | 3 | 24 | 43 | 6134 | Long | Sharp | 595 | 282 | -17.28 | 10.08 | 20.01 | |
| | 3 | 24 | 86 | 6147 | Long | Sharp | 718 | 297 | 2.4 | 7.68 | 8.05 | |
| | 3 | 24 | 86 | 6147 | Long | Sharp | 715 | 338 | 1.92 | 1.12 | 2.22 | |
| | 3 | 25 | 30 | 6160 | Short | Diffuse | 742 | 257 | 6.24 | 14.08 | 15.40 | |
| | 3 | 25 | 53 | 6167 | Long | Diffuse | 699 | 384 | -0.64 | -6.24 | 6.27 | |
| | 3 | 26 | 26 | 6189 | Long | Sharp | 725 | 306 | 3.52 | 6.24 | 7.16 | |
| | 3 | 26 | 53 | 6197 | Short | Diffuse | 741 | 368 | 6.08 | -3.68 | 7.11 | |
| | 3 | 27 | 23 | 6218 | Long | Sharp | 663 | 525 | -6.4 | -28.8 | 29.50 | |
| | 3 | 27 | 23 | 6218 | Long | Sharp | 687 | 418 | -2.56 | -11.68 | 11.96 | |
| | 3 | 27 | 53 | 6227 | Short | Diffuse | 673 | 384 | -4.8 | -6.24 | 7.87 | |
| | 3 | 27 | 53 | 6227 | Long | Sharp | 658 | 324 | -7.2 | 3.36 | 7.95 | |
| | 3 | 27 | 66 | 6231 | Long | Diffuse | 739 | 389 | 5.76 | -7.04 | 9.10 | |
| | 3 | 27 | 90 | 6238 | Short | Sharp | 758 | 322 | 8.8 | 3.68 | 9.54 | |
| | 3 | 28 | 0 | 6241 | Long | Sharp | 769 | 346 | 10.56 | -0.16 | 10.56 | |
| | 3 | 28 | 16 | 6246 | Long | Sharp | 702 | 306 | -0.16 | 6.24 | 6.24 | |
| | 3 | 28 | 63 | 6260 | Long | Sharp | 655 | 366 | -7.68 | -3.36 | 8.38 | |
| | 3 | 28 | 66 | 6261 | Short | Sharp | 734 | 320 | 4.96 | 4 | 6.37 | |
| | 3 | 29 | 6 | 6273 | Long | Sharp | 746 | 357 | 6.88 | -1.92 | 7.14 | |
| | 3 | 29 | 6 | 6273 | Long | Sharp | 767 | 371 | 10.24 | -4.16 | 11.05 | |
| | 3 | 29 | 23 | 6278 | Short | Diffuse | 656 | 317 | -7.52 | 4.48 | 8.75 | |
| | 3 | 29 | 60 | 6289 | Short | Diffuse | 723 | 349 | 3.2 | -0.64 | 3.26 | |
| | 3 | 30 | 23 | 6308 | Short | Diffuse | 703 | 310 | 0 | 5.6 | 5.60 | |
| | 3 | 30 | 23 | 6308 | Long | Sharp | 737 | 289 | 5.44 | 8.96 | 10.48 | |
| | 3 | 30 | 50 | 6316 | Long | Sharp | 748 | 367 | 7.2 | -3.52 | 8.01 | |
| | 3 | 30 | 56 | 6318 | Short | Diffuse | 680 | 385 | -3.68 | -6.4 | 7.38 | |
| | 3 | 31 | 10 | 6334 | Short | Diffuse | 669 | 291 | -5.44 | 8.64 | 10.21 | |
| | 3 | 31 | 10 | 6334 | Short | Diffuse | 655 | 314 | -7.68 | 4.96 | 9.14 | |
| | 3 | 32 | 0 | 6361 | Long | Sharp | 731 | 340 | 4.48 | 0.8 | 4.55 | |
| | 3 | 32 | 20 | 6367 | Short | Sharp | 641 | 305 | -9.92 | 6.4 | 11.81 | |
| | 3 | 32 | 83 | 6386 | Short | Diffuse | 686 | 303 | -2.72 | 6.72 | 7.25 | |
| | 3 | 32 | 96 | 6390 | Long | Sharp | 736 | 281 | 5.28 | 10.24 | 11.52 | |
| | 3 | 33 | 96 | 6420 | Long | Sharp | 723 | 286 | 3.2 | 9.44 | 9.97 | |
| | 3 | 34 | 10 | 6424 | Short | Sharp | 728 | 322 | 4 | 3.68 | 5.44 | |
| | 3 | 34 | 23 | 6428 | Short | Diffuse | 685 | 377 | -2.88 | -5.12 | 5.87 | |
| | 3 | 34 | 30 | 6430 | Long | Sharp | 686 | 315 | -2.72 | 4.8 | 5.52 | |
| | 3 | 34 | 33 | 6431 | Short | Diffuse | 705 | 310 | 0.32 | 5.6 | 5.61 | |
| | 3 | 34 | 63 | 6440 | Short | Sharp | 657 | 306 | -7.36 | 6.24 | 9.65 | |
| | 3 | 34 | 73 | 6443 | Long | Sharp | 640 | 302 | -10.08 | 6.88 | 12.20 | |
| | 3 | 35 | 20 | 6457 | Long | Sharp | 658 | 277 | -7.2 | 10.88 | 13.05 | |
| | 3 | 35 | 20 | 6457 | Long | Sharp | 743 | 319 | 6.4 | 4.16 | 7.63 | |
| | 3 | 35 | 96 | 6480 | Long | Sharp | 655 | 372 | -7.68 | -4.32 | 8.81 | |
| | 3 | 36 | 0 | 6481 | Short | Diffuse | 680 | 355 | -3.68 | -1.6 | 4.01 | |
| | 3 | 36 | 23 | 6488 | Long | Sharp | 739 | 266 | 5.76 | 12.64 | 13.89 | |
| | 3 | 36 | 36 | 6492 | Short | Diffuse | 698 | 410 | -0.8 | -10.4 | 10.43 | |
| | 3 | 36 | 43 | 6494 | Short | Diffuse | 748 | 368 | 7.2 | -3.68 | 8.09 | |
| | 3 | 36 | 56 | 6498 | Short | Diffuse | 762 | 368 | 9.44 | -3.68 | 10.13 | |
| | 3 | 36 | 90 | 6508 | Long | Sharp | 657 | 364 | -7.36 | -3.04 | 7.96 | |
| | 3 | 37 | 30 | 6520 | Long | Sharp | 732 | 357 | 4.64 | -1.92 | 5.02 | |
| | 3 | 37 | 60 | 6529 | Long | Sharp | 688 | 318 | -2.4 | 4.32 | 4.94 | |
| | 3 | 37 | 73 | 6533 | Long | Sharp | 661 | 374 | -6.72 | -4.64 | 8.17 | |
| | 3 | 37 | 80 | 6535 | Short | Sharp | 706 | 323 | 0.48 | 3.52 | 3.55 | |
| | 3 | 38 | 26 | 6549 | Long | Sharp | 723 | 332 | 3.2 | 2.08 | 3.82 | |
| | 3 | 38 | 30 | 6550 | Long | Sharp | 746 | 304 | 6.88 | 6.56 | 9.51 | |
| | 3 | 39 | 23 | 6578 | Long | Sharp | 738 | 351 | 5.6 | -0.96 | 5.68 | |
| | 3 | 39 | 30 | 6580 | Long | Sharp | 735 | 357 | 5.12 | -1.92 | 5.47 | |
| | 3 | 39 | 30 | 6580 | Short | Diffuse | 700 | 348 | -0.48 | -0.48 | 0.68 | |
| | 3 | 39 | 70 | 6592 | Long | Sharp | 726 | 318 | 3.68 | 4.32 | 5.67 | |
| | 3 | 40 | 23 | 6608 | Long | Sharp | 734 | 356 | 4.96 | -1.76 | 5.26 | |
| | 3 | 40 | 23 | 6608 | Short | Diffuse | 680 | 352 | -3.68 | -1.12 | 3.85 | |
| | 3 | 40 | 53 | 6617 | Long | Sharp | 676 | 325 | -4.32 | 3.2 | 5.38 | |
| | 3 | 40 | 93 | 6629 | Long | Diffuse | 746 | 318 | 6.88 | 4.32 | 8.12 | |
| | 3 | 41 | 46 | 6645 | Long | Sharp | 715 | 331 | 1.92 | 2.24 | 2.95 | |
| | 3 | 41 | 70 | 6652 | Long | Sharp | 671 | 361 | -5.12 | -2.56 | 5.72 | |
| | 3 | 42 | 36 | 6672 | Short | Diffuse | 745 | 365 | 6.72 | -3.2 | 7.44 | |
| | 3 | 43 | 23 | 6698 | Short | Sharp | 710 | 349 | 1.12 | -0.64 | 1.29 | |
| | 3 | 43 | 73 | 6713 | Long | Sharp | 754 | 281 | 8.16 | 10.24 | 13.09 | |
| | 3 | 43 | 93 | 6719 | Short | Diffuse | 715 | 336 | 1.92 | 1.44 | 2.40 | |
| | 3 | 43 | 96 | 6720 | Short | Sharp | 704 | 338 | 0.16 | 1.12 | 1.13 | |
| | 3 | 44 | 13 | 6725 | Long | Sharp | 663 | 348 | -6.4 | -0.48 | 6.42 | |
| | 3 | 44 | 16 | 6726 | Long | Sharp | 662 | 345 | -6.56 | 0 | 6.56 | |
| | 3 | 44 | 33 | 6731 | Short | Sharp | 688 | 331 | -2.4 | 2.24 | 3.28 | |
| | 3 | 44 | 46 | 6735 | Long | Sharp | 694 | 297 | -1.44 | 7.68 | 7.81 | |
| | 3 | 44 | 90 | 6748 | Long | Sharp | 663 | 324 | -6.4 | 3.36 | 7.23 | |
| | 3 | 45 | 23 | 6758 | Short | Sharp | 711 | 354 | 1.28 | -1.44 | 1.93 | |
| | 3 | 45 | 46 | 6765 | Long | Sharp | 694 | 337 | -1.44 | 1.28 | 1.93 | |
| | 3 | 46 | 36 | 6792 | Long | Sharp | 748 | 337 | 7.2 | 1.28 | 7.31 | |
| | 3 | 47 | 23 | 6818 | Short | Diffuse | 781 | 367 | 12.48 | -3.52 | 12.97 | |
| | 3 | 47 | 46 | 6825 | Long | Sharp | 712 | 332 | 1.44 | 2.08 | 2.53 | |
| | 3 | 47 | 80 | 6835 | Long | Sharp | 746 | 346 | 6.88 | -0.16 | 6.88 | |
| | 3 | 47 | 86 | 6837 | Long | Sharp | 662 | 309 | -6.56 | 5.76 | 8.73 | |
| | 3 | 47 | 96 | 6840 | Long | Sharp | 704 | 315 | 0.16 | 4.8 | 4.80 | |
| | 3 | 48 | 3 | 6842 | Short | Sharp | 757 | 342 | 8.64 | 0.48 | 8.65 | |
| | 3 | 48 | 40 | 6853 | Short | Diffuse | 748 | 296 | 7.2 | 7.84 | 10.64 | |
| | 3 | 48 | 76 | 6864 | Short | Sharp | 699 | 342 | -0.64 | 0.48 | 0.80 | |
| | 2 | 49 | 3 | 5072 | Long | Sharp | 682 | 326 | -3.36 | 3.04 | 4.53 | |
| | 3 | 49 | 23 | 6878 | Short | Diffuse | 730 | 351 | 4.32 | -0.96 | 4.43 | |
| | 3 | 50 | 3 | 6902 | Long | Sharp | 742 | 341 | 6.24 | 0.64 | 6.27 | |
| | 3 | 50 | 16 | 6906 | Short | Diffuse | 605 | 329 | -15.68 | 2.56 | 15.89 | |
| | 3 | 50 | 20 | 6907 | Short | Diffuse | 569 | 315 | -21.44 | 4.8 | 21.97 | |
| | 3 | 50 | 30 | 6910 | Short | Diffuse | 610 | 331 | -14.88 | 2.24 | 15.05 | |
| | 3 | 50 | 46 | 6915 | Short | Sharp | 682 | 332 | -3.36 | 2.08 | 3.95 | |
| | 3 | 50 | 80 | 6925 | Long | Sharp | 704 | 319 | 0.16 | 4.16 | 4.16 | |
| | 3 | 51 | 20 | 6937 | Long | Diffuse | 713 | 345 | 1.6 | 0 | 1.60 | |
| | 3 | 51 | 66 | 6951 | Short | Sharp | 713 | 345 | 1.6 | 0 | 1.60 | |
| | 3 | 52 | 36 | 6972 | Short | Sharp | 721 | 335 | 2.88 | 1.6 | 3.29 | |
| | 3 | 52 | 70 | 6982 | Short | Diffuse | 694 | 369 | -1.44 | -3.84 | 4.10 | |

| | | | | | | | | | | | |
|---|---|---|---|---|---|---|---|---|---|---|---|
| 3 | 53 | 6 | 6993 | Short | Diffuse | 752 | 344 | 7.84 | 0.16 | 7.84 | |
| 3 | 53 | 26 | 6999 | Short | Sharp | 685 | 325 | -2.88 | 3.2 | 4.31 | |
| 3 | 53 | 43 | 7004 | Short | Diffuse | 712 | 336 | 1.44 | 1.44 | 2.04 | |
| 3 | 55 | 3 | 7052 | Long | Sharp | 675 | 361 | -4.48 | -2.56 | 5.16 | |
| 3 | 55 | 56 | 7068 | Short | Diffuse | 726 | 346 | 3.68 | -0.16 | 3.68 | |
| 3 | 57 | 16 | 7116 | Short | Diffuse | 728 | 358 | 4 | -2.08 | 4.51 | |
| 3 | 57 | 43 | 7124 | Short | Diffuse | 729 | 362 | 4.16 | -2.72 | 4.97 | |
| 3 | 57 | 50 | 7126 | Short | Sharp | 721 | 354 | 2.88 | -1.44 | 3.22 | |
| 3 | 57 | 86 | 7137 | Long | Sharp | 688 | 358 | -2.4 | -2.08 | 3.18 | |
| 3 | 58 | 63 | 7160 | Long | Diffuse | 749 | 328 | 7.36 | 2.72 | 7.85 | |
| 3 | 59 | 60 | 7189 | Long | Sharp | 703 | 326 | 0 | 3.04 | 3.04 | |
| 3 | 59 | 90 | 7198 | Short | Diffuse | 717 | 337 | 2.24 | 1.28 | 2.58 | |
| 4 | 0 | 56 | 7218 | Short | Diffuse | 701 | 359 | -0.32 | -2.24 | 2.26 | |
| 4 | 1 | 40 | 7243 | Long | Sharp | 666 | 310 | -5.92 | 5.6 | 8.15 | |
| 4 | 2 | 0 | 7261 | Long | Sharp | 675 | 370 | -4.48 | -4 | 6.01 | |
| 4 | 2 | 56 | 7278 | Long | Sharp | 661 | 356 | -6.72 | -1.76 | 6.95 | |
| 4 | 2 | 86 | 7287 | Short | Diffuse | 746 | 363 | 6.88 | -2.88 | 7.46 | |
| 4 | 3 | 10 | 7294 | Short | Sharp | 723 | 339 | 3.2 | 0.96 | 3.34 | |
| 4 | 3 | 53 | 7307 | Short | Diffuse | 726 | 350 | 3.68 | -0.8 | 3.77 | |
| 4 | 3 | 73 | 7313 | Short | Sharp | 725 | 346 | 3.52 | -0.16 | 3.52 | |
| 4 | 4 | 33 | 7331 | Short | Diffuse | 705 | 336 | 0.32 | 1.44 | 1.48 | |
| 4 | 5 | 70 | 7372 | Short | Sharp | 705 | 336 | 0.32 | 1.44 | 1.48 | |
| 4 | 6 | 26 | 7389 | Long | Sharp | 755 | 336 | 8.32 | 1.44 | 8.44 | |
| 4 | 6 | 30 | 7390 | Long | Sharp | 753 | 337 | 8 | 1.28 | 8.10 | |
| 4 | 6 | 46 | 7395 | Short | Diffuse | 705 | 352 | 0.32 | -1.12 | 1.16 | |
| 4 | 6 | 56 | 7398 | Short | Diffuse | 741 | 355 | 6.08 | -1.6 | 6.29 | |
| 4 | 7 | 6 | 7413 | Long | Sharp | 687 | 322 | -2.56 | 3.68 | 4.48 | |
| 4 | 7 | 13 | 7415 | Long | Diffuse | 749 | 338 | 7.36 | 1.12 | 7.44 | |
| 4 | 8 | 0 | 7441 | Long | Sharp | 714 | 296 | 1.76 | 7.84 | 8.04 | |
| 4 | 8 | 66 | 7461 | Short | Sharp | 701 | 349 | -0.32 | -0.64 | 0.72 | |
| 4 | 9 | 0 | 7471 | Long | Sharp | 732 | 360 | 4.64 | -2.4 | 5.22 | |
| 4 | 9 | 33 | 7481 | Long | Sharp | 676 | 364 | -4.32 | -3.04 | 5.28 | |
| 4 | 9 | 83 | 7496 | Short | Diffuse | 700 | 351 | -0.48 | -0.96 | 1.07 | |
| 4 | 10 | 3 | 7502 | Long | Sharp | 658 | 317 | -7.2 | 4.48 | 8.48 | |
| 4 | 10 | 26 | 7509 | Long | Sharp | 714 | 340 | 1.76 | 0.8 | 1.93 | |
| 4 | 11 | 0 | 7531 | Short | Diffuse | 738 | 356 | 5.6 | -1.76 | 5.87 | |
| | 4 | 11 | 6 | 7533 | Long | Sharp | 662 | 350 | -6.56 | -0.8 | 6.61 | |
| | 4 | 11 | 76 | 7554 | Long | Sharp | 704 | 345 | 0.16 | 0 | 0.16 | |
| | 4 | 11 | 96 | 7560 | Long | Sharp | 638 | 306 | -10.4 | 6.24 | 12.13 | |
| | 4 | 12 | 6 | 7563 | Long | Sharp | 742 | 382 | 6.24 | -5.92 | 8.60 | |
| | 4 | 12 | 60 | 7579 | Long | Sharp | 692 | 317 | -1.76 | 4.48 | 4.81 | |
| | 4 | 13 | 23 | 7598 | Short | Diffuse | 704 | 338 | 0.16 | 1.12 | 1.13 | |
| | 4 | 13 | 93 | 7619 | Long | Sharp | 709 | 329 | 0.96 | 2.56 | 2.73 | |
| | 4 | 14 | 23 | 7628 | Long | Sharp | 675 | 334 | -4.48 | 1.76 | 4.81 | |
| | 4 | 14 | 26 | 7629 | Long | Sharp | 699 | 338 | -0.64 | 1.12 | 1.29 | |
| | 4 | 14 | 50 | 7636 | Long | Sharp | 661 | 325 | -6.72 | 3.2 | 7.44 | |
| | 4 | 14 | 63 | 7640 | Short | Sharp | 644 | 272 | -9.44 | 11.68 | 15.02 | |
| | 4 | 15 | 26 | 7668 | Long | Diffuse | 693 | 344 | -1.6 | 0.16 | 1.61 | |
| | 4 | 15 | 26 | 7673 | Long | Sharp | 710 | 330 | 1.12 | 2.4 | 2.65 | |
| | 4 | 15 | 56 | 7673 | Long | Diffuse | 722 | 347 | 3.04 | -0.32 | 3.06 | |
| | 4 | 15 | 73 | 7674 | Long | Sharp | 686 | 366 | -2.72 | -3.36 | 4.32 | |
| | 4 | 15 | 76 | 7679 | Short | Diffuse | 791 | 310 | 14.08 | 5.6 | 15.15 | |
| | 4 | 15 | 93 | 7679 | Long | Diffuse | 665 | 351 | -6.08 | -0.96 | 6.16 | |
| | 4 | 16 | 30 | 7690 | Long | Sharp | 708 | 331 | 0.8 | 2.24 | 2.38 | |
| | 4 | 17 | 23 | 7718 | Long | Sharp | 714 | 337 | 1.76 | 1.28 | 2.18 | |
| | 4 | 17 | 50 | 7726 | Short | Diffuse | 706 | 303 | 0.48 | 6.72 | 6.74 | |
| | 4 | 17 | 50 | 7726 | Short | Diffuse | 682 | 311 | -3.36 | 5.44 | 6.39 | |
| | 4 | 17 | 93 | 7739 | Long | Sharp | 654 | 316 | -7.84 | 4.64 | 9.11 | |
| | 4 | 18 | 53 | 7757 | Short | Diffuse | 626 | 236 | -12.32 | 17.44 | 21.35 | |
| | 4 | 19 | 36 | 7782 | Short | Diffuse | 710 | 337 | 1.12 | 1.28 | 1.70 | |
| | 4 | 19 | 46 | 7785 | Short | Diffuse | 713 | 337 | 1.6 | 1.28 | 2.05 | |
| | 4 | 20 | 20 | 7807 | Long | Sharp | 750 | 267 | 7.52 | 12.48 | 14.57 | |
| | 4 | 20 | 30 | 7810 | Long | Sharp | 733 | 278 | 4.8 | 10.72 | 11.75 | |
| | 4 | 21 | 86 | 7857 | Long | Sharp | 716 | 336 | 2.08 | 1.44 | 2.53 | |
| | 4 | 22 | 33 | 7871 | Short | Diffuse | 694 | 353 | -1.44 | -1.28 | 1.93 | |
| | 4 | 22 | 60 | 7879 | Short | Diffuse | 678 | 349 | -4 | -0.64 | 4.05 | |
| | 4 | 22 | 60 | 7879 | Short | Diffuse | 681 | 393 | -3.52 | -7.68 | 8.45 | |
| | 4 | 22 | 80 | 7885 | Short | Diffuse | 658 | 320 | -7.2 | 4 | 8.24 | |
| | 4 | 23 | 26 | 7899 | Short | Diffuse | 666 | 384 | -5.92 | -6.24 | 8.60 | |
| | 4 | 23 | 26 | 7899 | Long | Diffuse | 734 | 306 | 4.96 | 6.24 | 7.97 | |
| | 4 | 23 | 73 | 7913 | Long | Diffuse | 653 | 315 | -8 | 4.8 | 9.33 | |
| | 4 | 24 | 30 | 7930 | Short | Sharp | 698 | 330 | -0.8 | 2.4 | 2.53 | |

| | | | | | | | | | | | |
|---|---|---|---|---|---|---|---|---|---|---|---|
| | 4 | 24 | 43 | 7934 | Short | Diffuse | 702 | 327 | -0.16 | 2.88 | 2.88 |
| | 4 | 25 | 43 | 7964 | Short | Diffuse | 720 | 340 | 2.72 | 0.8 | 2.84 |
| | 4 | 25 | 73 | 7973 | Short | Sharp | 706 | 329 | 0.48 | 2.56 | 2.60 |
| | 4 | 26 | 40 | 7993 | Long | Diffuse | 567 | 372 | -21.76 | -4.32 | 22.18 |
| | 4 | 27 | 53 | 8027 | Long | Diffuse | 626 | 285 | -12.32 | 9.6 | 15.62 |
| | 4 | 27 | 76 | 8034 | Short | Sharp | 685 | 289 | -2.88 | 8.96 | 9.41 |
| | 4 | 28 | 43 | 8054 | Short | Sharp | 681 | 358 | -3.52 | -2.08 | 4.09 |
| | 4 | 28 | 50 | 8056 | Short | Sharp | 716 | 357 | 2.08 | -1.92 | 2.83 |
| | 4 | 29 | 10 | 8074 | Short | Diffuse | 710 | 338 | 1.12 | 1.12 | 1.58 |
| | 4 | 29 | 10 | 8074 | Long | Sharp | 716 | 345 | 2.08 | 0 | 2.08 |
| | 4 | 29 | 56 | 8088 | Short | Diffuse | 707 | 333 | 0.64 | 1.92 | 2.02 |
| | 4 | 31 | 66 | 8151 | Short | Sharp | 662 | 367 | -6.56 | -3.52 | 7.44 |
| | 4 | 32 | 50 | 8176 | Long | Sharp | 717 | 343 | 2.24 | 0.32 | 2.26 |
| | 4 | 33 | 6 | 8193 | Short | Diffuse | 606 | 305 | -15.52 | 6.4 | 16.79 |
| | 4 | 33 | 6 | 8193 | Long | Sharp | 661 | 320 | -6.72 | 4 | 7.82 |
| | 4 | 34 | 33 | 8231 | Long | Diffuse | 656 | 335 | -7.52 | 1.6 | 7.69 |
| | 4 | 35 | 0 | 8251 | Short | Sharp | 710 | 342 | 1.12 | 0.48 | 1.22 |
| | 4 | 35 | 23 | 8258 | Short | Diffuse | 644 | 359 | -9.44 | -2.24 | 9.70 |
| | 4 | 36 | 26 | 8289 | Short | Diffuse | 658 | 306 | -7.2 | 6.24 | 9.53 |
| | 4 | 36 | 30 | 8290 | Short | Sharp | 716 | 358 | 2.08 | -2.08 | 2.94 |
| | 4 | 41 | 16 | 8436 | Short | Diffuse | 712 | 351 | 1.44 | -0.96 | 1.73 |
| | 4 | 41 | 36 | 8442 | Short | Diffuse | 699 | 351 | -0.64 | -0.96 | 1.15 |
| | 4 | 41 | 73 | 8453 | Short | Diffuse | 627 | 361 | -12.16 | -2.56 | 12.43 |
| | 4 | 42 | 53 | 8477 | Long | Sharp | 724 | 341 | 3.36 | 0.64 | 3.42 |
| | 4 | 42 | 66 | 8481 | Short | Diffuse | 684 | 323 | -3.04 | 3.52 | 4.65 |
| | 4 | 44 | 36 | 8532 | Short | Sharp | 704 | 340 | 0.16 | 0.8 | 0.82 |

| | | | | | | | | |
|---|---|---|---|---|---|---|---|---|
| TAB #2: IDEAL DISTRIBUTION C(R,b)+C(R,a-b), DIVIDED BY PI | | | | | | | | |
| | | | | | | | | |
| PETRIE DISH SENSITIVE VOLUME PARAMETERS | | | | | | | | |
| radius=50mm | | | | | | | | |
| a= | 2.3 | mm | | | | | | |
| b= | 3.5 | mm | | | | | | |
| a-b= | -1 | mm | | | | | | |
| scale | 24 | | | | | | | |
| | | | | | | | | |
| n | R | R/b | Log term | Arctan term | Total b part | R/(a-b) | Log term | Arctan term | Total a-b part |
| 1 | 0.3 | 0.1 | 0.005089 | 0.214212695 | 0.767556174 | -0.2 | 0.039221 | 0.549360307 | -0.735726275 |
| 2 | 0.5 | 0.1 | 0.020203 | 0.408256935 | 1.499608748 | -0.4 | 0.14842 | 0.95223196 | -1.375814956 |
| 3 | 0.8 | 0.2 | 0.044895 | 0.582729854 | 2.19668811 | -0.6 | 0.307485 | 1.236452192 | -1.929921114 |
| 4 | 1 | 0.3 | 0.078472 | 0.738569524 | 2.85964399 | -0.8 | 0.494696 | 1.433688615 | -2.410481071 |
| 5 | 1.3 | 0.4 | 0.120048 | 0.876980276 | 3.489599114 | -1 | 0.693147 | 1.570796327 | -2.829929384 |
| 6 | 1.5 | 0.4 | 0.168623 | 0.999346749 | 4.087893115 | -1.2 | 0.891998 | 1.667371863 | -3.199212378 |
| 7 | 1.8 | 0.5 | 0.223144 | 1.107148718 | 4.656022942 | -1.4 | 1.085189 | 1.736698561 | -3.527359786 |
| 8 | 2 | 0.6 | 0.282567 | 1.201885957 | 5.195585251 | -1.6 | 1.269761 | 1.787517809 | -3.821597942 |
| 9 | 2.3 | 0.6 | 0.345903 | 1.285018518 | 5.708224776 | -1.8 | 1.444563 | 1.825554616 | -4.087647356 |
| 10 | 2.5 | 0.7 | 0.412245 | 1.357924058 | 6.195590987 | -2 | 1.609438 | 1.854590436 | -4.330035436 |
| 11 | 2.8 | 0.8 | 0.480787 | 1.421871141 | 6.659303893 | -2.2 | 1.764731 | 1.87716097 | -4.552364708 |
| 12 | 3 | 0.9 | 0.550831 | 1.478005808 | 7.100928682 | -2.4 | 1.911023 | 1.894997375 | -4.757525331 |
| 13 | 3.3 | 0.9 | 0.621783 | 1.527348233 | 7.521958245 | -2.6 | 2.048982 | 1.909303936 | -4.947857838 |
| 14 | 3.5 | 1 | 0.693147 | 1.570796327 | 7.923802276 | -2.8 | 2.179287 | 1.920934066 | -5.125276179 |
| 15 | 3.8 | 1.1 | 0.764518 | 1.609133705 | 8.307781579 | -3 | 2.302585 | 1.930503326 | -5.291360524 |
| 16 | 4 | 1.1 | 0.835568 | 1.643039999 | 8.675126319 | -3.2 | 2.419479 | 1.938463158 | -5.447427503 |
| 17 | 4.3 | 1.2 | 0.906034 | 1.673102086 | 9.026977111 | -3.4 | 2.530517 | 1.945149804 | -5.594583706 |
| 18 | 4.5 | 1.3 | 0.975714 | 1.699825291 | 9.364388053 | -3.6 | 2.636196 | 1.950817322 | -5.733766775 |
| 19 | 4.8 | 1.4 | 1.044451 | 1.723643996 | 9.688331019 | -3.8 | 2.736962 | 1.955660234 | -5.865777223 |
| 20 | 5 | 1.4 | 1.112126 | 1.744931327 | 9.999700671 | -4 | 2.833213 | 1.959829305 | -5.991303311 |
| 21 | 5.3 | 1.5 | 1.178655 | 1.764007811 | 10.29931982 | -4.2 | 2.92531 | 1.963442719 | -6.110940661 |
| 22 | 5.5 | 1.6 | 1.243978 | 1.781148969 | 10.58794488 | -4.4 | 3.013572 | 1.96659409 | -6.225207852 |
| 23 | 5.8 | 1.6 | 1.308057 | 1.796591906 | 10.86627115 | -4.6 | 3.098289 | 1.969358289 | -6.334558938 |
| 24 | 6 | 1.7 | 1.37087 | 1.810540966 | 11.134938 | -4.8 | 3.179719 | 1.971795736 | -6.439393557 |
| 25 | 6.3 | 1.8 | 1.432408 | 1.823172578 | 11.3945336 | -5 | 3.258097 | 1.973955598 | -6.540065171 |
| 26 | 6.5 | 1.9 | 1.492675 | 1.83463937 | 11.64559947 | -5.2 | 3.333632 | 1.975878194 | -6.636887821 |
| 27 | 6.8 | 1.9 | 1.551679 | 1.845073664 | 11.8886346 | -5.4 | 3.406517 | 1.977596826 | -6.730141712 |
| 28 | 7 | 2 | 1.609438 | 1.854590436 | 12.12409922 | -5.6 | 3.476923 | 1.979139188 | -6.82007785 |
| 29 | 7.3 | 2.1 | 1.665973 | 1.863289826 | 12.35241831 | -5.8 | 3.545009 | 1.980528453 | -6.906921921 |
| 30 | 7.5 | 2.1 | 1.721308 | 1.871259256 | 12.57398473 | -6 | 3.610918 | 1.981784129 | -6.990877552 |
| 31 | 7.8 | 2.2 | 1.775471 | 1.878575235 | 12.78916206 | -6.2 | 3.674781 | 1.982922727 | -7.072129071 |
| 32 | 8 | 2.3 | 1.828491 | 1.885304876 | 12.99828724 | -6.4 | 3.736717 | 1.983958297 | -7.15084385 |
| 33 | 8.3 | 2.4 | 1.880399 | 1.891507195 | 13.20167287 | -6.6 | 3.796837 | 1.98490285 | -7.227174307 |
| 34 | 8.5 | 2.4 | 1.931226 | 1.897234212 | 13.39960933 | -6.8 | 3.855241 | 1.985766705 | -7.30125962 |
| 35 | 8.8 | 2.5 | 1.981001 | 1.902531886 | 13.59236674 | -7 | 3.912023 | 1.986558764 | -7.373227212 |
| 36 | 9 | 2.6 | 2.029758 | 1.907440914 | 13.78019662 | -7.2 | 3.967268 | 1.987286741 | -7.443194025 |
| 37 | 9.3 | 2.6 | 2.077526 | 1.911997419 | 13.96333352 | -7.4 | 4.021057 | 1.987957341 | -7.511267634 |

| 38 | 9.5 | 2.7 | 2.124337 | 1.916233534 | 14.14199638 | -7.6 | 4.073461 | 1.98857642 | -7.577547215 |
| 39 | 9.8 | 2.8 | 2.170219 | 1.920177901 | 14.31638986 | -7.8 | 4.12455 | 1.989149108 | -7.642124391 |
| 40 | 10 | 2.9 | 2.215203 | 1.923856111 | 14.48670545 | -8 | 4.174387 | 1.989679913 | -7.705083978 |
| 41 | 10 | 2.9 | 2.259315 | 1.92729107 | 14.65312256 | -8.2 | 4.223031 | 1.990172809 | -7.766504641 |
| 42 | 11 | 3 | 2.302585 | 1.930503326 | 14.81580947 | -8.4 | 4.270536 | 1.990631314 | -7.826459466 |
| 43 | 11 | 3.1 | 2.345038 | 1.933511349 | 14.97492415 | -8.6 | 4.316955 | 1.991058545 | -7.885016478 |
| 44 | 11 | 3.1 | 2.386701 | 1.936331766 | 15.13061512 | -8.8 | 4.362334 | 1.991457276 | -7.942239097 |
| 45 | 11 | 3.2 | 2.427598 | 1.938979583 | 15.28302212 | -9 | 4.406719 | 1.991829981 | -7.998186535 |
| 46 | 12 | 3.3 | 2.467754 | 1.941468358 | 15.43227675 | -9.2 | 4.450152 | 1.992178872 | -8.052914169 |
| 47 | 12 | 3.4 | 2.507191 | 1.943810369 | 15.57850312 | -9.4 | 4.492673 | 1.99250593 | -8.106473856 |
| 48 | 12 | 3.4 | 2.545931 | 1.946016749 | 15.72181835 | -9.6 | 4.534318 | 1.992812937 | -8.158914227 |
| 49 | 12 | 3.5 | 2.583998 | 1.948097613 | 15.86233308 | -9.8 | 4.575123 | 1.993101495 | -8.210280949 |
| 50 | 13 | 3.6 | 2.62141 | 1.950062165 | 16.00015192 | -10 | 4.615121 | 1.99337305 | -8.260616958 |
| 51 | 13 | 3.6 | 2.658188 | 1.951918792 | 16.13537389 | -10.2 | 4.654341 | 1.99362891 | -8.309962675 |
| 52 | 13 | 3.7 | 2.694351 | 1.95367515 | 16.26809277 | -10.4 | 4.692815 | 1.99387026 | -8.358356195 |
| 53 | 13 | 3.8 | 2.729918 | 1.955338236 | 16.39839748 | -10.6 | 4.730569 | 1.994098174 | -8.405833462 |
| 54 | 14 | 3.9 | 2.764906 | 1.956914456 | 16.52637239 | -10.8 | 4.767629 | 1.99431363 | -8.45242843 |
| 55 | 14 | 3.9 | 2.799332 | 1.958409682 | 16.6520976 | -11 | 4.804021 | 1.994517518 | -8.498173204 |
| 56 | 14 | 4 | 2.833213 | 1.959829305 | 16.77564927 | -11.2 | 4.839768 | 1.994710651 | -8.543098173 |
| 57 | 14 | 4.1 | 2.866565 | 1.961178278 | 16.89709981 | -11.4 | 4.874892 | 1.99489377 | -8.587232129 |
| 58 | 15 | 4.1 | 2.899401 | 1.962461162 | 17.01651814 | -11.6 | 4.909414 | 1.995067556 | -8.630602378 |
| 59 | 15 | 4.2 | 2.931738 | 1.963682158 | 17.13396993 | -11.8 | 4.943355 | 1.99523263 | -8.673234838 |
| 60 | 15 | 4.3 | 2.963589 | 1.964845142 | 17.24951775 | -12 | 4.976734 | 1.995389565 | -8.715154135 |
| 61 | 15 | 4.4 | 2.994967 | 1.965953696 | 17.3632213 | -12.2 | 5.009568 | 1.995538887 | -8.756383682 |
| 62 | 16 | 4.4 | 3.025885 | 1.967011131 | 17.47513755 | -12.4 | 5.041876 | 1.995681079 | -8.796945762 |
| 63 | 16 | 4.5 | 3.056357 | 1.968020513 | 17.58532093 | -12.6 | 5.073673 | 1.995816588 | -8.836861597 |
| 64 | 16 | 4.6 | 3.086393 | 1.968984683 | 17.69382345 | -12.8 | 5.104975 | 1.995945826 | -8.876151416 |
| 65 | 16 | 4.6 | 3.116007 | 1.969906278 | 17.80069484 | -13 | 5.135798 | 1.996069173 | -8.914834513 |
| 66 | 17 | 4.7 | 3.145207 | 1.970787746 | 17.90598272 | -13.2 | 5.166156 | 1.996186981 | -8.952929305 |
| 67 | 17 | 4.8 | 3.174007 | 1.971631365 | 18.00973265 | -13.4 | 5.196063 | 1.996299578 | -8.990453388 |
| 68 | 17 | 4.9 | 3.202415 | 1.972439253 | 18.1119883 | -13.6 | 5.225532 | 1.996407265 | -9.027423577 |
| 69 | 17 | 4.9 | 3.230441 | 1.973213384 | 18.21279153 | -13.8 | 5.254574 | 1.996510323 | -9.063855958 |
| 70 | 18 | 5 | 3.258097 | 1.973955598 | 18.31218248 | -14 | 5.283204 | 1.996609014 | -9.099765928 |
| 71 | 18 | 5.1 | 3.285389 | 1.974667614 | 18.41019967 | -14.2 | 5.311431 | 1.996703581 | -9.135168233 |
| 72 | 18 | 5.1 | 3.312329 | 1.975351034 | 18.5068801 | -14.4 | 5.339267 | 1.99679425 | -9.170077004 |
| 73 | 18 | 5.2 | 3.338924 | 1.976007361 | 18.6022593 | -14.6 | 5.366723 | 1.996881233 | -9.20450579 |
| 74 | 19 | 5.3 | 3.365182 | 1.976637996 | 18.69637142 | -14.8 | 5.393809 | 1.996964726 | -9.238467592 |
| 75 | 19 | 5.4 | 3.391113 | 1.977244256 | 18.7892493 | -15 | 5.420535 | 1.997044913 | -9.271974891 |
| 76 | 19 | 5.4 | 3.416722 | 1.977827371 | 18.88092455 | -15.2 | 5.44691 | 1.997121967 | -9.305039672 |
| 77 | 19 | 5.5 | 3.442019 | 1.978388498 | 18.97142756 | -15.4 | 5.472943 | 1.997196049 | -9.337673457 |
| 78 | 20 | 5.6 | 3.467011 | 1.978928721 | 19.06078762 | -15.6 | 5.498643 | 1.997267308 | -9.36988732 |
| 79 | 20 | 5.6 | 3.491703 | 1.97944906 | 19.14903296 | -15.8 | 5.524018 | 1.997335888 | -9.401691918 |
| 80 | 20 | 5.7 | 3.516104 | 1.979950474 | 19.23619074 | -16 | 5.549076 | 1.99740192 | -9.433097506 |
| 81 | 20 | 5.8 | 3.54022 | 1.980433863 | 19.3222872 | -16.2 | 5.573826 | 1.997465529 | -9.464113958 |
| 82 | 21 | 5.9 | 3.564056 | 1.980900076 | 19.40734763 | -16.4 | 5.598274 | 1.997526832 | -9.494750789 |
| 83 | 21 | 5.9 | 3.58762 | 1.981349913 | 19.49139642 | -16.6 | 5.622428 | 1.997585938 | -9.525017167 |
| 84 | 21 | 6 | 3.610918 | 1.981784129 | 19.57445715 | -16.8 | 5.646295 | 1.997642953 | -9.554921934 |
| 85 | 21 | 6.1 | 3.633954 | 1.982203434 | 19.65655255 | -17 | 5.669881 | 1.997697972 | -9.584473619 |
| 86 | 22 | 6.1 | 3.656736 | 1.982608499 | 19.73770463 | -17.2 | 5.693193 | 1.997751089 | -9.613680453 |
| 87 | 22 | 6.2 | 3.679267 | 1.98299996 | 19.81793463 | -17.4 | 5.716238 | 1.997802389 | -9.642550382 |
| 88 | 22 | 6.3 | 3.701554 | 1.983378415 | 19.89726309 | -17.6 | 5.739021 | 1.997851955 | -9.671091082 |
| 89 | 22 | 6.4 | 3.723601 | 1.983744432 | 19.97570988 | -17.8 | 5.761548 | 1.997899864 | -9.69930997 |
| 90 | 23 | 6.4 | 3.745414 | 1.984098548 | 20.05329423 | -18 | 5.783825 | 1.997946189 | -9.727214214 |
| 91 | 23 | 6.5 | 3.766997 | 1.984441269 | 20.13003476 | -18.2 | 5.805858 | 1.997990998 | -9.754810747 |
| 92 | 23 | 6.6 | 3.788355 | 1.984773078 | 20.20594947 | -18.4 | 5.827651 | 1.998034358 | -9.782106275 |
| 93 | 23 | 6.6 | 3.809493 | 1.985094432 | 20.28105583 | -18.6 | 5.849209 | 1.99807633 | -9.809107287 |
| 94 | 24 | 6.7 | 3.830414 | 1.985405762 | 20.35537073 | -18.8 | 5.870539 | 1.998116973 | -9.835820065 |
| 95 | 24 | 6.8 | 3.851124 | 1.98570748 | 20.42891057 | -19 | 5.891644 | 1.998156341 | -9.862250691 |
| 96 | 24 | 6.9 | 3.871626 | 1.985999976 | 20.50169123 | -19.2 | 5.91253 | 1.998194489 | -9.88840506 |
| 97 | 24 | 6.9 | 3.891924 | 1.98628362 | 20.57372813 | -19.4 | 5.9332 | 1.998231465 | -9.91428888 |
| 98 | 25 | 7 | 3.912023 | 1.986558764 | 20.64503619 | -19.6 | 5.953659 | 1.998267317 | -9.939907687 |
| 99 | 25 | 7.1 | 3.931926 | 1.986825744 | 20.71562994 | -19.8 | 5.973911 | 1.998302091 | -9.965266849 |
| 100 | 25 | 7.1 | 3.951636 | 1.987084878 | 20.78552345 | -20 | 5.993961 | 1.998335829 | -9.99037157 |
| 101 | 25 | 7.2 | 3.971158 | 1.98733647 | 20.85473037 | -20.2 | 6.013813 | 1.998368572 | -10.0152269 |
| 102 | 26 | 7.3 | 3.990495 | 1.987580807 | 20.923264 | -20.4 | 6.03347 | 1.998400358 | -10.03983775 |
| 103 | 26 | 7.4 | 4.00965 | 1.987818166 | 20.99113721 | -20.6 | 6.052936 | 1.998431224 | -10.06420887 |
| 104 | 26 | 7.4 | 4.028626 | 1.988048808 | 21.05836255 | -20.8 | 6.072215 | 1.998461206 | -10.08834488 |
| 105 | 26 | 7.5 | 4.047428 | 1.988272984 | 21.12495219 | -21 | 6.09131 | 1.998490338 | -10.11225027 |
| 106 | 27 | 7.6 | 4.066057 | 1.988490933 | 21.19091798 | -21.2 | 6.110225 | 1.99851865 | -10.13592942 |
| 107 | 27 | 7.6 | 4.084518 | 1.988702882 | 21.25627142 | -21.4 | 6.128963 | 1.998546173 | -10.15938654 |
| 108 | 27 | 7.7 | 4.102812 | 1.988909049 | 21.32102372 | -21.6 | 6.147528 | 1.998572937 | -10.18262577 |
| 109 | 27 | 7.8 | 4.120943 | 1.989109641 | 21.38518577 | -21.8 | 6.165922 | 1.998598968 | -10.20565112 |
| 110 | 28 | 7.9 | 4.138915 | 1.989304857 | 21.44876819 | -22 | 6.184149 | 1.998624295 | -10.22846648 |
| 111 | 28 | 7.9 | 4.156728 | 1.989494887 | 21.5117813 | -22.2 | 6.202212 | 1.99864894 | -10.25107565 |
| 112 | 28 | 8 | 4.174387 | 1.989679913 | 21.57423514 | -22.4 | 6.220113 | 1.99867293 | -10.27348231 |
| 113 | 28 | 8.1 | 4.191894 | 1.989860107 | 21.63613951 | -22.6 | 6.237856 | 1.998696287 | -10.29569006 |
| 114 | 29 | 8.1 | 4.209251 | 1.990035637 | 21.69750395 | -22.8 | 6.255443 | 1.998719032 | -10.31770241 |
| 115 | 29 | 8.2 | 4.226461 | 1.990206661 | 21.75833774 | -23 | 6.272877 | 1.998741188 | -10.33952274 |
| 116 | 29 | 8.3 | 4.243527 | 1.990373332 | 21.81864994 | -23.2 | 6.290161 | 1.998762774 | -10.36115439 |
| 117 | 29 | 8.4 | 4.26045 | 1.990535796 | 21.87844938 | -23.4 | 6.307297 | 1.99878381 | -10.38260059 |
| 118 | 30 | 8.4 | 4.277233 | 1.990694193 | 21.93774466 | -23.6 | 6.324287 | 1.998804314 | -10.40386449 |
| 119 | 30 | 8.5 | 4.293878 | 1.990848658 | 21.99654417 | -23.8 | 6.341135 | 1.998824304 | -10.42494915 |
| 120 | 30 | 8.6 | 4.310388 | 1.990999318 | 22.0548561 | -24 | 6.357842 | 1.998843797 | -10.44585758 |
| 121 | 30 | 8.6 | 4.326765 | 1.991146298 | 22.11268842 | -24.2 | 6.374411 | 1.998862809 | -10.46659269 |
| 122 | 31 | 8.7 | 4.34301 | 1.991289716 | 22.17004894 | -24.4 | 6.390845 | 1.998881356 | -10.48715734 |
| 123 | 31 | 8.8 | 4.359126 | 1.991429686 | 22.22694526 | -24.6 | 6.407144 | 1.998899454 | -10.50755429 |
| 124 | 31 | 8.9 | 4.375115 | 1.991566318 | 22.28338478 | -24.8 | 6.423312 | 1.998917116 | -10.52778627 |
| 125 | 31 | 8.9 | 4.390979 | 1.991699716 | 22.33937477 | -25 | 6.43935 | 1.998934356 | -10.54785591 |
| 126 | 32 | 9 | 4.406719 | 1.991829981 | 22.3949223 | -25.2 | 6.455261 | 1.998951188 | -10.5677658 |
| 127 | 32 | 9.1 | 4.422338 | 1.991957211 | 22.45003427 | -25.4 | 6.471047 | 1.998967625 | -10.58751847 |
| 128 | 32 | 9.1 | 4.437838 | 1.992081499 | 22.50471744 | -25.6 | 6.486709 | 1.998983678 | -10.60711637 |

| Comment | Min | Sec | 1/100 | Frame # | Track length | Track thickness | X (pixels) | Y (pixels) | X source offset (mm) | Y source offset (mm) | R | | |
|---|---|---|---|---|---|---|---|---|---|---|---|---|---|
| **TAB #3: DATA SORTED** | | | | | | | | | | | | | |
| | | | | | | | | | | | | | |
| **CALIBRATION** | | | | | | | | | | | | | |
| Pinpoint (source) | 3 | 28 | 3 | 6242 | N/A | N/A | 703 | 345 | | | | | |
| Boundary below point | 3 | 28 | 3 | 6242 | N/A | N/A | 705 | 631 | | | | | |
| Boundary above point | 3 | 28 | 3 | 6242 | N/A | N/A | 699 | 6 | | | Diameter = | 625 | pixels |
| Boundary left of point | 3 | 28 | 3 | 6242 | N/A | N/A | 405 | 340 | | | Pixel = | 0.16 | mm |
| Boundary right of point | 3 | 28 | 3 | 6242 | N/A | N/A | 1049 | 342 | | | Height (a) = | 2.25 | mm |
| Last data frame | 4 | 45 | 46 | 8565 | N/A | N/A | N/A | N/A | | | Height (b) = | 3.5 | mm |
| First data frame | 3 | 15 | 3 | 5852 | N/A | N/A | N/A | N/A | | | | | |
| | | | | | | | | | | | | | |
| **SORTED TRACK STARTS** | | | | | | | | | | | | | |
| | 4 | 11 | 76 | 7554 | Long | Sharp | 704 | 345 | 0.16 | 0 | 0.16 | 1 | |
| | 3 | 39 | 30 | 6580 | Short | Diffuse | 700 | 348 | -0.48 | -0.48 | 0.68 | 2 | |
| | 4 | 8 | 66 | 7461 | Short | Sharp | 701 | 349 | -0.32 | -0.64 | 0.72 | 3 | |
| | 3 | 48 | 76 | 6864 | Short | Sharp | 699 | 342 | -0.64 | 0.48 | 0.80 | 4 | |
| | 4 | 44 | 36 | 8532 | Short | Sharp | 704 | 340 | 0.16 | 0.8 | 0.82 | 5 | |
| | 4 | 9 | 83 | 7496 | Short | Diffuse | 700 | 351 | -0.48 | -0.96 | 1.07 | 6 | |
| | 3 | 43 | 96 | 6720 | Short | Sharp | 704 | 338 | 0.16 | 1.12 | 1.13 | 7 | |
| | 4 | 13 | 23 | 7598 | Short | Diffuse | 704 | 338 | 0.16 | 1.12 | 1.13 | 8 | |
| | 4 | 41 | 36 | 8442 | Short | Diffuse | 699 | 351 | -0.64 | -0.96 | 1.15 | 9 | |
| | 4 | 6 | 46 | 7395 | Short | Diffuse | 705 | 352 | 0.32 | -1.12 | 1.16 | 10 | |
| | 4 | 35 | 0 | 8251 | Short | Sharp | 710 | 342 | 1.12 | 0.48 | 1.22 | 11 | |
| | 3 | 43 | 23 | 6698 | Short | Sharp | 710 | 349 | 1.12 | -0.64 | 1.29 | 12 | |
| | 4 | 14 | 26 | 7629 | Long | Sharp | 699 | 338 | -0.64 | 1.12 | 1.29 | 13 | |
| | 4 | 4 | 33 | 7331 | Short | Diffuse | 705 | 336 | 0.32 | 1.44 | 1.48 | 14 | |
| | 4 | 5 | 70 | 7372 | Short | Sharp | 705 | 336 | 0.32 | 1.44 | 1.48 | 15 | |
| | 4 | 29 | 10 | 8074 | Short | Diffuse | 710 | 338 | 1.12 | 1.12 | 1.58 | 16 | |
| | 3 | 51 | 20 | 6937 | Long | Diffuse | 713 | 345 | 1.6 | 0 | 1.60 | 17 | |
| | 3 | 51 | 66 | 6951 | Short | Sharp | 713 | 345 | 1.6 | 0 | 1.60 | 18 | |
| | 4 | 15 | 26 | 7656 | Long | Diffuse | 693 | 344 | -1.6 | 0.16 | 1.61 | 19 | |
| | 4 | 19 | 36 | 7782 | Short | Diffuse | 710 | 337 | 1.12 | 1.28 | 1.70 | 20 | |
| | 4 | 41 | 16 | 8436 | Short | Diffuse | 712 | 351 | 1.44 | -0.96 | 1.73 | 21 | |
| | 3 | 45 | 23 | 6758 | Short | Sharp | 711 | 354 | 1.28 | -1.44 | 1.93 | 22 | |

| | | | | | | | | | | | |
|---|---|---|---|---|---|---|---|---|---|---|---|
| 3 | 45 | 46 | 6765 | Long | Sharp | 694 | 337 | -1.44 | 1.28 | 1.93 | 23 |
| 4 | 22 | 33 | 7871 | Short | Diffuse | 694 | 353 | -1.44 | -1.28 | 1.93 | 24 |
| 4 | 10 | 26 | 7509 | Long | Sharp | 714 | 340 | 1.76 | 0.8 | 1.93 | 25 |
| 4 | 29 | 56 | 8088 | Short | Diffuse | 707 | 333 | 0.64 | 1.92 | 2.02 | 26 |
| 3 | 53 | 43 | 7004 | Short | Diffuse | 712 | 336 | 1.44 | 1.44 | 2.04 | 27 |
| 4 | 19 | 46 | 7785 | Short | Diffuse | 713 | 337 | 1.6 | 1.28 | 2.05 | 28 |
| 4 | 29 | 10 | 8074 | Long | Sharp | 716 | 345 | 2.08 | 0 | 2.08 | 29 |
| 4 | 17 | 23 | 7718 | Long | Sharp | 714 | 337 | 1.76 | 1.28 | 2.18 | 30 |
| 3 | 24 | 86 | 6147 | Long | Sharp | 715 | 338 | 1.92 | 1.12 | 2.22 | 31 |
| 4 | 0 | 56 | 7218 | Short | Diffuse | 701 | 359 | -0.32 | -2.24 | 2.26 | 32 |
| 4 | 32 | 50 | 8176 | Long | Sharp | 717 | 343 | 2.24 | 0.32 | 2.26 | 33 |
| 4 | 16 | 30 | 7690 | Long | Sharp | 708 | 331 | 0.8 | 2.24 | 2.38 | 34 |
| 3 | 43 | 93 | 6719 | Short | Diffuse | 715 | 336 | 1.92 | 1.44 | 2.40 | 35 |
| 3 | 47 | 46 | 6825 | Long | Sharp | 712 | 332 | 1.44 | 2.08 | 2.53 | 36 |
| 4 | 21 | 86 | 7857 | Long | Sharp | 716 | 336 | 2.08 | 1.44 | 2.53 | 37 |
| 4 | 24 | 30 | 7930 | Short | Sharp | 698 | 330 | -0.8 | 2.4 | 2.53 | 38 |
| 3 | 59 | 90 | 7198 | Short | Diffuse | 717 | 337 | 2.24 | 1.28 | 2.58 | 39 |
| 4 | 25 | 73 | 7973 | Short | Sharp | 706 | 329 | 0.48 | 2.56 | 2.60 | 40 |
| 4 | 15 | 26 | 7666 | Long | Sharp | 710 | 330 | 1.12 | 2.4 | 2.65 | 41 |
| 4 | 13 | 93 | 7619 | Long | Sharp | 709 | 329 | 0.96 | 2.56 | 2.73 | 42 |
| 4 | 28 | 50 | 8056 | Short | Sharp | 716 | 357 | 2.08 | -1.92 | 2.83 | 43 |
| 4 | 25 | 43 | 7964 | Short | Diffuse | 720 | 340 | 2.72 | 0.8 | 2.84 | 44 |
| 4 | 24 | 43 | 7934 | Short | Diffuse | 702 | 327 | -0.16 | 2.88 | 2.88 | 45 |
| 4 | 36 | 30 | 8290 | Short | Sharp | 716 | 358 | 2.08 | -2.08 | 2.94 | 46 |
| 3 | 41 | 46 | 6645 | Long | Sharp | 715 | 331 | 1.92 | 2.24 | 2.95 | 47 |
| 3 | 59 | 60 | 7189 | Long | Sharp | 703 | 326 | 0 | 3.04 | 3.04 | 48 |
| 4 | 15 | 56 | 7677 | Long | Diffuse | 722 | 347 | 3.04 | -0.32 | 3.06 | 49 |
| 3 | 57 | 86 | 7137 | Long | Sharp | 688 | 358 | -2.4 | -2.08 | 3.18 | 50 |
| 3 | 57 | 50 | 7126 | Short | Sharp | 721 | 354 | 2.88 | -1.44 | 3.22 | 51 |
| 3 | 29 | 60 | 6289 | Short | Diffuse | 723 | 349 | 3.2 | -0.64 | 3.26 | 52 |
| 3 | 44 | 33 | 6731 | Short | Sharp | 688 | 331 | -2.4 | 2.24 | 3.28 | 53 |
| 3 | 52 | 36 | 6972 | Short | Sharp | 721 | 335 | 2.88 | 1.6 | 3.29 | 54 |
| 4 | 3 | 10 | 7294 | Short | Sharp | 723 | 339 | 3.2 | 0.96 | 3.34 | 55 |
| 4 | 42 | 53 | 8477 | Long | Sharp | 724 | 341 | 3.36 | 0.64 | 3.42 | 56 |
| 4 | 3 | 73 | 7313 | Short | Sharp | 725 | 346 | 3.52 | -0.16 | 3.52 | 57 |
| 3 | 37 | 80 | 6535 | Short | Sharp | 706 | 323 | 0.48 | 3.52 | 3.55 | 58 |

| | | | | | | | | | | | | |
|---|---|---|---|---|---|---|---|---|---|---|---|---|
| | 3 | 55 | 56 | 7068 | Short | Diffuse | 726 | 346 | 3.68 | -0.16 | 3.68 | 59 |
| | 4 | 3 | 53 | 7307 | Short | Diffuse | 726 | 350 | 3.68 | -0.8 | 3.77 | 60 |
| | 3 | 38 | 26 | 6549 | Long | Sharp | 723 | 332 | 3.2 | 2.08 | 3.82 | 61 |
| | 3 | 40 | 23 | 6608 | Short | Diffuse | 680 | 352 | -3.68 | -1.12 | 3.85 | 62 |
| | 3 | 50 | 46 | 6915 | Short | Sharp | 682 | 332 | -3.36 | 2.08 | 3.95 | 63 |
| | 3 | 36 | 0 | 6481 | Short | Diffuse | 680 | 355 | -3.68 | -1.6 | 4.01 | 64 |
| | 4 | 22 | 60 | 7879 | Short | Diffuse | 678 | 349 | -4 | -0.64 | 4.05 | 65 |
| | 4 | 28 | 43 | 8054 | Short | Sharp | 681 | 358 | -3.52 | -2.08 | 4.09 | 66 |
| | 3 | 52 | 70 | 6982 | Short | Diffuse | 694 | 369 | -1.44 | -3.84 | 4.10 | 67 |
| | 3 | 50 | 80 | 6925 | Long | Sharp | 704 | 319 | 0.16 | 4.16 | 4.16 | 68 |
| | 3 | 53 | 26 | 6999 | Short | Sharp | 685 | 325 | -2.88 | 3.2 | 4.31 | 69 |
| | 4 | 15 | 73 | 7658 | Long | Sharp | 686 | 366 | -2.72 | -3.36 | 4.32 | 70 |
| | 3 | 49 | 23 | 6878 | Short | Diffuse | 730 | 351 | 4.32 | -0.96 | 4.43 | 71 |
| | 4 | 7 | 6 | 7413 | Long | Sharp | 687 | 322 | -2.56 | 3.68 | 4.48 | 72 |
| | 3 | 57 | 16 | 7116 | Short | Diffuse | 728 | 358 | 4 | -2.08 | 4.51 | 73 |
| | 2 | 49 | 3 | 5072 | Long | Sharp | 682 | 326 | -3.36 | 3.04 | 4.53 | 74 |
| | 3 | 32 | 0 | 6361 | Long | Sharp | 731 | 340 | 4.48 | 0.8 | 4.55 | 75 |
| | 4 | 42 | 66 | 8481 | Short | Diffuse | 684 | 323 | -3.04 | 3.52 | 4.65 | 76 |
| | 3 | 47 | 96 | 6840 | Long | Sharp | 704 | 315 | 0.16 | 4.8 | 4.80 | 77 |
| | 4 | 12 | 60 | 7579 | Long | Sharp | 692 | 317 | -1.76 | 4.48 | 4.81 | 78 |
| | 4 | 14 | 23 | 7628 | Long | Sharp | 675 | 334 | -4.48 | 1.76 | 4.81 | 79 |
| | 3 | 37 | 60 | 6529 | Long | Sharp | 688 | 318 | -2.4 | 4.32 | 4.94 | 80 |
| | 3 | 57 | 43 | 7124 | Short | Diffuse | 729 | 362 | 4.16 | -2.72 | 4.97 | 81 |
| | 3 | 37 | 30 | 6520 | Long | Sharp | 732 | 357 | 4.64 | -1.92 | 5.02 | 82 |
| | 3 | 55 | 3 | 7052 | Long | Sharp | 675 | 361 | -4.48 | -2.56 | 5.16 | 83 |
| | 4 | 9 | 0 | 7471 | Long | Sharp | 732 | 360 | 4.64 | -2.4 | 5.22 | 84 |
| | 3 | 40 | 23 | 6608 | Long | Sharp | 734 | 356 | 4.96 | -1.76 | 5.26 | 85 |
| | 4 | 9 | 33 | 7481 | Long | Sharp | 676 | 364 | -4.32 | -3.04 | 5.28 | 86 |
| | 3 | 40 | 53 | 6617 | Long | Sharp | 676 | 325 | -4.32 | 3.2 | 5.38 | 87 |
| | 3 | 22 | 53 | 6077 | Short | Diffuse | 735 | 356 | 5.12 | -1.76 | 5.41 | 88 |
| | 3 | 34 | 10 | 6424 | Short | Sharp | 728 | 322 | 4 | 3.68 | 5.44 | 89 |
| | 3 | 39 | 30 | 6580 | Long | Sharp | 735 | 357 | 5.12 | -1.92 | 5.47 | 90 |
| | 3 | 34 | 30 | 6430 | Long | Sharp | 686 | 315 | -2.72 | 4.8 | 5.52 | 91 |
| | 3 | 17 | 56 | 5928 | Short | Diffuse | 726 | 371 | 3.68 | -4.16 | 5.55 | 92 |
| | 3 | 30 | 23 | 6308 | Short | Diffuse | 703 | 310 | 0 | 5.6 | 5.60 | 93 |
| | 3 | 34 | 33 | 6431 | Short | Diffuse | 705 | 310 | 0.32 | 5.6 | 5.61 | 94 |

| | | | | | | | | | | | |
|---|---|---|---|---|---|---|---|---|---|---|---|
| 3 | 39 | 70 | 6592 | Long | Sharp | 726 | 318 | 3.68 | 4.32 | 5.67 | 95 |
| 3 | 39 | 23 | 6578 | Long | Sharp | 738 | 351 | 5.6 | -0.96 | 5.68 | 96 |
| 3 | 41 | 70 | 6652 | Long | Sharp | 671 | 361 | -5.12 | -2.56 | 5.72 | 97 |
| 3 | 16 | 33 | 5891 | Short | Diffuse | 729 | 370 | 4.16 | -4 | 5.77 | 98 |
| 4 | 11 | 0 | 7531 | Short | Diffuse | 738 | 356 | 5.6 | -1.76 | 5.87 | 99 |
| 3 | 34 | 23 | 6428 | Short | Diffuse | 685 | 377 | -2.88 | -5.12 | 5.87 | 100 |
| 4 | 2 | 0 | 7261 | Long | Sharp | 675 | 370 | -4.48 | -4 | 6.01 | 101 |
| 3 | 16 | 30 | 5890 | Long | Sharp | 712 | 382 | 1.44 | -5.92 | 6.09 | 102 |
| 4 | 15 | 93 | 7679 | Long | Diffuse | 665 | 351 | -6.08 | -0.96 | 6.16 | 103 |
| 3 | 21 | 3 | 6032 | Short | Diffuse | 717 | 381 | 2.24 | -5.76 | 6.18 | 104 |
| 3 | 28 | 16 | 6246 | Long | Sharp | 702 | 306 | -0.16 | 6.24 | 6.24 | 105 |
| 3 | 17 | 56 | 5928 | Short | Sharp | 727 | 376 | 3.84 | -4.96 | 6.27 | 106 |
| 3 | 25 | 53 | 6167 | Long | Diffuse | 699 | 384 | -0.64 | -6.24 | 6.27 | 107 |
| 3 | 50 | 3 | 6902 | Long | Sharp | 742 | 341 | 6.24 | 0.64 | 6.27 | 108 |
| 4 | 6 | 56 | 7398 | Short | Diffuse | 741 | 355 | 6.08 | -1.6 | 6.29 | 109 |
| 3 | 28 | 66 | 6261 | Short | Sharp | 734 | 320 | 4.96 | 4 | 6.37 | 110 |
| 4 | 17 | 50 | 7726 | Short | Diffuse | 682 | 311 | -3.36 | 5.44 | 6.39 | 111 |
| 3 | 44 | 13 | 6725 | Long | Sharp | 663 | 348 | -6.4 | -0.48 | 6.42 | 112 |
| 3 | 44 | 16 | 6726 | Long | Sharp | 662 | 345 | -6.56 | 0 | 6.56 | 113 |
| 4 | 11 | 6 | 7533 | Long | Sharp | 662 | 350 | -6.56 | -0.8 | 6.61 | 114 |
| 4 | 17 | 50 | 7726 | Short | Diffuse | 706 | 303 | 0.48 | 6.72 | 6.74 | 115 |
| 3 | 47 | 80 | 6835 | Long | Sharp | 746 | 346 | 6.88 | -0.16 | 6.88 | 116 |
| 4 | 2 | 56 | 7278 | Long | Sharp | 661 | 356 | -6.72 | -1.76 | 6.95 | 117 |
| 3 | 26 | 53 | 6197 | Short | Diffuse | 741 | 368 | 6.08 | -3.68 | 7.11 | 118 |
| 3 | 29 | 6 | 6273 | Long | Sharp | 746 | 357 | 6.88 | -1.92 | 7.14 | 119 |
| 3 | 26 | 26 | 6189 | Long | Sharp | 725 | 306 | 3.52 | 6.24 | 7.16 | 120 |
| 3 | 44 | 90 | 6748 | Long | Sharp | 663 | 324 | -6.4 | 3.36 | 7.23 | 121 |
| 3 | 32 | 83 | 6386 | Short | Diffuse | 686 | 303 | -2.72 | 6.72 | 7.25 | 122 |
| 3 | 46 | 36 | 6792 | Long | Sharp | 748 | 337 | 7.2 | 1.28 | 7.31 | 123 |
| 3 | 30 | 56 | 6318 | Short | Diffuse | 680 | 385 | -3.68 | -6.4 | 7.38 | 124 |
| 3 | 42 | 36 | 6672 | Short | Diffuse | 745 | 365 | 6.72 | -3.2 | 7.44 | 125 |
| 4 | 14 | 50 | 7636 | Long | Sharp | 661 | 325 | -6.72 | 3.2 | 7.44 | 126 |
| 4 | 7 | 13 | 7415 | Long | Diffuse | 749 | 338 | 7.36 | 1.12 | 7.44 | 127 |
| 4 | 31 | 66 | 8151 | Short | Sharp | 662 | 367 | -6.56 | -3.52 | 7.44 | 128 |
| 4 | 2 | 86 | 7287 | Short | Diffuse | 746 | 363 | 6.88 | -2.88 | 7.46 | 129 |
| 3 | 35 | 20 | 6457 | Long | Sharp | 743 | 319 | 6.4 | 4.16 | 7.63 | 130 |
| 4 | 34 | 33 | 8231 | Long | Diffuse | 656 | 335 | -7.52 | 1.6 | 7.69 | 131 |
| 3 | 19 | 26 | 5979 | Short | Diffuse | 674 | 384 | -4.64 | -6.24 | 7.78 | 132 |
| 3 | 44 | 46 | 6735 | Long | Sharp | 694 | 297 | -1.44 | 7.68 | 7.81 | 133 |
| 4 | 33 | 6 | 8193 | Long | Sharp | 661 | 320 | -6.72 | 4 | 7.82 | 134 |
| 3 | 53 | 6 | 6993 | Short | Diffuse | 752 | 344 | 7.84 | 0.16 | 7.84 | 135 |
| 3 | 58 | 63 | 7160 | Long | Diffuse | 749 | 328 | 7.36 | 2.72 | 7.85 | 136 |
| 3 | 27 | 53 | 6227 | Short | Diffuse | 673 | 384 | -4.8 | -6.24 | 7.87 | 137 |
| 3 | 27 | 53 | 6227 | Long | Sharp | 658 | 324 | -7.2 | 3.36 | 7.95 | 138 |
| 3 | 36 | 90 | 6508 | Long | Sharp | 657 | 364 | -7.36 | -3.04 | 7.96 | 139 |
| 4 | 23 | 26 | 7899 | Long | Diffuse | 734 | 306 | 4.96 | 6.24 | 7.97 | 140 |
| 3 | 30 | 50 | 6316 | Long | Sharp | 748 | 367 | 7.2 | -3.52 | 8.01 | 141 |
| 4 | 8 | 0 | 7441 | Long | Sharp | 714 | 296 | 1.76 | 7.84 | 8.04 | 142 |
| 3 | 24 | 86 | 6147 | Long | Sharp | 718 | 297 | 2.4 | 7.68 | 8.05 | 143 |
| 3 | 36 | 43 | 6494 | Short | Diffuse | 748 | 368 | 7.2 | -3.68 | 8.09 | 144 |
| 4 | 6 | 30 | 7390 | Long | Sharp | 753 | 337 | 8 | 1.28 | 8.10 | 145 |
| 3 | 40 | 93 | 6629 | Long | Diffuse | 746 | 318 | 6.88 | 4.32 | 8.12 | 146 |
| 4 | 1 | 40 | 7243 | Long | Sharp | 666 | 310 | -5.92 | 5.6 | 8.15 | 147 |
| 3 | 37 | 73 | 6533 | Long | Sharp | 661 | 374 | -6.72 | -4.64 | 8.17 | 148 |
| 4 | 22 | 80 | 7885 | Short | Diffuse | 658 | 320 | -7.2 | 4 | 8.24 | 149 |
| 3 | 28 | 63 | 6260 | Long | Sharp | 655 | 366 | -7.68 | -3.36 | 8.38 | 150 |
| 4 | 6 | 26 | 7389 | Long | Sharp | 755 | 336 | 8.32 | 1.44 | 8.44 | 151 |
| 4 | 22 | 60 | 7879 | Short | Diffuse | 681 | 393 | -3.52 | -7.68 | 8.45 | 152 |
| 4 | 10 | 3 | 7502 | Long | Sharp | 658 | 317 | -7.2 | 4.48 | 8.48 | 153 |
| 4 | 12 | 6 | 7563 | Long | Sharp | 742 | 382 | 6.24 | -5.92 | 8.60 | 154 |
| 4 | 23 | 26 | 7899 | Short | Diffuse | 666 | 384 | -5.92 | -6.24 | 8.60 | 155 |
| 3 | 48 | 3 | 6842 | Short | Sharp | 757 | 342 | 8.64 | 0.48 | 8.65 | 156 |
| 3 | 47 | 86 | 6837 | Long | Sharp | 662 | 309 | -6.56 | 5.76 | 8.73 | 157 |
| 3 | 29 | 23 | 6278 | Short | Diffuse | 656 | 317 | -7.52 | 4.48 | 8.75 | 158 |
| 3 | 35 | 96 | 6480 | Long | Sharp | 655 | 372 | -7.68 | -4.32 | 8.81 | 159 |
| 3 | 17 | 46 | 5925 | Short | Diffuse | 735 | 390 | 5.12 | -7.2 | 8.83 | 160 |
| 3 | 15 | 20 | 5857 | Long | Sharp | 731 | 393 | 4.48 | -7.68 | 8.89 | 161 |
| 3 | 18 | 93 | 5969 | Short | Sharp | 729 | 295 | 4.16 | 8 | 9.02 | 162 |
| 3 | 18 | 43 | 5954 | Long | Sharp | 759 | 352 | 8.96 | -1.12 | 9.03 | 163 |
| 3 | 27 | 66 | 6231 | Long | Diffuse | 739 | 389 | 5.76 | -7.04 | 9.10 | 164 |
| 4 | 17 | 93 | 7739 | Long | Sharp | 654 | 316 | -7.84 | 4.64 | 9.11 | 165 |
| 3 | 31 | 10 | 6334 | Short | Diffuse | 655 | 314 | -7.68 | 4.96 | 9.14 | 166 |
| 3 | 23 | 73 | 6113 | Long | Sharp | 653 | 375 | -8 | -4.8 | 9.33 | 167 |
| 4 | 23 | 73 | 7913 | Long | Diffuse | 653 | 315 | -8 | 4.8 | 9.33 | 168 |
| 3 | 19 | 93 | 5999 | Long | Sharp | 720 | 289 | 2.72 | 8.96 | 9.36 | 169 |
| 4 | 27 | 76 | 8034 | Short | Sharp | 685 | 289 | -2.88 | 8.96 | 9.41 | 170 |
| 3 | 38 | 30 | 6550 | Long | Sharp | 746 | 304 | 6.88 | 6.56 | 9.51 | 171 |
| 4 | 36 | 26 | 8289 | Short | Diffuse | 658 | 306 | -7.2 | 6.24 | 9.53 | 172 |
| 3 | 27 | 90 | 6238 | Short | Sharp | 758 | 322 | 8.8 | 3.68 | 9.54 | 173 |
| 3 | 34 | 63 | 6440 | Short | Sharp | 657 | 306 | -7.36 | 6.24 | 9.65 | 174 |
| 3 | 20 | 43 | 6014 | Long | Sharp | 686 | 287 | -2.72 | 9.28 | 9.67 | 175 |
| 4 | 35 | 23 | 8258 | Short | Diffuse | 644 | 359 | -9.44 | -2.24 | 9.70 | 176 |
| 3 | 33 | 96 | 6420 | Long | Sharp | 723 | 286 | 3.2 | 9.44 | 9.97 | 177 |
| 3 | 36 | 56 | 6498 | Short | Diffuse | 762 | 368 | 9.44 | -3.68 | 10.13 | 178 |
| 3 | 31 | 10 | 6334 | Short | Diffuse | 669 | 291 | -5.44 | 8.64 | 10.21 | 179 |
| 3 | 36 | 36 | 6492 | Short | Diffuse | 698 | 410 | -0.8 | -10.4 | 10.43 | 180 |
| 3 | 21 | 96 | 6060 | Short | Diffuse | 720 | 282 | 2.72 | 10.08 | 10.44 | 181 |
| 3 | 30 | 23 | 6308 | Long | Sharp | 737 | 289 | 5.44 | 8.96 | 10.48 | 182 |
| 3 | 28 | 0 | 6241 | Long | Sharp | 769 | 346 | 10.56 | -0.16 | 10.56 | 183 |
| 3 | 22 | 86 | 6087 | Long | Sharp | 715 | 280 | 1.92 | 10.4 | 10.58 | 184 |
| 3 | 48 | 40 | 6853 | Short | Diffuse | 748 | 296 | 7.2 | 7.84 | 10.64 | 185 |
| 3 | 17 | 90 | 5938 | Long | Sharp | 700 | 278 | -0.48 | 10.72 | 10.73 | 186 |
| 3 | 29 | 6 | 6273 | Long | Sharp | 767 | 371 | 10.24 | -4.16 | 11.05 | 187 |
| 3 | 32 | 96 | 6390 | Long | Sharp | 736 | 281 | 5.28 | 10.24 | 11.52 | 188 |
| 4 | 20 | 30 | 7810 | Long | Sharp | 733 | 278 | 4.8 | 10.72 | 11.75 | 189 |
| 3 | 32 | 20 | 6367 | Short | Sharp | 641 | 305 | -9.92 | 6.4 | 11.81 | 190 |
| 3 | 27 | 23 | 6218 | Long | Sharp | 687 | 418 | -2.56 | -11.68 | 11.96 | 191 |
| 4 | 11 | 96 | 7560 | Long | Sharp | 638 | 306 | -10.4 | 6.24 | 12.13 | 192 |
| 3 | 34 | 73 | 6443 | Long | Sharp | 640 | 302 | -10.08 | 6.88 | 12.20 | 193 |
| 4 | 41 | 73 | 8453 | Short | Diffuse | 627 | 361 | -12.16 | -2.56 | 12.43 | 194 |
| 3 | 15 | 56 | 5868 | Short | Diffuse | 757 | 285 | 8.64 | 9.6 | 12.92 | 195 |
| 3 | 47 | 23 | 6818 | Short | Diffuse | 781 | 367 | 12.48 | -3.52 | 12.97 | 196 |
| 3 | 17 | 20 | 5917 | Long | Diffuse | 758 | 285 | 8.8 | 9.6 | 13.02 | 197 |
| 3 | 35 | 20 | 6457 | Long | Sharp | 658 | 277 | -7.2 | 10.88 | 13.05 | 198 |
| 3 | 43 | 73 | 6713 | Long | Sharp | 754 | 281 | 8.16 | 10.24 | 13.09 | 199 |
| 3 | 15 | 70 | 5872 | Short | Diffuse | 762 | 286 | 9.44 | 9.44 | 13.35 | 200 |
| 3 | 19 | 60 | 5989 | Short | Diffuse | 769 | 397 | 10.56 | -8.32 | 13.44 | 201 |
| 3 | 18 | 26 | 5949 | Long | Sharp | 769 | 292 | 10.56 | 8.48 | 13.54 | 202 |
| | 3 | 17 | 26 | 5919 | Short | Diffuse | 653 | 276 | -8 | 11.04 | 13.63 | 203 |
| | 3 | 15 | 20 | 5857 | Short | Diffuse | 735 | 424 | 5.12 | -12.64 | 13.64 | 204 |
| | 3 | 16 | 43 | 5894 | Long | Diffuse | 788 | 334 | 13.6 | 1.76 | 13.71 | 205 |
| | 3 | 36 | 23 | 6488 | Long | Sharp | 739 | 266 | 5.76 | 12.64 | 13.89 | 206 |
| | 4 | 20 | 20 | 7807 | Long | Sharp | 750 | 267 | 7.52 | 12.48 | 14.57 | 207 |
| | 3 | 17 | 86 | 5937 | Short | Sharp | 789 | 376 | 13.76 | -4.96 | 14.63 | 208 |
| | 4 | 14 | 63 | 7640 | Short | Sharp | 644 | 272 | -9.44 | 11.68 | 15.02 | 209 |
| | 3 | 50 | 30 | 6910 | Short | Diffuse | 610 | 331 | -14.88 | 2.24 | 15.05 | 210 |
| | 4 | 15 | 76 | 7672 | Short | Diffuse | 791 | 310 | 14.08 | 5.6 | 15.15 | 211 |
| | 3 | 23 | 70 | 6112 | Long | Diffuse | 625 | 291 | -12.48 | 8.64 | 15.18 | 212 |
| | 3 | 25 | 30 | 6160 | Short | Diffuse | 742 | 257 | 6.24 | 14.08 | 15.40 | 213 |
| | 3 | 23 | 23 | 6098 | Long | Sharp | 743 | 256 | 6.4 | 14.24 | 15.61 | 214 |
| | 4 | 27 | 53 | 8027 | Long | Diffuse | 626 | 285 | -12.32 | 9.6 | 15.62 | 215 |
| | 3 | 50 | 16 | 6906 | Short | Diffuse | 605 | 329 | -15.68 | 2.56 | 15.89 | 216 |
| | 3 | 16 | 23 | 5888 | Short | Sharp | 619 | 404 | -13.44 | -9.44 | 16.42 | 217 |
| | 3 | 16 | 53 | 5897 | Short | Diffuse | 608 | 304 | -15.2 | 6.56 | 16.56 | 218 |
| | 4 | 33 | 6 | 8193 | Short | Diffuse | 606 | 305 | -15.52 | 6.4 | 16.79 | 219 |
| | 3 | 21 | 23 | 6038 | Long | Diffuse | 800 | 294 | 15.52 | 8.16 | 17.53 | 220 |
| | 3 | 18 | 66 | 5961 | Long | Sharp | 818 | 312 | 18.4 | 5.28 | 19.14 | 221 |
| | 3 | 24 | 43 | 6134 | Long | Sharp | 595 | 282 | -17.28 | 10.08 | 20.01 | 222 |
| | 3 | 15 | 93 | 5879 | Long | Diffuse | 827 | 380 | 19.84 | -5.6 | 20.62 | 223 |
| | 4 | 18 | 53 | 7757 | Short | Diffuse | 626 | 236 | -12.32 | 17.44 | 21.35 | 224 |
| | 3 | 50 | 20 | 6907 | Short | Diffuse | 569 | 315 | -21.44 | 4.8 | 21.97 | 225 |
| | 4 | 26 | 40 | 7993 | Long | Diffuse | 567 | 372 | -21.76 | -4.32 | 22.18 | 226 |
| | 3 | 27 | 23 | 6218 | Long | Sharp | 663 | 525 | -6.4 | -28.8 | 29.50 | 227 |
| | 3 | 23 | 70 | 6112 | Short | Diffuse | 515 | 335 | -30.08 | 1.6 | 30.12 | 228 |
| **CORRESPONDING IDEAL CUMULATIVE TRACK DISTRIBUTION - NO CUTOFF** | | | | | | | | | | | | |
| **COPIED FROM PRECEDING TAB** | | | | | | | | | | | | |
| | | | | | | | | | | | 0.25 | 0.78 |
| | | | | | | | | | | | 0.50 | 3.02 |
| | | | | | | | | | | | 0.75 | 6.51 |
| | | | | | | | | | | | 1.00 | 10.96 |
| | | | | | | | | | | | 1.25 | 16.09 |
| | | | | | | | | | | | 1.50 | 21.68 |
| | | | | | | | | | | | 1.75 | 27.53 |
| | | | | | | | | | | | 2.00 | 33.51 |

| | | | | | | | | | | | |
|---|---|---|---|---|---|---|---|---|---|---|---|
| | | | | | | | | | | 2.25 | 39.53 |
| | | | | | | | | | | 2.50 | 45.50 |
| | | | | | | | | | | 2.75 | 51.39 |
| | | | | | | | | | | 3.00 | 57.16 |
| | | | | | | | | | | 3.25 | 62.79 |
| | | | | | | | | | | 3.50 | 68.26 |
| | | | | | | | | | | 3.75 | 73.58 |
| | | | | | | | | | | 4.00 | 78.73 |
| | | | | | | | | | | 4.25 | 83.72 |
| | | | | | | | | | | 4.50 | 88.56 |
| | | | | | | | | | | 4.75 | 93.24 |
| | | | | | | | | | | 5.00 | 97.77 |
| | | | | | | | | | | 5.25 | 102.16 |
| | | | | | | | | | | 5.50 | 106.42 |
| | | | | | | | | | | 5.75 | 110.54 |
| | | | | | | | | | | 6.00 | 114.53 |
| | | | | | | | | | | 6.25 | 118.41 |

**TAB #4: CUTOFF MODELS, DIVIDED BY PI**

| | | | | |
|---|---|---|---|---|
| a = | 2.25 mm | | h= | 1.25 |
| b = | 3.5 mm | | Active region scale | 1.05 |
| a-b= | -1.25 mm | | | |
| K = | 22.5 mm | | | |
| K/b= | 6.43 | | | |
| sqrt(K^2-b^2)= | 22.23 mm | | | |
| K/(a-b)= | -18.00 | | | |
| sqrt(K^2-(a-b)^2)= | 22.47 mm | | | |
| simple scale= | 22.14 | | | |
| subtractive scale= | 28.23 | | | |

| R | R/b | Simple cutoff b part (C1) | R/(a-b) | Simple cutoff a-b part (C1) | Simple scaled (C1) |
|---|---|---|---|---|---|
| 0.25 | 0.07 | 0.77 | -0.20 | -0.74 | 0.70 |
| 0.50 | 0.14 | 1.50 | -0.40 | -1.38 | 2.74 |
| 0.75 | 0.21 | 2.20 | -0.60 | -1.93 | 5.91 |
| 1.00 | 0.29 | 2.86 | -0.80 | -2.41 | 9.94 |
| 1.25 | 0.36 | 3.49 | -1.00 | -2.83 | 14.60 |
| 1.50 | 0.43 | 4.09 | -1.20 | -3.20 | 19.67 |
| 1.75 | 0.50 | 4.66 | -1.40 | -3.53 | 24.99 |
| 2.00 | 0.57 | 5.20 | -1.60 | -3.82 | 30.42 |
| 2.25 | 0.64 | 5.71 | -1.80 | -4.09 | 35.88 |
| 2.50 | 0.71 | 6.20 | -2.00 | -4.33 | 41.30 |
| 2.75 | 0.79 | 6.66 | -2.20 | -4.55 | 46.64 |
| 3.00 | 0.86 | 7.10 | -2.40 | -4.76 | 51.88 |
| 3.25 | 0.93 | 7.52 | -2.60 | -4.95 | 56.98 |
| 3.50 | 1.00 | 7.92 | -2.80 | -5.13 | 61.95 |
| 3.75 | 1.07 | 8.31 | -3.00 | -5.29 | 66.78 |
| 4.00 | 1.14 | 8.68 | -3.20 | -5.45 | 71.45 |
| 4.25 | 1.21 | 9.03 | -3.40 | -5.59 | 75.98 |
| 4.50 | 1.29 | 9.36 | -3.60 | -5.73 | 80.37 |
| 4.75 | 1.36 | 9.69 | -3.80 | -5.87 | 84.62 |
| 5.00 | 1.43 | 10.00 | -4.00 | -5.99 | 88.74 |
| 5.25 | 1.50 | 10.30 | -4.20 | -6.11 | 92.72 |
| 5.50 | 1.57 | 10.59 | -4.40 | -6.23 | 96.58 |
| 5.75 | 1.64 | 10.87 | -4.60 | -6.33 | 100.32 |
| 6.00 | 1.71 | 11.13 | -4.80 | -6.44 | 103.95 |
| 6.25 | 1.79 | 11.39 | -5.00 | -6.54 | 107.47 |
| 6.50 | 1.86 | 11.65 | -5.20 | -6.64 | 110.88 |
| 6.75 | 1.93 | 11.89 | -5.40 | -6.73 | 114.20 |
| 7.00 | 2.00 | 12.12 | -5.60 | -6.82 | 117.42 |
| 7.25 | 2.07 | 12.35 | -5.80 | -6.91 | 120.55 |
| 7.50 | 2.14 | 12.57 | -6.00 | -6.99 | 123.60 |
| 7.75 | 2.21 | 12.79 | -6.20 | -7.07 | 126.56 |
| 8.00 | 2.29 | 13.00 | -6.40 | -7.15 | 129.45 |
| 8.25 | 2.36 | 13.20 | -6.60 | -7.23 | 132.26 |
| 8.50 | 2.43 | 13.40 | -6.80 | -7.30 | 135.00 |
| 8.75 | 2.50 | 13.59 | -7.00 | -7.37 | 137.68 |
| 9.00 | 2.57 | 13.78 | -7.20 | -7.44 | 140.28 |
| 9.25 | 2.64 | 13.96 | -7.40 | -7.51 | 142.83 |
| 9.50 | 2.71 | 14.14 | -7.60 | -7.58 | 145.32 |
| 9.75 | 2.79 | 14.32 | -7.80 | -7.64 | 147.75 |
| 10.00 | 2.86 | 14.49 | -8.00 | -7.71 | 150.13 |
| 10.25 | 2.93 | 14.65 | -8.20 | -7.77 | 152.45 |
| 10.50 | 3.00 | 14.82 | -8.40 | -7.83 | 154.73 |
| 10.75 | 3.07 | 14.97 | -8.60 | -7.89 | 156.95 |
| 11.00 | 3.14 | 15.13 | -8.80 | -7.94 | 159.13 |
| 11.25 | 3.21 | 15.28 | -9.00 | -8.00 | 161.27 |
| 11.50 | 3.29 | 15.43 | -9.20 | -8.05 | 163.36 |
| 11.75 | 3.36 | 15.58 | -9.40 | -8.11 | 165.41 |
| 12.00 | 3.43 | 15.72 | -9.60 | -8.16 | 167.42 |
| 12.25 | 3.50 | 15.86 | -9.80 | -8.21 | 169.40 |
| 12.50 | 3.57 | 16.00 | -10.00 | -8.26 | 171.33 |
| 12.75 | 3.64 | 16.14 | -10.20 | -8.31 | 173.23 |
| 13.00 | 3.71 | 16.27 | -10.40 | -8.36 | 175.10 |
| 13.25 | 3.79 | 16.40 | -10.60 | -8.41 | 176.93 |
| 13.50 | 3.86 | 16.53 | -10.80 | -8.45 | 178.74 |
| 13.75 | 3.93 | 16.65 | -11.00 | -8.50 | 180.51 |
| 14.00 | 4.00 | 16.78 | -11.20 | -8.54 | 182.25 |
| 14.25 | 4.07 | 16.90 | -11.40 | -8.59 | 183.96 |
| 14.50 | 4.14 | 17.02 | -11.60 | -8.63 | 185.64 |
| 14.75 | 4.21 | 17.13 | -11.80 | -8.67 | 187.30 |
| 15.00 | 4.29 | 17.25 | -12.00 | -8.72 | 188.93 |
| 15.25 | 4.36 | 17.36 | -12.20 | -8.76 | 190.53 |
| 15.50 | 4.43 | 17.48 | -12.40 | -8.80 | 192.11 |
| 15.75 | 4.50 | 17.59 | -12.60 | -8.84 | 193.67 |
| 16.00 | 4.57 | 17.69 | -12.80 | -8.88 | 195.20 |
| 16.25 | 4.64 | 17.80 | -13.00 | -8.91 | 196.71 |
| 16.50 | 4.71 | 17.91 | -13.20 | -8.95 | 198.20 |
| 16.75 | 4.79 | 18.01 | -13.40 | -8.99 | 199.66 |
| 17.00 | 4.86 | 18.11 | -13.60 | -9.03 | 201.11 |
| 17.25 | 4.93 | 18.21 | -13.80 | -9.06 | 202.53 |
| 17.50 | 5.00 | 18.31 | -14.00 | -9.10 | 203.94 |
| 17.75 | 5.07 | 18.41 | -14.20 | -9.14 | 205.32 |
| 18.00 | 5.14 | 18.51 | -14.40 | -9.17 | 206.69 |
| 18.25 | 5.21 | 18.60 | -14.60 | -9.20 | 208.04 |
| 18.50 | 5.29 | 18.70 | -14.80 | -9.24 | 209.37 |
| 18.75 | 5.36 | 18.79 | -15.00 | -9.27 | 210.69 |
| 19.00 | 5.43 | 18.88 | -15.20 | -9.31 | 211.98 |
| 19.25 | 5.50 | 18.97 | -15.40 | -9.34 | 213.27 |
| 19.50 | 5.57 | 19.06 | -15.60 | -9.37 | 214.53 |
| 19.75 | 5.64 | 19.15 | -15.80 | -9.40 | 215.78 |
| 20.00 | 5.71 | 19.24 | -16.00 | -9.43 | 217.01 |
| 20.25 | 5.79 | 19.32 | -16.20 | -9.46 | 218.23 |
| 20.50 | 5.86 | 19.41 | -16.40 | -9.49 | 219.44 |
| 20.75 | 5.93 | 19.49 | -16.60 | -9.53 | 220.63 |
| 21.00 | 6.00 | 19.57 | -16.80 | -9.55 | 221.81 |
| 21.25 | 6.07 | 19.66 | -17.00 | -9.58 | 222.97 |
| 21.50 | 6.14 | 19.74 | -17.20 | -9.61 | 224.12 |
| 21.75 | 6.21 | 19.82 | -17.40 | -9.64 | 225.26 |
| 22.00 | 6.29 | 19.90 | -17.60 | -9.67 | 226.38 |
| 22.25 | 6.36 | 19.98 | -17.80 | -9.70 | 227.49 |
| 22.50 | 6.43 | 20.03 | -18.00 | -9.73 | 228.00 |
| 22.75 | 6.50 | 20.03 | -18.20 | -9.73 | 228.00 |
| 23.00 | 6.57 | 20.03 | -18.40 | -9.73 | 228.00 |
| 23.25 | 6.64 | 20.03 | -18.60 | -9.73 | 228.00 |
| 23.50 | 6.71 | 20.03 | -18.80 | -9.73 | 228.00 |
| 23.75 | 6.79 | 20.03 | -19.00 | -9.73 | 228.00 |
| 24.00 | 6.86 | 20.03 | -19.20 | -9.73 | 228.00 |
| 24.25 | 6.93 | 20.03 | -19.40 | -9.73 | 228.00 |
| 24.50 | 7.00 | 20.03 | -19.60 | -9.73 | 228.00 |
| 24.75 | 7.07 | 20.03 | -19.80 | -9.73 | 228.00 |
| 25.00 | 7.14 | 20.03 | -20.00 | -9.73 | 228.00 |
| 25.25 | 7.21 | 20.03 | -20.20 | -9.73 | 228.00 |
| 25.50 | 7.29 | 20.03 | -20.40 | -9.73 | 228.00 |
| 25.75 | 7.36 | 20.03 | -20.60 | -9.73 | 228.00 |
| 26.00 | 7.43 | 20.03 | -20.80 | -9.73 | 228.00 |
| 26.25 | 7.50 | 20.03 | -21.00 | -9.73 | 228.00 |
| 26.50 | 7.57 | 20.03 | -21.20 | -9.73 | 228.00 |
| 26.75 | 7.64 | 20.03 | -21.40 | -9.73 | 228.00 |
| 27.00 | 7.71 | 20.03 | -21.60 | -9.73 | 228.00 |
| 27.25 | 7.79 | 20.03 | -21.80 | -9.73 | 228.00 |
| 27.50 | 7.86 | 20.03 | -22.00 | -9.73 | 228.00 |
| 27.75 | 7.93 | 20.03 | -22.20 | -9.73 | 228.00 |
| 28.00 | 8.00 | 20.03 | -22.40 | -9.73 | 228.00 |
| 28.25 | 8.07 | 20.03 | -22.60 | -9.73 | 228.00 |
| 28.50 | 8.14 | 20.03 | -22.80 | -9.73 | 228.00 |
| 28.75 | 8.21 | 20.03 | -23.00 | -9.73 | 228.00 |
| 29.00 | 8.29 | 20.03 | -23.20 | -9.73 | 228.00 |
| 29.25 | 8.36 | 20.03 | -23.40 | -9.73 | 228.00 |
| 29.50 | 8.43 | 20.03 | -23.60 | -9.73 | 228.00 |
| 29.75 | 8.50 | 20.03 | -23.80 | -9.73 | 228.00 |
| 30.00 | 8.57 | 20.03 | -24.00 | -9.73 | 228.00 |
| 30.25 | 8.64 | 20.03 | -24.20 | -9.73 | 228.00 |
| 30.50 | 8.71 | 20.03 | -24.40 | -9.73 | 228.00 |
| 30.75 | 8.79 | 20.03 | -24.60 | -9.73 | 228.00 |
| 31.00 | 8.86 | 20.03 | -24.80 | -9.73 | 228.00 |
| 31.25 | 8.93 | 20.03 | -25.00 | -9.73 | 228.00 |
| 31.50 | 9.00 | 20.03 | -25.20 | -9.73 | 228.00 |
| 31.75 | 9.07 | 20.03 | -25.40 | -9.73 | 228.00 |
| 32.00 | 9.14 | 20.03 | -25.60 | -9.73 | 228.00 |
| 32.25 | 9.21 | 20.03 | -25.80 | -9.73 | 228.00 |
| 32.50 | 9.29 | 20.03 | -26.00 | -9.73 | 228.00 |
| 32.75 | 9.36 | 20.03 | -26.20 | -9.73 | 228.00 |
| 33.00 | 9.43 | 20.03 | -26.40 | -9.73 | 228.00 |
| 33.25 | 9.50 | 20.03 | -26.60 | -9.73 | 228.00 |
| 33.50 | 9.57 | 20.03 | -26.80 | -9.73 | 228.00 |
| 33.75 | 9.64 | 20.03 | -27.00 | -9.73 | 228.00 |
| 34.00 | 9.71 | 20.03 | -27.20 | -9.73 | 228.00 |
| 34.25 | 9.79 | 20.03 | -27.40 | -9.73 | 228.00 |
| 34.50 | 9.86 | 20.03 | -27.60 | -9.73 | 228.00 |
| 34.75 | 9.93 | 20.03 | -27.80 | -9.73 | 228.00 |
| 35.00 | 10.00 | 20.03 | -28.00 | -9.73 | 228.00 |
| 35.25 | 10.07 | 20.03 | -28.20 | -9.73 | 228.00 |
| 35.50 | 10.14 | 20.03 | -28.40 | -9.73 | 228.00 |
| 35.75 | 10.21 | 20.03 | -28.60 | -9.73 | 228.00 |
| 36.00 | 10.29 | 20.03 | -28.80 | -9.73 | 228.00 |
| 36.25 | 10.36 | 20.03 | -29.00 | -9.73 | 228.00 |
| 36.50 | 10.43 | 20.03 | -29.20 | -9.73 | 228.00 |
| 36.75 | 10.50 | 20.03 | -29.40 | -9.73 | 228.00 |
| 37.00 | 10.57 | 20.03 | -29.60 | -9.73 | 228.00 |
| 37.25 | 10.64 | 20.03 | -29.80 | -9.73 | 228.00 |
| 37.50 | 10.71 | 20.03 | -30.00 | -9.73 | 228.00 |
| 37.75 | 10.79 | 20.03 | -30.20 | -9.73 | 228.00 |
| 38.00 | 10.86 | 20.03 | -30.40 | -9.73 | 228.00 |
| 38.25 | 10.93 | 20.03 | -30.60 | -9.73 | 228.00 |
| 38.50 | 11.00 | 20.03 | -30.80 | -9.73 | 228.00 |
| 38.75 | 11.07 | 20.03 | -31.00 | -9.73 | 228.00 |
| 39.00 | 11.14 | 20.03 | -31.20 | -9.73 | 228.00 |
| 39.25 | 11.21 | 20.03 | -31.40 | -9.73 | 228.00 |
| 39.50 | 11.29 | 20.03 | -31.60 | -9.73 | 228.00 |
| 39.75 | 11.36 | 20.03 | -31.80 | -9.73 | 228.00 |
| 40.00 | 11.43 | 20.03 | -32.00 | -9.73 | 228.00 |
| 40.25 | 11.50 | 20.03 | -32.20 | -9.73 | 228.00 |
| 40.50 | 11.57 | 20.03 | -32.40 | -9.73 | 228.00 |
| 40.75 | 11.64 | 20.03 | -32.60 | -9.73 | 228.00 |
| 41.00 | 11.71 | 20.03 | -32.80 | -9.73 | 228.00 |
| 41.25 | 11.79 | 20.03 | -33.00 | -9.73 | 228.00 |
| 41.50 | 11.86 | 20.03 | -33.20 | -9.73 | 228.00 |
| 41.75 | 11.93 | 20.03 | -33.40 | -9.73 | 228.00 |
| 42.00 | 12.00 | 20.03 | -33.60 | -9.73 | 228.00 |
| 42.25 | 12.07 | 20.03 | -33.80 | -9.73 | 228.00 |
| 42.50 | 12.14 | 20.03 | -34.00 | -9.73 | 228.00 |
| 42.75 | 12.21 | 20.03 | -34.20 | -9.73 | 228.00 |
| 43.00 | 12.29 | 20.03 | -34.40 | -9.73 | 228.00 |
| 43.25 | 12.36 | 20.03 | -34.60 | -9.73 | 228.00 |
| 43.50 | 12.43 | 20.03 | -34.80 | -9.73 | 228.00 |
| 43.75 | 12.50 | 20.03 | -35.00 | -9.73 | 228.00 |
| 44.00 | 12.57 | 20.03 | -35.20 | -9.73 | 228.00 |
| 44.25 | 12.64 | 20.03 | -35.40 | -9.73 | 228.00 |
| 44.50 | 12.71 | 20.03 | -35.60 | -9.73 | 228.00 |
| 44.75 | 12.79 | 20.03 | -35.80 | -9.73 | 228.00 |
| 45.00 | 12.86 | 20.03 | -36.00 | -9.73 | 228.00 |
| 45.25 | 12.93 | 20.03 | -36.20 | -9.73 | 228.00 |
| 45.50 | 13.00 | 20.03 | -36.40 | -9.73 | 228.00 |
| 45.75 | 13.07 | 20.03 | -36.60 | -9.73 | 228.00 |
| 46.00 | 13.14 | 20.03 | -36.80 | -9.73 | 228.00 |
| 46.25 | 13.21 | 20.03 | -37.00 | -9.73 | 228.00 |
| 46.50 | 13.29 | 20.03 | -37.20 | -9.73 | 228.00 |
| 46.75 | 13.36 | 20.03 | -37.40 | -9.73 | 228.00 |
| 47.00 | 13.43 | 20.03 | -37.60 | -9.73 | 228.00 |
| 47.25 | 13.50 | 20.03 | -37.80 | -9.73 | 228.00 |
| 47.50 | 13.57 | 20.03 | -38.00 | -9.73 | 228.00 |
| 47.75 | 13.64 | 20.03 | -38.20 | -9.73 | 228.00 |
| 48.00 | 13.71 | 20.03 | -38.40 | -9.73 | 228.00 |
| 48.25 | 13.79 | 20.03 | -38.60 | -9.73 | 228.00 |
| 48.50 | 13.86 | 20.03 | -38.80 | -9.73 | 228.00 |
| 48.75 | 13.93 | 20.03 | -39.00 | -9.73 | 228.00 |
| 49.00 | 14.00 | 20.03 | -39.20 | -9.73 | 228.00 |
| 49.25 | 14.07 | 20.03 | -39.40 | -9.73 | 228.00 |
| 49.50 | 14.14 | 20.03 | -39.60 | -9.73 | 228.00 |
| 49.75 | 14.21 | 20.03 | -39.80 | -9.73 | 228.00 |
| 50.00 | 14.29 | 20.03 | -40.00 | -9.73 | 228.00 |

mm        height of source above hypothetical active region

| Subtraction b part (C1-C2) | Subtraction a-b part (C1-C2) | Subtractive scaled (C2) | Active region solid angle |
|---|---|---|---|
| 0.00 | 0.00 | 0.89 | 4.65 |
| 0.00 | 0.00 | 3.46 | 17.12 |
| 0.00 | 0.00 | 7.46 | 34.12 |
| 0.01 | 0.00 | 12.55 | 52.46 |
| 0.01 | 0.00 | 18.43 | 70.12 |
| 0.02 | -0.01 | 24.81 | 86.14 |
| 0.02 | -0.01 | 31.48 | 100.25 |
| 0.03 | -0.01 | 38.29 | 112.52 |
| 0.04 | -0.01 | 45.12 | 123.14 |
| 0.04 | -0.02 | 51.88 | 132.34 |
| 0.05 | -0.02 | 58.53 | 140.34 |
| 0.06 | -0.02 | 65.03 | 147.32 |
| 0.07 | -0.03 | 71.34 | 153.46 |
| 0.08 | -0.03 | 77.47 | 158.88 |
| 0.10 | -0.03 | 83.39 | 163.70 |
| 0.11 | -0.04 | 89.11 | 167.99 |
| 0.12 | -0.04 | 94.63 | 171.85 |
| 0.14 | -0.05 | 99.95 | 175.33 |
| 0.16 | -0.06 | 105.08 | 178.47 |
| 0.17 | -0.06 | 110.02 | 181.34 |
| 0.19 | -0.07 | 114.78 | 183.95 |
| 0.21 | -0.07 | 119.37 | 186.34 |
| 0.23 | -0.08 | 123.79 | 188.54 |
| 0.25 | -0.09 | 128.04 | 190.57 |
| 0.27 | -0.10 | 132.14 | 192.45 |
| 0.29 | -0.10 | 136.10 | 194.19 |
| 0.32 | -0.11 | 139.91 | 195.81 |
| 0.34 | -0.12 | 143.59 | 197.32 |
| 0.36 | -0.13 | 147.14 | 198.72 |
| 0.39 | -0.14 | 150.56 | 200.04 |
| 0.42 | -0.15 | 153.86 | 201.28 |
| 0.44 | -0.16 | 157.05 | 202.44 |
| 0.47 | -0.17 | 160.13 | 203.54 |
| 0.50 | -0.18 | 163.10 | 204.57 |
| 0.53 | -0.19 | 165.97 | 205.54 |
| 0.56 | -0.20 | 168.74 | 206.47 |
| 0.59 | -0.21 | 171.41 | 207.34 |
| 0.62 | -0.22 | 174.00 | 208.17 |
| 0.66 | -0.23 | 176.49 | 208.96 |
| 0.69 | -0.25 | 178.90 | 209.71 |
| 0.73 | -0.26 | 181.23 | 210.42 |
| 0.76 | -0.27 | 183.48 | 211.10 |
| 0.80 | -0.29 | 185.65 | 211.75 |
| 0.84 | -0.30 | 187.75 | 212.37 |
| 0.88 | -0.31 | 189.78 | 212.96 |
| 0.91 | -0.33 | 191.73 | 213.53 |
| 0.95 | -0.34 | 193.62 | 214.07 |
| 1.00 | -0.36 | 195.44 | 214.60 |
| 1.04 | -0.37 | 197.20 | 215.10 |
| 1.08 | -0.39 | 198.89 | 215.58 |
| 1.12 | -0.40 | 200.52 | 216.04 |
| 1.17 | -0.42 | 202.09 | 216.49 |
| 1.21 | -0.43 | 203.61 | 216.91 |
| 1.26 | -0.45 | 205.07 | 217.33 |
| 1.31 | -0.47 | 206.47 | 217.73 |
| 1.36 | -0.48 | 207.82 | 218.11 |
| 1.40 | -0.50 | 209.12 | 218.48 |
| 1.45 | -0.52 | 210.36 | 218.84 |
| 1.50 | -0.54 | 211.56 | 219.18 |
| 1.56 | -0.56 | 212.70 | 219.52 |
| 1.61 | -0.57 | 213.80 | 219.84 |
| 1.66 | -0.59 | 214.85 | 220.16 |
| 1.72 | -0.61 | 215.85 | 220.46 |
| 1.77 | -0.63 | 216.81 | 220.75 |
| 1.83 | -0.65 | 217.72 | 221.04 |
| 1.88 | -0.67 | 218.59 | 221.32 |
| 1.94 | -0.69 | 219.42 | 221.58 |
| 2.00 | -0.71 | 220.20 | 221.84 |
| 2.06 | -0.73 | 220.95 | 222.10 |
| 2.12 | -0.76 | 221.65 | 222.34 |
| 2.18 | -0.78 | 222.31 | 222.58 |
| 2.24 | -0.80 | 222.93 | 222.81 |
| 2.30 | -0.82 | 223.52 | 223.04 |
| 2.37 | -0.85 | 224.06 | 223.26 |
| 2.43 | -0.87 | 224.57 | 223.48 |
| 2.50 | -0.89 | 225.04 | 223.68 |
| 2.56 | -0.91 | 225.47 | 223.89 |
| 2.63 | -0.94 | 225.87 | 224.09 |
| 2.70 | -0.96 | 226.23 | 224.28 |
| 2.77 | -0.99 | 226.56 | 224.47 |
| 2.84 | -1.01 | 226.85 | 224.65 |
| 2.91 | -1.04 | 227.11 | 224.83 |
| 2.98 | -1.06 | 227.34 | 225.00 |
| 3.05 | -1.09 | 227.53 | 225.18 |
| 3.12 | -1.11 | 227.69 | 225.34 |
| 3.20 | -1.14 | 227.81 | 225.50 |
| 3.27 | -1.17 | 227.90 | 225.66 |
| 3.35 | -1.20 | 227.97 | 225.82 |
| 3.42 | -1.22 | 228.00 | 225.97 |
| 3.47 | -1.25 | 228.00 | 226.12 |
| 3.47 | -1.25 | 228.00 | 226.27 |
| 3.47 | -1.25 | 228.00 | 226.41 |
| 3.47 | -1.25 | 228.00 | 226.55 |
| 3.47 | -1.25 | 228.00 | 226.68 |
| 3.47 | -1.25 | 228.00 | 226.82 |
| 3.47 | -1.25 | 228.00 | 226.95 |
| 3.47 | -1.25 | 228.00 | 227.08 |
| 3.47 | -1.25 | 228.00 | 227.20 |
| 3.47 | -1.25 | 228.00 | 227.32 |
| 3.47 | -1.25 | 228.00 | 227.44 |
| 3.47 | -1.25 | 228.00 | 227.56 |
| 3.47 | -1.25 | 228.00 | 227.68 |
| 3.47 | -1.25 | 228.00 | 227.79 |
| 3.47 | -1.25 | 228.00 | 227.90 |
| 3.47 | -1.25 | 228.00 | 228.01 |
| 3.47 | -1.25 | 228.00 | 228.12 |
| 3.47 | -1.25 | 228.00 | 228.23 |
| 3.47 | -1.25 | 228.00 | 228.33 |
| 3.47 | -1.25 | 228.00 | 228.43 |
| 3.47 | -1.25 | 228.00 | 228.53 |
| 3.47 | -1.25 | 228.00 | 228.63 |
| 3.47 | -1.25 | 228.00 | 228.72 |
| 3.47 | -1.25 | 228.00 | 228.82 |
| 3.47 | -1.25 | 228.00 | 228.91 |
| 3.47 | -1.25 | 228.00 | 229.00 |
| 3.47 | -1.25 | 228.00 | 229.09 |
| 3.47 | -1.25 | 228.00 | 229.18 |
| 3.47 | -1.25 | 228.00 | 229.27 |
| 3.47 | -1.25 | 228.00 | 229.35 |
| 3.47 | -1.25 | 228.00 | 229.43 |
| 3.47 | -1.25 | 228.00 | 229.52 |
| 3.47 | -1.25 | 228.00 | 229.60 |
| 3.47 | -1.25 | 228.00 | 229.68 |
| 3.47 | -1.25 | 228.00 | 229.75 |
| 3.47 | -1.25 | 228.00 | 229.83 |
| 3.47 | -1.25 | 228.00 | 229.91 |
| 3.47 | -1.25 | 228.00 | 229.98 |
| 3.47 | -1.25 | 228.00 | 230.06 |
| 3.47 | -1.25 | 228.00 | 230.13 |
| 3.47 | -1.25 | 228.00 | 230.20 |
| 3.47 | -1.25 | 228.00 | 230.27 |
| 3.47 | -1.25 | 228.00 | 230.34 |
| 3.47 | -1.25 | 228.00 | 230.41 |
| 3.47 | -1.25 | 228.00 | 230.47 |
| 3.47 | -1.25 | 228.00 | 230.54 |
| 3.47 | -1.25 | 228.00 | 230.60 |
| 3.47 | -1.25 | 228.00 | 230.67 |
| 3.47 | -1.25 | 228.00 | 230.73 |
| 3.47 | -1.25 | 228.00 | 230.79 |
| 3.47 | -1.25 | 228.00 | 230.86 |
| 3.47 | -1.25 | 228.00 | 230.92 |
| 3.47 | -1.25 | 228.00 | 230.98 |
| 3.47 | -1.25 | 228.00 | 231.03 |
| 3.47 | -1.25 | 228.00 | 231.09 |
| 3.47 | -1.25 | 228.00 | 231.15 |
| 3.47 | -1.25 | 228.00 | 231.21 |
| 3.47 | -1.25 | 228.00 | 231.26 |
| 3.47 | -1.25 | 228.00 | 231.32 |
| 3.47 | -1.25 | 228.00 | 231.37 |
| 3.47 | -1.25 | 228.00 | 231.42 |
| 3.47 | -1.25 | 228.00 | 231.48 |
| 3.47 | -1.25 | 228.00 | 231.53 |
| 3.47 | -1.25 | 228.00 | 231.58 |
| 3.47 | -1.25 | 228.00 | 231.63 |
| 3.47 | -1.25 | 228.00 | 231.68 |
| 3.47 | -1.25 | 228.00 | 231.73 |
| 3.47 | -1.25 | 228.00 | 231.78 |
| 3.47 | -1.25 | 228.00 | 231.83 |
| 3.47 | -1.25 | 228.00 | 231.88 |
| 3.47 | -1.25 | 228.00 | 231.92 |
| 3.47 | -1.25 | 228.00 | 231.97 |
| 3.47 | -1.25 | 228.00 | 232.01 |
| 3.47 | -1.25 | 228.00 | 232.06 |
| 3.47 | -1.25 | 228.00 | 232.10 |
| 3.47 | -1.25 | 228.00 | 232.15 |
| 3.47 | -1.25 | 228.00 | 232.19 |
| 3.47 | -1.25 | 228.00 | 232.24 |
| 3.47 | -1.25 | 228.00 | 232.28 |
| 3.47 | -1.25 | 228.00 | 232.32 |
| 3.47 | -1.25 | 228.00 | 232.36 |
| 3.47 | -1.25 | 228.00 | 232.40 |
| 3.47 | -1.25 | 228.00 | 232.44 |
| 3.47 | -1.25 | 228.00 | 232.48 |
| 3.47 | -1.25 | 228.00 | 232.52 |
| 3.47 | -1.25 | 228.00 | 232.56 |
| 3.47 | -1.25 | 228.00 | 232.60 |
| 3.47 | -1.25 | 228.00 | 232.64 |
| 3.47 | -1.25 | 228.00 | 232.68 |
| 3.47 | -1.25 | 228.00 | 232.72 |
| 3.47 | -1.25 | 228.00 | 232.75 |
| 3.47 | -1.25 | 228.00 | 232.79 |
| 3.47 | -1.25 | 228.00 | 232.83 |
| 3.47 | -1.25 | 228.00 | 232.86 |
| 3.47 | -1.25 | 228.00 | 232.90 |
| 3.47 | -1.25 | 228.00 | 232.93 |
| 3.47 | -1.25 | 228.00 | 232.97 |
| 3.47 | -1.25 | 228.00 | 233.00 |
| 3.47 | -1.25 | 228.00 | 233.04 |
| 3.47 | -1.25 | 228.00 | 233.07 |
| 3.47 | -1.25 | 228.00 | 233.10 |
| 3.47 | -1.25 | 228.00 | 233.14 |
| 3.47 | -1.25 | 228.00 | 233.17 |
| 3.47 | -1.25 | 228.00 | 233.20 |
| 3.47 | -1.25 | 228.00 | 233.23 |
| 3.47 | -1.25 | 228.00 | 233.26 |
| 3.47 | -1.25 | 228.00 | 233.29 |
| 3.47 | -1.25 | 228.00 | 233.33 |
| 3.47 | -1.25 | 228.00 | 233.36 |
| 3.47 | -1.25 | 228.00 | 233.39 |
| 3.47 | -1.25 | 228.00 | 233.42 |

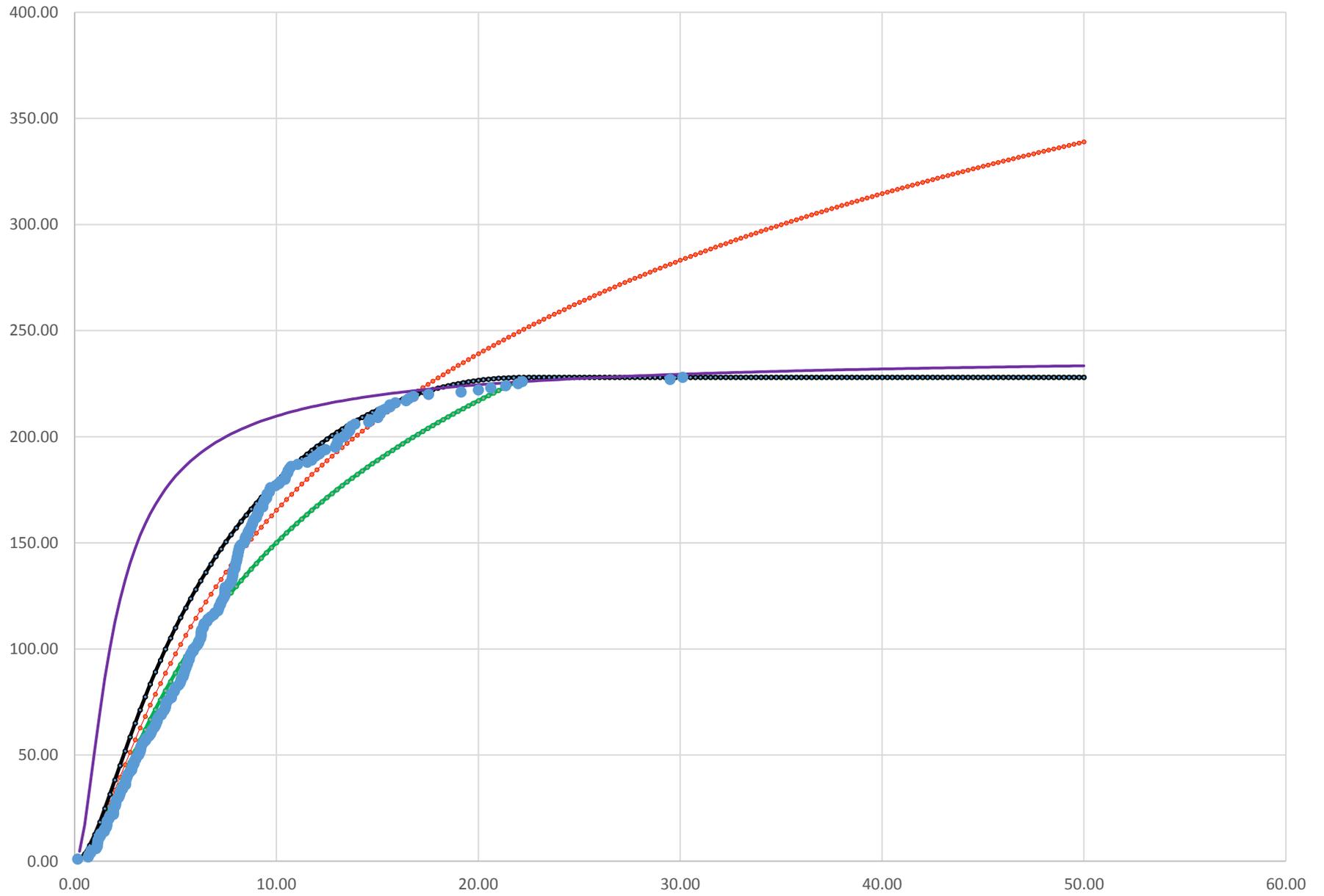

| Comment | Min | Sec | 1/100 | Frame # | Track length | Track thickness | X (pixels) | Y (pixels) | X source offset (mm) | Y source offset (mm) | R | | |
|---|---|---|---|---|---|---|---|---|---|---|---|---|---|
| **TAB #1: VIDEO DATA** | | | | | | | | | | | | | |
| | | | | | | | | | | | | | |
| **CALIBRATION** | | | | | | | | | | | | | |
| Pinpoint (source) | 3 | 28 | 3 | 6242 | N/A | N/A | 703 | 345 | | | | | |
| Boundary below point | 3 | 28 | 3 | 6242 | N/A | N/A | 705 | 631 | | | | | |
| Boundary above point | 3 | 28 | 3 | 6242 | N/A | N/A | 699 | 6 | | | Diameter = | 625 | pixels |
| Boundary left of point | 3 | 28 | 3 | 6242 | N/A | N/A | 405 | 340 | | | Pixel = | 0.16 | mm |
| Boundary right of point | 3 | 28 | 3 | 6242 | N/A | N/A | 1049 | 342 | | | Height (a) = | 4 | mm |
| Last data frame | 4 | 45 | 46 | 8565 | N/A | N/A | N/A | N/A | | | Height (b) = | 3.5 | mm |
| First data frame | 3 | 15 | 3 | 5852 | N/A | N/A | N/A | N/A | | | | | |
| | | | | | | | | | | | | | |
| **TRACK STARTS** | | | | | | | | | | | | | |
| | 3 | 15 | 20 | 5857 | Long | Sharp | 731 | 393 | 4.48 | -7.68 | 8.89 | | |
| | 3 | 15 | 20 | 5857 | Short | Diffuse | 735 | 424 | 5.12 | -12.64 | 13.64 | | |
| | 3 | 15 | 56 | 5868 | Short | Diffuse | 757 | 285 | 8.64 | 9.6 | 12.92 | | |
| | 3 | 15 | 70 | 5872 | Short | Diffuse | 762 | 286 | 9.44 | 9.44 | 13.35 | | |
| | 3 | 15 | 93 | 5879 | Long | Diffuse | 827 | 380 | 19.84 | -5.6 | 20.62 | | |
| | 3 | 16 | 23 | 5888 | Short | Sharp | 619 | 404 | -13.44 | -9.44 | 16.42 | | |
| | 3 | 16 | 30 | 5890 | Long | Sharp | 712 | 382 | 1.44 | -5.92 | 6.09 | | |
| | 3 | 16 | 33 | 5891 | Short | Diffuse | 729 | 370 | 4.16 | -4 | 5.77 | | |
| | 3 | 16 | 43 | 5894 | Long | Diffuse | 788 | 334 | 13.6 | 1.76 | 13.71 | | |
| | 3 | 16 | 53 | 5897 | Short | Diffuse | 608 | 304 | -15.2 | 6.56 | 16.56 | | |
| | 3 | 17 | 20 | 5917 | Long | Diffuse | 758 | 285 | 8.8 | 9.6 | 13.02 | | |
| | 3 | 17 | 26 | 5919 | Short | Diffuse | 653 | 276 | -8 | 11.04 | 13.63 | | |
| | 3 | 17 | 46 | 5925 | Short | Diffuse | 735 | 390 | 5.12 | -7.2 | 8.83 | | |
| | 3 | 17 | 56 | 5928 | Short | Diffuse | 726 | 371 | 3.68 | -4.16 | 5.55 | | |
| | 3 | 17 | 56 | 5928 | Short | Sharp | 727 | 376 | 3.84 | -4.96 | 6.27 | | |
| | 3 | 17 | 86 | 5937 | Short | Sharp | 789 | 376 | 13.76 | -4.96 | 14.63 | | |
| | 3 | 17 | 90 | 5938 | Long | Sharp | 700 | 278 | -0.48 | 10.72 | 10.73 | | |
| | 3 | 18 | 26 | 5949 | Long | Sharp | 769 | 292 | 10.56 | 8.48 | 13.54 | | |
| | 3 | 18 | 43 | 5954 | Long | Sharp | 759 | 352 | 8.96 | -1.12 | 9.03 | | |
| | 3 | 18 | 66 | 5961 | Long | Sharp | 818 | 312 | 18.4 | 5.28 | 19.14 | | |
| | 3 | 18 | 93 | 5969 | Short | Sharp | 729 | 295 | 4.16 | 8 | 9.02 | | |
| | 3 | 19 | 26 | 5979 | Short | Diffuse | 674 | 384 | -4.64 | -6.24 | 7.78 | | |

| | | | | | | | | | | | | |
|---|---|---|---|---|---|---|---|---|---|---|---|---|
| | 3 | 19 | 60 | 5989 | Short | Diffuse | 769 | 397 | 10.56 | -8.32 | 13.44 | |
| | 3 | 19 | 93 | 5999 | Long | Sharp | 720 | 289 | 2.72 | 8.96 | 9.36 | |
| | 3 | 20 | 43 | 6014 | Long | Sharp | 686 | 287 | -2.72 | 9.28 | 9.67 | |
| | 3 | 21 | 3 | 6032 | Short | Diffuse | 717 | 381 | 2.24 | -5.76 | 6.18 | |
| | 3 | 21 | 23 | 6038 | Long | Diffuse | 800 | 294 | 15.52 | 8.16 | 17.53 | |
| | 3 | 21 | 96 | 6060 | Short | Diffuse | 720 | 282 | 2.72 | 10.08 | 10.44 | |
| | 3 | 22 | 53 | 6077 | Short | Diffuse | 735 | 356 | 5.12 | -1.76 | 5.41 | |
| | 3 | 22 | 86 | 6087 | Long | Sharp | 715 | 280 | 1.92 | 10.4 | 10.58 | |
| | 3 | 23 | 23 | 6098 | Long | Sharp | 743 | 256 | 6.4 | 14.24 | 15.61 | |
| | 3 | 23 | 70 | 6112 | Short | Diffuse | 515 | 335 | -30.08 | 1.6 | 30.12 | |
| | 3 | 23 | 70 | 6112 | Long | Diffuse | 625 | 291 | -12.48 | 8.64 | 15.18 | |
| | 3 | 23 | 73 | 6113 | Long | Sharp | 653 | 375 | -8 | -4.8 | 9.33 | |
| | 3 | 24 | 43 | 6134 | Long | Sharp | 595 | 282 | -17.28 | 10.08 | 20.01 | |
| | 3 | 24 | 86 | 6147 | Long | Sharp | 718 | 297 | 2.4 | 7.68 | 8.05 | |
| | 3 | 24 | 86 | 6147 | Long | Sharp | 715 | 338 | 1.92 | 1.12 | 2.22 | |
| | 3 | 25 | 30 | 6160 | Short | Diffuse | 742 | 257 | 6.24 | 14.08 | 15.40 | |
| | 3 | 25 | 53 | 6167 | Long | Diffuse | 699 | 384 | -0.64 | -6.24 | 6.27 | |
| | 3 | 26 | 26 | 6189 | Long | Sharp | 725 | 306 | 3.52 | 6.24 | 7.16 | |
| | 3 | 26 | 53 | 6197 | Short | Diffuse | 741 | 368 | 6.08 | -3.68 | 7.11 | |
| | 3 | 27 | 23 | 6218 | Long | Sharp | 663 | 525 | -6.4 | -28.8 | 29.50 | |
| | 3 | 27 | 23 | 6218 | Long | Sharp | 687 | 418 | -2.56 | -11.68 | 11.96 | |
| | 3 | 27 | 53 | 6227 | Short | Diffuse | 673 | 384 | -4.8 | -6.24 | 7.87 | |
| | 3 | 27 | 53 | 6227 | Long | Sharp | 658 | 324 | -7.2 | 3.36 | 7.95 | |
| | 3 | 27 | 66 | 6231 | Long | Diffuse | 739 | 389 | 5.76 | -7.04 | 9.10 | |
| | 3 | 27 | 90 | 6238 | Short | Sharp | 758 | 322 | 8.8 | 3.68 | 9.54 | |
| | 3 | 28 | 0 | 6241 | Long | Sharp | 769 | 346 | 10.56 | -0.16 | 10.56 | |
| | 3 | 28 | 16 | 6246 | Long | Sharp | 702 | 306 | -0.16 | 6.24 | 6.24 | |
| | 3 | 28 | 63 | 6260 | Long | Sharp | 655 | 366 | -7.68 | -3.36 | 8.38 | |
| | 3 | 28 | 66 | 6261 | Short | Sharp | 734 | 320 | 4.96 | 4 | 6.37 | |
| | 3 | 29 | 6 | 6273 | Long | Sharp | 746 | 357 | 6.88 | -1.92 | 7.14 | |
| | 3 | 29 | 6 | 6273 | Long | Sharp | 767 | 371 | 10.24 | -4.16 | 11.05 | |
| | 3 | 29 | 23 | 6278 | Short | Diffuse | 656 | 317 | -7.52 | 4.48 | 8.75 | |
| | 3 | 29 | 60 | 6289 | Short | Diffuse | 723 | 349 | 3.2 | -0.64 | 3.26 | |
| | 3 | 30 | 23 | 6308 | Short | Diffuse | 703 | 310 | 0 | 5.6 | 5.60 | |
| | 3 | 30 | 23 | 6308 | Long | Sharp | 737 | 289 | 5.44 | 8.96 | 10.48 | |
| | 3 | 30 | 50 | 6316 | Long | Sharp | 748 | 367 | 7.2 | -3.52 | 8.01 | |
| | 3 | 30 | 56 | 6318 | Short | Diffuse | 680 | 385 | -3.68 | -6.4 | 7.38 | |
| | 3 | 31 | 10 | 6334 | Short | Diffuse | 669 | 291 | -5.44 | 8.64 | 10.21 | |
| | 3 | 31 | 10 | 6334 | Short | Diffuse | 655 | 314 | -7.68 | 4.96 | 9.14 | |
| | 3 | 32 | 0 | 6361 | Long | Sharp | 731 | 340 | 4.48 | 0.8 | 4.55 | |
| | 3 | 32 | 20 | 6367 | Short | Sharp | 641 | 305 | -9.92 | 6.4 | 11.81 | |
| | 3 | 32 | 83 | 6386 | Short | Diffuse | 686 | 303 | -2.72 | 6.72 | 7.25 | |
| | 3 | 32 | 96 | 6390 | Long | Sharp | 736 | 281 | 5.28 | 10.24 | 11.52 | |
| | 3 | 33 | 96 | 6420 | Long | Sharp | 723 | 286 | 3.2 | 9.44 | 9.97 | |
| | 3 | 34 | 10 | 6424 | Short | Sharp | 728 | 322 | 4 | 3.68 | 5.44 | |
| | 3 | 34 | 23 | 6428 | Short | Diffuse | 685 | 377 | -2.88 | -5.12 | 5.87 | |
| | 3 | 34 | 30 | 6430 | Long | Sharp | 686 | 315 | -2.72 | 4.8 | 5.52 | |
| | 3 | 34 | 33 | 6431 | Short | Diffuse | 705 | 310 | 0.32 | 5.6 | 5.61 | |
| | 3 | 34 | 63 | 6440 | Short | Sharp | 657 | 306 | -7.36 | 6.24 | 9.65 | |
| | 3 | 34 | 73 | 6443 | Long | Sharp | 640 | 302 | -10.08 | 6.88 | 12.20 | |
| | 3 | 35 | 20 | 6457 | Long | Sharp | 658 | 277 | -7.2 | 10.88 | 13.05 | |
| | 3 | 35 | 20 | 6457 | Long | Sharp | 743 | 319 | 6.4 | 4.16 | 7.63 | |
| | 3 | 35 | 96 | 6480 | Long | Sharp | 655 | 372 | -7.68 | -4.32 | 8.81 | |
| | 3 | 36 | 0 | 6481 | Short | Diffuse | 680 | 355 | -3.68 | -1.6 | 4.01 | |
| | 3 | 36 | 23 | 6488 | Long | Sharp | 739 | 266 | 5.76 | 12.64 | 13.89 | |
| | 3 | 36 | 36 | 6492 | Short | Diffuse | 698 | 410 | -0.8 | -10.4 | 10.43 | |
| | 3 | 36 | 43 | 6494 | Short | Diffuse | 748 | 368 | 7.2 | -3.68 | 8.09 | |
| | 3 | 36 | 56 | 6498 | Short | Diffuse | 762 | 368 | 9.44 | -3.68 | 10.13 | |
| | 3 | 36 | 90 | 6508 | Long | Sharp | 657 | 364 | -7.36 | -3.04 | 7.96 | |
| | 3 | 37 | 30 | 6520 | Long | Sharp | 732 | 357 | 4.64 | -1.92 | 5.02 | |
| | 3 | 37 | 60 | 6529 | Long | Sharp | 688 | 318 | -2.4 | 4.32 | 4.94 | |
| | 3 | 37 | 73 | 6533 | Long | Sharp | 661 | 374 | -6.72 | -4.64 | 8.17 | |
| | 3 | 37 | 80 | 6535 | Short | Sharp | 706 | 323 | 0.48 | 3.52 | 3.55 | |
| | 3 | 38 | 26 | 6549 | Long | Sharp | 723 | 332 | 3.2 | 2.08 | 3.82 | |
| | 3 | 38 | 30 | 6550 | Long | Sharp | 746 | 304 | 6.88 | 6.56 | 9.51 | |
| | 3 | 39 | 23 | 6578 | Long | Sharp | 738 | 351 | 5.6 | -0.96 | 5.68 | |
| | 3 | 39 | 30 | 6580 | Long | Sharp | 735 | 357 | 5.12 | -1.92 | 5.47 | |
| | 3 | 39 | 30 | 6580 | Short | Diffuse | 700 | 348 | -0.48 | -0.48 | 0.68 | |
| | 3 | 39 | 70 | 6592 | Long | Sharp | 726 | 318 | 3.68 | 4.32 | 5.67 | |
| | 3 | 40 | 23 | 6608 | Long | Sharp | 734 | 356 | 4.96 | -1.76 | 5.26 | |
| | 3 | 40 | 23 | 6608 | Short | Diffuse | 680 | 352 | -3.68 | -1.12 | 3.85 | |
| | 3 | 40 | 53 | 6617 | Long | Sharp | 676 | 325 | -4.32 | 3.2 | 5.38 | |
| | 3 | 40 | 93 | 6629 | Long | Diffuse | 746 | 318 | 6.88 | 4.32 | 8.12 | |
| | 3 | 41 | 46 | 6645 | Long | Sharp | 715 | 331 | 1.92 | 2.24 | 2.95 | |
| | 3 | 41 | 70 | 6652 | Long | Sharp | 671 | 361 | -5.12 | -2.56 | 5.72 | |
| | 3 | 42 | 36 | 6672 | Short | Diffuse | 745 | 365 | 6.72 | -3.2 | 7.44 | |
| | 3 | 43 | 23 | 6698 | Short | Sharp | 710 | 349 | 1.12 | -0.64 | 1.29 | |
| | 3 | 43 | 73 | 6713 | Long | Sharp | 754 | 281 | 8.16 | 10.24 | 13.09 | |
| | 3 | 43 | 93 | 6719 | Short | Diffuse | 715 | 336 | 1.92 | 1.44 | 2.40 | |
| | 3 | 43 | 96 | 6720 | Short | Sharp | 704 | 338 | 0.16 | 1.12 | 1.13 | |
| | 3 | 44 | 13 | 6725 | Long | Sharp | 663 | 348 | -6.4 | -0.48 | 6.42 | |
| | 3 | 44 | 16 | 6726 | Long | Sharp | 662 | 345 | -6.56 | 0 | 6.56 | |
| | 3 | 44 | 33 | 6731 | Short | Sharp | 688 | 331 | -2.4 | 2.24 | 3.28 | |
| | 3 | 44 | 46 | 6735 | Long | Sharp | 694 | 297 | -1.44 | 7.68 | 7.81 | |
| | 3 | 44 | 90 | 6748 | Long | Sharp | 663 | 324 | -6.4 | 3.36 | 7.23 | |
| | 3 | 45 | 23 | 6758 | Short | Sharp | 711 | 354 | 1.28 | -1.44 | 1.93 | |
| | 3 | 45 | 46 | 6765 | Long | Sharp | 694 | 337 | -1.44 | 1.28 | 1.93 | |
| | 3 | 46 | 36 | 6792 | Long | Sharp | 748 | 337 | 7.2 | 1.28 | 7.31 | |
| | 3 | 47 | 23 | 6818 | Short | Diffuse | 781 | 367 | 12.48 | -3.52 | 12.97 | |
| | 3 | 47 | 46 | 6825 | Long | Sharp | 712 | 332 | 1.44 | 2.08 | 2.53 | |
| | 3 | 47 | 80 | 6835 | Long | Sharp | 746 | 346 | 6.88 | -0.16 | 6.88 | |
| | 3 | 47 | 86 | 6837 | Long | Sharp | 662 | 309 | -6.56 | 5.76 | 8.73 | |
| | 3 | 47 | 96 | 6840 | Long | Sharp | 704 | 315 | 0.16 | 4.8 | 4.80 | |
| | 3 | 48 | 3 | 6842 | Short | Sharp | 757 | 342 | 8.64 | 0.48 | 8.65 | |
| | 3 | 48 | 40 | 6853 | Short | Diffuse | 748 | 296 | 7.2 | 7.84 | 10.64 | |
| | 3 | 48 | 76 | 6864 | Short | Sharp | 699 | 342 | -0.64 | 0.48 | 0.80 | |
| | 2 | 49 | 3 | 5072 | Long | Sharp | 682 | 326 | -3.36 | 3.04 | 4.53 | |
| | 3 | 49 | 23 | 6878 | Short | Diffuse | 730 | 351 | 4.32 | -0.96 | 4.43 | |
| | 3 | 50 | 3 | 6902 | Long | Sharp | 742 | 341 | 6.24 | 0.64 | 6.27 | |
| | 3 | 50 | 16 | 6906 | Short | Diffuse | 605 | 329 | -15.68 | 2.56 | 15.89 | |
| | 3 | 50 | 20 | 6907 | Short | Diffuse | 569 | 315 | -21.44 | 4.8 | 21.97 | |
| | 3 | 50 | 30 | 6910 | Short | Diffuse | 610 | 331 | -14.88 | 2.24 | 15.05 | |
| | 3 | 50 | 46 | 6915 | Short | Sharp | 682 | 332 | -3.36 | 2.08 | 3.95 | |
| | 3 | 50 | 80 | 6925 | Long | Sharp | 704 | 319 | 0.16 | 4.16 | 4.16 | |
| | 3 | 51 | 20 | 6937 | Long | Diffuse | 713 | 345 | 1.6 | 0 | 1.60 | |
| | 3 | 51 | 66 | 6951 | Short | Sharp | 713 | 345 | 1.6 | 0 | 1.60 | |
| | 3 | 52 | 36 | 6972 | Short | Sharp | 721 | 335 | 2.88 | 1.6 | 3.29 | |
| | 3 | 52 | 70 | 6982 | Short | Diffuse | 694 | 369 | -1.44 | -3.84 | 4.10 | |

| | | | | | | | | | | | |
|---|---|---|---|---|---|---|---|---|---|---|---|
| 3 | 53 | 6 | 6993 | Short | Diffuse | 752 | 344 | 7.84 | 0.16 | 7.84 | |
| 3 | 53 | 26 | 6999 | Short | Sharp | 685 | 325 | -2.88 | 3.2 | 4.31 | |
| 3 | 53 | 43 | 7004 | Short | Diffuse | 712 | 336 | 1.44 | 1.44 | 2.04 | |
| 3 | 55 | 3 | 7052 | Long | Sharp | 675 | 361 | -4.48 | -2.56 | 5.16 | |
| 3 | 55 | 56 | 7068 | Short | Diffuse | 726 | 346 | 3.68 | -0.16 | 3.68 | |
| 3 | 57 | 16 | 7116 | Short | Diffuse | 728 | 358 | 4 | -2.08 | 4.51 | |
| 3 | 57 | 43 | 7124 | Short | Diffuse | 729 | 362 | 4.16 | -2.72 | 4.97 | |
| 3 | 57 | 50 | 7126 | Short | Sharp | 721 | 354 | 2.88 | -1.44 | 3.22 | |
| 3 | 57 | 86 | 7137 | Long | Sharp | 688 | 358 | -2.4 | -2.08 | 3.18 | |
| 3 | 58 | 63 | 7160 | Long | Diffuse | 749 | 328 | 7.36 | 2.72 | 7.85 | |
| 3 | 59 | 60 | 7189 | Long | Sharp | 703 | 326 | 0 | 3.04 | 3.04 | |
| 3 | 59 | 90 | 7198 | Short | Diffuse | 717 | 337 | 2.24 | 1.28 | 2.58 | |
| 4 | 0 | 56 | 7218 | Short | Diffuse | 701 | 359 | -0.32 | -2.24 | 2.26 | |
| 4 | 1 | 40 | 7243 | Long | Sharp | 666 | 310 | -5.92 | 5.6 | 8.15 | |
| 4 | 2 | 0 | 7261 | Long | Sharp | 675 | 370 | -4.48 | -4 | 6.01 | |
| 4 | 2 | 56 | 7278 | Long | Sharp | 661 | 356 | -6.72 | -1.76 | 6.95 | |
| 4 | 2 | 86 | 7287 | Short | Diffuse | 746 | 363 | 6.88 | -2.88 | 7.46 | |
| 4 | 3 | 10 | 7294 | Short | Sharp | 723 | 339 | 3.2 | 0.96 | 3.34 | |
| 4 | 3 | 53 | 7307 | Short | Diffuse | 726 | 350 | 3.68 | -0.8 | 3.77 | |
| 4 | 3 | 73 | 7313 | Short | Sharp | 725 | 346 | 3.52 | -0.16 | 3.52 | |
| 4 | 4 | 33 | 7331 | Short | Diffuse | 705 | 336 | 0.32 | 1.44 | 1.48 | |
| 4 | 5 | 70 | 7372 | Short | Sharp | 705 | 336 | 0.32 | 1.44 | 1.48 | |
| 4 | 6 | 26 | 7389 | Long | Sharp | 755 | 336 | 8.32 | 1.44 | 8.44 | |
| 4 | 6 | 30 | 7390 | Long | Sharp | 753 | 337 | 8 | 1.28 | 8.10 | |
| 4 | 6 | 46 | 7395 | Short | Diffuse | 705 | 352 | 0.32 | -1.12 | 1.16 | |
| 4 | 6 | 56 | 7398 | Short | Diffuse | 741 | 355 | 6.08 | -1.6 | 6.29 | |
| 4 | 7 | 6 | 7413 | Long | Sharp | 687 | 322 | -2.56 | 3.68 | 4.48 | |
| 4 | 7 | 13 | 7415 | Long | Diffuse | 749 | 338 | 7.36 | 1.12 | 7.44 | |
| 4 | 8 | 0 | 7441 | Long | Sharp | 714 | 296 | 1.76 | 7.84 | 8.04 | |
| 4 | 8 | 66 | 7461 | Short | Sharp | 701 | 349 | -0.32 | -0.64 | 0.72 | |
| 4 | 9 | 0 | 7471 | Long | Sharp | 732 | 360 | 4.64 | -2.4 | 5.22 | |
| 4 | 9 | 33 | 7481 | Long | Sharp | 676 | 364 | -4.32 | -3.04 | 5.28 | |
| 4 | 9 | 83 | 7496 | Short | Diffuse | 700 | 351 | -0.48 | -0.96 | 1.07 | |
| 4 | 10 | 3 | 7502 | Long | Sharp | 658 | 317 | -7.2 | 4.48 | 8.48 | |
| 4 | 10 | 26 | 7509 | Long | Sharp | 714 | 340 | 1.76 | 0.8 | 1.93 | |
| 4 | 11 | 0 | 7531 | Short | Diffuse | 738 | 356 | 5.6 | -1.76 | 5.87 | |
| | 4 | 11 | 6 | 7533 | Long | Sharp | 662 | 350 | -6.56 | -0.8 | 6.61 | |
| | 4 | 11 | 76 | 7554 | Long | Sharp | 704 | 345 | 0.16 | 0 | 0.16 | |
| | 4 | 11 | 96 | 7560 | Long | Sharp | 638 | 306 | -10.4 | 6.24 | 12.13 | |
| | 4 | 12 | 6 | 7563 | Long | Sharp | 742 | 382 | 6.24 | -5.92 | 8.60 | |
| | 4 | 12 | 60 | 7579 | Long | Sharp | 692 | 317 | -1.76 | 4.48 | 4.81 | |
| | 4 | 13 | 23 | 7598 | Short | Diffuse | 704 | 338 | 0.16 | 1.12 | 1.13 | |
| | 4 | 13 | 93 | 7619 | Long | Sharp | 709 | 329 | 0.96 | 2.56 | 2.73 | |
| | 4 | 14 | 23 | 7628 | Long | Sharp | 675 | 334 | -4.48 | 1.76 | 4.81 | |
| | 4 | 14 | 26 | 7629 | Long | Sharp | 699 | 338 | -0.64 | 1.12 | 1.29 | |
| | 4 | 14 | 50 | 7636 | Long | Sharp | 661 | 325 | -6.72 | 3.2 | 7.44 | |
| | 4 | 14 | 63 | 7640 | Short | Sharp | 644 | 272 | -9.44 | 11.68 | 15.02 | |
| | 4 | 15 | 26 | 7668 | Long | Diffuse | 693 | 344 | -1.6 | 0.16 | 1.61 | |
| | 4 | 15 | 26 | 7673 | Long | Sharp | 710 | 330 | 1.12 | 2.4 | 2.65 | |
| | 4 | 15 | 56 | 7673 | Long | Diffuse | 722 | 347 | 3.04 | -0.32 | 3.06 | |
| | 4 | 15 | 73 | 7674 | Long | Sharp | 686 | 366 | -2.72 | -3.36 | 4.32 | |
| | 4 | 15 | 76 | 7679 | Short | Diffuse | 791 | 310 | 14.08 | 5.6 | 15.15 | |
| | 4 | 15 | 93 | 7679 | Long | Diffuse | 665 | 351 | -6.08 | -0.96 | 6.16 | |
| | 4 | 16 | 30 | 7690 | Long | Sharp | 708 | 331 | 0.8 | 2.24 | 2.38 | |
| | 4 | 17 | 23 | 7718 | Long | Sharp | 714 | 337 | 1.76 | 1.28 | 2.18 | |
| | 4 | 17 | 50 | 7726 | Short | Diffuse | 706 | 303 | 0.48 | 6.72 | 6.74 | |
| | 4 | 17 | 50 | 7726 | Short | Diffuse | 682 | 311 | -3.36 | 5.44 | 6.39 | |
| | 4 | 17 | 93 | 7739 | Long | Sharp | 654 | 316 | -7.84 | 4.64 | 9.11 | |
| | 4 | 18 | 53 | 7757 | Short | Diffuse | 626 | 236 | -12.32 | 17.44 | 21.35 | |
| | 4 | 19 | 36 | 7782 | Short | Diffuse | 710 | 337 | 1.12 | 1.28 | 1.70 | |
| | 4 | 19 | 46 | 7785 | Short | Diffuse | 713 | 337 | 1.6 | 1.28 | 2.05 | |
| | 4 | 20 | 20 | 7807 | Long | Sharp | 750 | 267 | 7.52 | 12.48 | 14.57 | |
| | 4 | 20 | 30 | 7810 | Long | Sharp | 733 | 278 | 4.8 | 10.72 | 11.75 | |
| | 4 | 21 | 86 | 7857 | Long | Sharp | 716 | 336 | 2.08 | 1.44 | 2.53 | |
| | 4 | 22 | 33 | 7871 | Short | Diffuse | 694 | 353 | -1.44 | -1.28 | 1.93 | |
| | 4 | 22 | 60 | 7879 | Short | Diffuse | 678 | 349 | -4 | -0.64 | 4.05 | |
| | 4 | 22 | 60 | 7879 | Short | Diffuse | 681 | 393 | -3.52 | -7.68 | 8.45 | |
| | 4 | 22 | 80 | 7885 | Short | Diffuse | 658 | 320 | -7.2 | 4 | 8.24 | |
| | 4 | 23 | 26 | 7899 | Short | Diffuse | 666 | 384 | -5.92 | -6.24 | 8.60 | |
| | 4 | 23 | 26 | 7899 | Long | Diffuse | 734 | 306 | 4.96 | 6.24 | 7.97 | |
| | 4 | 23 | 73 | 7913 | Long | Diffuse | 653 | 315 | -8 | 4.8 | 9.33 | |
| | 4 | 24 | 30 | 7930 | Short | Sharp | 698 | 330 | -0.8 | 2.4 | 2.53 | |

| | | | | | | | | | | | |
|---|---|---|---|---|---|---|---|---|---|---|---|
| | 4 | 24 | 43 | 7934 | Short | Diffuse | 702 | 327 | -0.16 | 2.88 | 2.88 | |
| | 4 | 25 | 43 | 7964 | Short | Diffuse | 720 | 340 | 2.72 | 0.8 | 2.84 | |
| | 4 | 25 | 73 | 7973 | Short | Sharp | 706 | 329 | 0.48 | 2.56 | 2.60 | |
| | 4 | 26 | 40 | 7993 | Long | Diffuse | 567 | 372 | -21.76 | -4.32 | 22.18 | |
| | 4 | 27 | 53 | 8027 | Long | Diffuse | 626 | 285 | -12.32 | 9.6 | 15.62 | |
| | 4 | 27 | 76 | 8034 | Short | Sharp | 685 | 289 | -2.88 | 8.96 | 9.41 | |
| | 4 | 28 | 43 | 8054 | Short | Sharp | 681 | 358 | -3.52 | -2.08 | 4.09 | |
| | 4 | 28 | 50 | 8056 | Short | Sharp | 716 | 357 | 2.08 | -1.92 | 2.83 | |
| | 4 | 29 | 10 | 8074 | Short | Diffuse | 710 | 338 | 1.12 | 1.12 | 1.58 | |
| | 4 | 29 | 10 | 8074 | Long | Sharp | 716 | 345 | 2.08 | 0 | 2.08 | |
| | 4 | 29 | 56 | 8088 | Short | Diffuse | 707 | 333 | 0.64 | 1.92 | 2.02 | |
| | 4 | 31 | 66 | 8151 | Short | Sharp | 662 | 367 | -6.56 | -3.52 | 7.44 | |
| | 4 | 32 | 50 | 8176 | Long | Sharp | 717 | 343 | 2.24 | 0.32 | 2.26 | |
| | 4 | 33 | 6 | 8193 | Short | Diffuse | 606 | 305 | -15.52 | 6.4 | 16.79 | |
| | 4 | 33 | 6 | 8193 | Long | Sharp | 661 | 320 | -6.72 | 4 | 7.82 | |
| | 4 | 34 | 33 | 8231 | Long | Diffuse | 656 | 335 | -7.52 | 1.6 | 7.69 | |
| | 4 | 35 | 0 | 8251 | Short | Sharp | 710 | 342 | 1.12 | 0.48 | 1.22 | |
| | 4 | 35 | 23 | 8258 | Short | Diffuse | 644 | 359 | -9.44 | -2.24 | 9.70 | |
| | 4 | 36 | 26 | 8289 | Short | Diffuse | 658 | 306 | -7.2 | 6.24 | 9.53 | |
| | 4 | 36 | 30 | 8290 | Short | Sharp | 716 | 358 | 2.08 | -2.08 | 2.94 | |
| | 4 | 41 | 16 | 8436 | Short | Diffuse | 712 | 351 | 1.44 | -0.96 | 1.73 | |
| | 4 | 41 | 36 | 8442 | Short | Diffuse | 699 | 351 | -0.64 | -0.96 | 1.15 | |
| | 4 | 41 | 73 | 8453 | Short | Diffuse | 627 | 361 | -12.16 | -2.56 | 12.43 | |
| | 4 | 42 | 53 | 8477 | Long | Sharp | 724 | 341 | 3.36 | 0.64 | 3.42 | |
| | 4 | 42 | 66 | 8481 | Short | Diffuse | 684 | 323 | -3.04 | 3.52 | 4.65 | |
| | 4 | 44 | 36 | 8532 | Short | Sharp | 704 | 340 | 0.16 | 0.8 | 0.82 | |

| TAB #2: IDEAL DISTRIBUTION C(R,b)+C(R,a-b), DIVIDED BY PI | | | | | | | | |
|---|---|---|---|---|---|---|---|---|
| | | | | | | | | |
| PETRIE DISH SENSITIVE VOLUME PARAMETERS | | | | | | | | |
| radius=50mm | | | | | | | | |
| a= | 4 | mm | | | | | | |
| b= | 3.5 | mm | | | | | | |
| a-b= | 0.5 | mm | | | | | | |
| scale | 9.8 | | | | | | | |
| | | | | | | | | |
| n | R | R/b | Log term | Arctan term | Total b part | R/(a-b) | Log term | Arctan term | Total a-b part |
| 1 | 0.3 | 0.1 | 0.005089 | 0.214212695 | 0.767556174 | 0.5 | 0.223144 | 1.107148718 | 0.665146135 |
| 2 | 0.5 | 0.1 | 0.020203 | 0.408256935 | 1.499608748 | 1 | 0.693147 | 1.570796327 | 1.131971754 |
| 3 | 0.8 | 0.2 | 0.044895 | 0.582729854 | 2.19668811 | 1.5 | 1.178655 | 1.764007811 | 1.471331403 |
| 4 | 1 | 0.3 | 0.078472 | 0.738569524 | 2.85964399 | 2 | 1.609438 | 1.854590436 | 1.732014174 |
| 5 | 1.3 | 0.4 | 0.120048 | 0.876980276 | 3.489599114 | 2.5 | 1.981001 | 1.902531886 | 1.941766677 |
| 6 | 1.5 | 0.4 | 0.168623 | 0.999346749 | 4.087893115 | 3 | 2.302585 | 1.930503326 | 2.11654421 |
| 7 | 1.8 | 0.5 | 0.223144 | 1.107148718 | 4.656022942 | 3.5 | 2.583998 | 1.948097613 | 2.266047583 |
| 8 | 2 | 0.6 | 0.282567 | 1.201885957 | 5.195585251 | 4 | 2.833213 | 1.959829305 | 2.396521325 |
| 9 | 2.3 | 0.6 | 0.345903 | 1.285018518 | 5.708224776 | 4.5 | 3.056357 | 1.968020513 | 2.512188704 |
| 10 | 2.5 | 0.7 | 0.412245 | 1.357924058 | 6.195590987 | 5 | 3.258097 | 1.973955598 | 2.616026068 |
| 11 | 2.8 | 0.8 | 0.480787 | 1.421871141 | 6.659303893 | 5.5 | 3.442019 | 1.978388498 | 2.710203937 |
| 12 | 3 | 0.9 | 0.550831 | 1.478005808 | 7.100928682 | 6 | 3.610918 | 1.981784129 | 2.796351021 |
| 13 | 3.3 | 0.9 | 0.621783 | 1.527348233 | 7.521958245 | 6.5 | 3.766997 | 1.984441269 | 2.875719251 |
| 14 | 3.5 | 1 | 0.693147 | 1.570796327 | 7.923802276 | 7 | 3.912023 | 1.986558764 | 2.949290885 |
| 15 | 3.8 | 1.1 | 0.764518 | 1.609133705 | 8.307781579 | 7.5 | 4.047428 | 1.988272984 | 3.017850313 |
| 16 | 4 | 1.1 | 0.835568 | 1.643039999 | 8.675126319 | 8 | 4.174387 | 1.989679913 | 3.082033591 |
| 17 | 4.3 | 1.2 | 0.906034 | 1.673102086 | 9.026977111 | 8.5 | 4.293878 | 1.990848658 | 3.142363453 |
| 18 | 4.5 | 1.3 | 0.975714 | 1.699825291 | 9.364388053 | 9 | 4.406719 | 1.991829981 | 3.199274614 |
| 19 | 4.8 | 1.4 | 1.044451 | 1.723643996 | 9.688331019 | 9.5 | 4.513603 | 1.992661836 | 3.253132414 |
| 20 | 5 | 1.4 | 1.112126 | 1.744931327 | 9.999700671 | 10 | 4.615121 | 1.99337305 | 3.304246783 |
| 21 | 5.3 | 1.5 | 1.178655 | 1.764007811 | 10.29931982 | 10.5 | 4.71178 | 1.993985833 | 3.352882877 |
| 22 | 5.5 | 1.6 | 1.243978 | 1.781148969 | 10.58794488 | 11 | 4.804021 | 1.994517518 | 3.399269282 |
| 23 | 5.8 | 1.6 | 1.308057 | 1.796591906 | 10.86627115 | 11.5 | 4.892227 | 1.99498179 | 3.443604426 |
| 24 | 6 | 1.7 | 1.37087 | 1.810540966 | 11.134938 | 12 | 4.976734 | 1.995389565 | 3.486061654 |
| 25 | 6.3 | 1.8 | 1.432408 | 1.823172578 | 11.3945336 | 12.5 | 5.057837 | 1.995749643 | 3.526793269 |
| 26 | 6.5 | 1.9 | 1.492675 | 1.83463937 | 11.64559947 | 13 | 5.135798 | 1.996069173 | 3.565933805 |
| 27 | 6.8 | 1.9 | 1.551679 | 1.845073664 | 11.8886346 | 13.5 | 5.210851 | 1.996354017 | 3.603602679 |
| 28 | 7 | 2 | 1.609438 | 1.854590436 | 12.12409922 | 14 | 5.283204 | 1.996609014 | 3.639906371 |
| 29 | 7.3 | 2.1 | 1.665973 | 1.863289826 | 12.35241831 | 14.5 | 5.353042 | 1.99683819 | 3.674940228 |
| 30 | 7.5 | 2.1 | 1.721308 | 1.871259256 | 12.57398473 | 15 | 5.420535 | 1.997044913 | 3.708789956 |
| 31 | 7.8 | 2.2 | 1.775471 | 1.878575235 | 12.78916206 | 15.5 | 5.485834 | 1.997232022 | 3.741532881 |
| 32 | 8 | 2.3 | 1.828491 | 1.885304876 | 12.99828724 | 16 | 5.549076 | 1.99740192 | 3.773239002 |
| 33 | 8.3 | 2.4 | 1.880399 | 1.891507195 | 13.20167287 | 16.5 | 5.610387 | 1.997556653 | 3.80397189 |
| 34 | 8.5 | 2.4 | 1.931226 | 1.897234212 | 13.39960933 | 17 | 5.669881 | 1.997697972 | 3.833789448 |
| 35 | 8.8 | 2.5 | 1.981001 | 1.902531886 | 13.59236674 | 17.5 | 5.727662 | 1.997827384 | 3.862744566 |
| 36 | 9 | 2.6 | 2.029758 | 1.907440914 | 13.78019662 | 18 | 5.783825 | 1.997946189 | 3.890885686 |
| 37 | 9.3 | 2.6 | 2.077526 | 1.911997419 | 13.96333352 | 18.5 | 5.838459 | 1.998055514 | 3.918257279 |
| 38 | 9.5 | 2.7 | 2.124337 | 1.916233534 | 14.14199638 | 19 | 5.891644 | 1.998156341 | 3.944900277 |
| 39 | 9.8 | 2.8 | 2.170219 | 1.920177901 | 14.31638986 | 19.5 | 5.943455 | 1.998249529 | 3.970852428 |
| 40 | 10 | 2.9 | 2.215203 | 1.923856111 | 14.48670545 | 20 | 5.993961 | 1.998335829 | 3.996148628 |
| 41 | 10 | 2.9 | 2.259315 | 1.92729107 | 14.65312256 | 20.5 | 6.043226 | 1.998415904 | 4.020821193 |
| 42 | 11 | 3 | 2.302585 | 1.930503326 | 14.81580947 | 21 | 6.09131 | 1.998490338 | 4.04490011 |
| 43 | 11 | 3.1 | 2.345038 | 1.933511349 | 14.97492415 | 21.5 | 6.138267 | 1.998559648 | 4.068413257 |
| 44 | 11 | 3.1 | 2.386701 | 1.936331766 | 15.13061512 | 22 | 6.184149 | 1.998624295 | 4.091386593 |
| 45 | 11 | 3.2 | 2.427598 | 1.938979583 | 15.28302212 | 22.5 | 6.229004 | 1.998684686 | 4.113844332 |
| 46 | 12 | 3.3 | 2.467754 | 1.941468358 | 15.43227675 | 23 | 6.272877 | 1.998741188 | 4.135809097 |
| 47 | 12 | 3.4 | 2.507191 | 1.943810369 | 15.57850312 | 23.5 | 6.31581 | 1.998794127 | 4.157302053 |
| 48 | 12 | 3.4 | 2.545931 | 1.946016749 | 15.72181835 | 24 | 6.357842 | 1.998843797 | 4.178343032 |
| 49 | 12 | 3.5 | 2.583998 | 1.948097613 | 15.86233308 | 24.5 | 6.399011 | 1.998890461 | 4.198950641 |
| 50 | 13 | 3.6 | 2.62141 | 1.950062165 | 16.00015192 | 25 | 6.43935 | 1.998934356 | 4.219142364 |
| 51 | 13 | 3.6 | 2.658188 | 1.951918792 | 16.13537389 | 25.5 | 6.478894 | 1.998975698 | 4.238934646 |
| 52 | 13 | 3.7 | 2.694351 | 1.95367515 | 16.26809277 | 26 | 6.517671 | 1.999014681 | 4.258342977 |
| 53 | 13 | 3.8 | 2.729918 | 1.955338236 | 16.39839748 | 26.5 | 6.555712 | 1.999051481 | 4.277381964 |
| 54 | 14 | 3.9 | 2.764906 | 1.956914456 | 16.52637239 | 27 | 6.593045 | 1.999086257 | 4.296065396 |
| 55 | 14 | 3.9 | 2.799332 | 1.958409682 | 16.6520976 | 27.5 | 6.629693 | 1.999119156 | 4.314406303 |
| 56 | 14 | 4 | 2.833213 | 1.959829305 | 16.77564927 | 28 | 6.665684 | 1.99915031 | 4.332417014 |
| 57 | 14 | 4.1 | 2.866565 | 1.961178278 | 16.89709981 | 28.5 | 6.701039 | 1.99917984 | 4.350109203 |
| 58 | 15 | 4.1 | 2.899401 | 1.962461162 | 17.01651814 | 29 | 6.73578 | 1.999207858 | 4.367493936 |
| 59 | 15 | 4.2 | 2.931738 | 1.963682158 | 17.13396993 | 29.5 | 6.769929 | 1.999234464 | 4.384581713 |
| 60 | 15 | 4.3 | 2.963589 | 1.964845142 | 17.24951775 | 30 | 6.803505 | 1.999259753 | 4.401382505 |
| 61 | 15 | 4.4 | 2.994967 | 1.965953696 | 17.3632213 | 30.5 | 6.836528 | 1.999283809 | 4.417905789 |
| 62 | 16 | 4.4 | 3.025885 | 1.967011131 | 17.47513755 | 31 | 6.869014 | 1.999306711 | 4.434160581 |
| 63 | 16 | 4.5 | 3.056357 | 1.968020513 | 17.58532093 | 31.5 | 6.900982 | 1.999328532 | 4.450155463 |
| 64 | 16 | 4.6 | 3.086393 | 1.968984683 | 17.69382345 | 32 | 6.932448 | 1.99934934 | 4.465898616 |
| 65 | 16 | 4.6 | 3.116007 | 1.969906278 | 17.80069484 | 32.5 | 6.963426 | 1.999369195 | 4.481397835 |
| 66 | 17 | 4.7 | 3.145207 | 1.970787746 | 17.90598272 | 33 | 6.993933 | 1.999388155 | 4.496660565 |
| 67 | 17 | 4.8 | 3.174007 | 1.971631365 | 18.00973265 | 33.5 | 7.023982 | 1.999406273 | 4.51169391 |
| 68 | 17 | 4.9 | 3.202415 | 1.972439253 | 18.1119883 | 34 | 7.053586 | 1.999423598 | 4.526504663 |
| 69 | 17 | 4.9 | 3.230441 | 1.973213384 | 18.21279153 | 34.5 | 7.082758 | 1.999440176 | 4.541099315 |
| 70 | 18 | 5 | 3.258097 | 1.973955598 | 18.31218248 | 35 | 7.111512 | 1.999456049 | 4.555484083 |
| 71 | 18 | 5.1 | 3.285389 | 1.974667614 | 18.41019967 | 35.5 | 7.139859 | 1.999471256 | 4.569664914 |
| 72 | 18 | 5.1 | 3.312329 | 1.975351034 | 18.5068801 | 36 | 7.167809 | 1.999485835 | 4.58364751 |
| 73 | 18 | 5.2 | 3.338924 | 1.976007361 | 18.6022593 | 36.5 | 7.195375 | 1.999499819 | 4.597437334 |
| 74 | 19 | 5.3 | 3.365182 | 1.976637996 | 18.69637142 | 37 | 7.222566 | 1.99951324 | 4.611039629 |
| 75 | 19 | 5.4 | 3.391113 | 1.977244256 | 18.7892493 | 37.5 | 7.249393 | 1.999526128 | 4.624459426 |
| 76 | 19 | 5.4 | 3.416722 | 1.977827371 | 18.88092455 | 38 | 7.275865 | 1.999538511 | 4.637701556 |
| 77 | 19 | 5.5 | 3.442019 | 1.978388498 | 18.97142756 | 38.5 | 7.301991 | 1.999550415 | 4.65077066 |
| 78 | 20 | 5.6 | 3.467011 | 1.978928721 | 19.06078762 | 39 | 7.327781 | 1.999561865 | 4.663671202 |
| 79 | 20 | 5.6 | 3.491703 | 1.97944906 | 19.14903296 | 39.5 | 7.353242 | 1.999572882 | 4.676407472 |
| 80 | 20 | 5.7 | 3.516104 | 1.979950474 | 19.23619074 | 40 | 7.378384 | 1.99958349 | 4.688983601 |
| 81 | 20 | 5.8 | 3.54022 | 1.980433863 | 19.3222872 | 40.5 | 7.403213 | 1.999593707 | 4.701403566 |
| 82 | 21 | 5.9 | 3.564056 | 1.980900076 | 19.40734763 | 41 | 7.427739 | 1.999603552 | 4.713671196 |
| 83 | 21 | 5.9 | 3.58762 | 1.981349913 | 19.49139642 | 41.5 | 7.451967 | 1.999613044 | 4.725790183 |
| 84 | 21 | 6 | 3.610918 | 1.981784129 | 19.57445715 | 42 | 7.475906 | 1.9996222 | 4.737764084 |
| 85 | 21 | 6.1 | 3.633954 | 1.982203434 | 19.65655255 | 42.5 | 7.499562 | 1.999631034 | 4.749596333 |
| 86 | 22 | 6.1 | 3.656736 | 1.982608499 | 19.73770463 | 43 | 7.522941 | 1.999639562 | 4.76129024 |
| 87 | 22 | 6.2 | 3.679267 | 1.98299996 | 19.81793463 | 43.5 | 7.54605 | 1.999647797 | 4.772849003 |
| 88 | 22 | 6.3 | 3.701554 | 1.983378415 | 19.89726309 | 44 | 7.568896 | 1.999655754 | 4.784275709 |
| 89 | 22 | 6.4 | 3.723601 | 1.983744432 | 19.97570988 | 44.5 | 7.591483 | 1.999663444 | 4.795573341 |
| 90 | 23 | 6.4 | 3.745414 | 1.984098548 | 20.05329423 | 45 | 7.613819 | 1.999670879 | 4.806744782 |
| 91 | 23 | 6.5 | 3.766997 | 1.984441269 | 20.13003476 | 45.5 | 7.635908 | 1.999678071 | 4.81779282 |
| 92 | 23 | 6.6 | 3.788355 | 1.984773078 | 20.20594947 | 46 | 7.657755 | 1.999685029 | 4.82872015 |
| 93 | 23 | 6.6 | 3.809493 | 1.985094432 | 20.28105583 | 46.5 | 7.679367 | 1.999691765 | 4.839529382 |
| 94 | 24 | 6.7 | 3.830414 | 1.985405762 | 20.35537073 | 47 | 7.700748 | 1.999698286 | 4.85022304 |
| 95 | 24 | 6.8 | 3.851124 | 1.98570748 | 20.42891057 | 47.5 | 7.721903 | 1.999704603 | 4.86080357 |
| 96 | 24 | 6.9 | 3.871626 | 1.985999976 | 20.50169123 | 48 | 7.742836 | 1.999710723 | 4.871273339 |
| 97 | 24 | 6.9 | 3.891924 | 1.98628362 | 20.57372813 | 48.5 | 7.763553 | 1.999716656 | 4.881634643 |
| 98 | 25 | 7 | 3.912023 | 1.986558764 | 20.64503619 | 49 | 7.784057 | 1.999722407 | 4.891889705 |
| 99 | 25 | 7.1 | 3.931926 | 1.986825744 | 20.71562994 | 49.5 | 7.804353 | 1.999727986 | 4.902040682 |
| 100 | 25 | 7.1 | 3.951636 | 1.987084878 | 20.78552345 | 50 | 7.824446 | 1.999733397 | 4.912089664 |
| 101 | 25 | 7.2 | 3.971158 | 1.98733647 | 20.85473037 | 50.5 | 7.844339 | 1.999738649 | 4.922038682 |
| 102 | 26 | 7.3 | 3.990495 | 1.987580807 | 20.923264 | 51 | 7.864036 | 1.999743747 | 4.931889703 |
| 103 | 26 | 7.4 | 4.00965 | 1.987818166 | 20.99113721 | 51.5 | 7.883541 | 1.999748698 | 4.94164464 |
| 104 | 26 | 7.4 | 4.028626 | 1.988048808 | 21.05836255 | 52 | 7.902857 | 1.999753506 | 4.951305349 |
| 105 | 26 | 7.5 | 4.047428 | 1.988272984 | 21.12495219 | 52.5 | 7.921989 | 1.999758178 | 4.960873632 |
| 106 | 27 | 7.6 | 4.066057 | 1.988490933 | 21.19091798 | 53 | 7.94094 | 1.999762718 | 4.97035124 |
| 107 | 27 | 7.6 | 4.084518 | 1.988702882 | 21.25627142 | 53.5 | 7.959713 | 1.999767132 | 4.979739877 |
| 108 | 27 | 7.7 | 4.102812 | 1.988909049 | 21.32102372 | 54 | 7.978311 | 1.999771423 | 4.989041197 |
| 109 | 27 | 7.8 | 4.120943 | 1.989109641 | 21.38518577 | 54.5 | 7.996738 | 1.999775597 | 4.998256808 |
| 110 | 28 | 7.9 | 4.138915 | 1.989304857 | 21.44876819 | 55 | 8.014997 | 1.999779658 | 5.007388276 |
| 111 | 28 | 7.9 | 4.156728 | 1.989494887 | 21.5117813 | 55.5 | 8.033091 | 1.99978361 | 5.016437124 |
| 112 | 28 | 8 | 4.174387 | 1.989679913 | 21.57423514 | 56 | 8.051022 | 1.999787456 | 5.025404832 |
| 113 | 28 | 8.1 | 4.191894 | 1.989860107 | 21.63613951 | 56.5 | 8.068794 | 1.9997912 | 5.034292843 |
| 114 | 29 | 8.1 | 4.209251 | 1.990035637 | 21.69750395 | 57 | 8.08641 | 1.999794847 | 5.043102561 |
| 115 | 29 | 8.2 | 4.226461 | 1.990206661 | 21.75833774 | 57.5 | 8.103872 | 1.999798398 | 5.051835353 |
| 116 | 29 | 8.3 | 4.243527 | 1.990373332 | 21.81864994 | 58 | 8.121183 | 1.999801859 | 5.06049255 |
| 117 | 29 | 8.4 | 4.26045 | 1.990535796 | 21.87844938 | 58.5 | 8.138346 | 1.999805231 | 5.069075451 |
| 118 | 30 | 8.4 | 4.277233 | 1.990694193 | 21.93774466 | 59 | 8.155362 | 1.999808517 | 5.077585319 |
| 119 | 30 | 8.5 | 4.293878 | 1.990848658 | 21.99654417 | 59.5 | 8.172235 | 1.999811721 | 5.086023386 |
| 120 | 30 | 8.6 | 4.310388 | 1.990999318 | 22.0548561 | 60 | 8.188967 | 1.999814846 | 5.094390855 |
| 121 | 30 | 8.6 | 4.326765 | 1.991146298 | 22.11268842 | 60.5 | 8.20556 | 1.999817893 | 5.102688896 |
| 122 | 31 | 8.7 | 4.34301 | 1.991289716 | 22.17004894 | 61 | 8.222016 | 1.999820866 | 5.110918651 |
| 123 | 31 | 8.8 | 4.359126 | 1.991429686 | 22.22694526 | 61.5 | 8.238339 | 1.999823766 | 5.119081237 |
| 124 | 31 | 8.9 | 4.375115 | 1.991566318 | 22.28338478 | 62 | 8.254529 | 1.999826597 | 5.127177739 |
| 125 | 31 | 8.9 | 4.390979 | 1.991699716 | 22.33937477 | 62.5 | 8.270589 | 1.99982936 | 5.13520922 |
| 126 | 32 | 9 | 4.406719 | 1.991829981 | 22.3949223 | 63 | 8.286521 | 1.999832057 | 5.143176715 |
| 127 | 32 | 9.1 | 4.422338 | 1.991957211 | 22.45003427 | 63.5 | 8.302328 | 1.999834691 | 5.151081236 |
| 128 | 32 | 9.1 | 4.437838 | 1.992081499 | 22.50471744 | 64 | 8.31801 | 1.999837263 | 5.15892377 |

| Comment | Min | Sec | 1/100 | Frame # | Track length | Track thickness | X (pixels) | Y (pixels) | X source offset (mm) | Y source offset (mm) | R | | | |
|---|---|---|---|---|---|---|---|---|---|---|---|---|---|---|
| TAB #3: DATA SORTED | | | | | | | | | | | | | | |
| CALIBRATION | | | | | | | | | | | | | | |
| Pinpoint (source) | 3 | 28 | 3 | 6242 | N/A | N/A | 703 | 345 | | | | | | |
| Boundary below point | 3 | 28 | 3 | 6242 | N/A | N/A | 705 | 631 | | | | | | |
| Boundary above point | 3 | 28 | 3 | 6242 | N/A | N/A | 699 | 6 | | | | Diameter = | 625 | pixels |
| Boundary left of point | 3 | 28 | 3 | 6242 | N/A | N/A | 405 | 340 | | | | Pixel = | 0.16 | mm |
| Boundary right of point | 3 | 28 | 3 | 6242 | N/A | N/A | 1049 | 342 | | | | Height (a) = | 4 | mm |
| Last data frame | 4 | 45 | 46 | 8565 | N/A | N/A | N/A | N/A | | | | Height (b) = | 3.5 | mm |
| First data frame | 3 | 15 | 3 | 5852 | N/A | N/A | N/A | N/A | | | | | | |
| | | | | | | | | | | | | | | |
| SORTED TRACK STARTS | | | | | | | | | | | | | | |
| | 4 | 11 | 76 | 7554 | Long | Sharp | 704 | 345 | 0.16 | 0 | 0.16 | 1 | | |
| | 3 | 39 | 30 | 6580 | Short | Diffuse | 700 | 348 | -0.48 | -0.48 | 0.68 | 2 | | |
| | 4 | 8 | 66 | 7461 | Short | Sharp | 701 | 349 | -0.32 | -0.64 | 0.72 | 3 | | |
| | 3 | 48 | 76 | 6864 | Short | Sharp | 699 | 342 | -0.64 | 0.48 | 0.80 | 4 | | |
| | 4 | 44 | 36 | 8532 | Short | Sharp | 704 | 340 | 0.16 | 0.8 | 0.82 | 5 | | |
| | 4 | 9 | 83 | 7496 | Short | Diffuse | 700 | 351 | -0.48 | -0.96 | 1.07 | 6 | | |
| | 3 | 43 | 96 | 6720 | Short | Sharp | 704 | 338 | 0.16 | 1.12 | 1.13 | 7 | | |
| | 4 | 13 | 23 | 7598 | Short | Diffuse | 704 | 338 | 0.16 | 1.12 | 1.13 | 8 | | |
| | 4 | 41 | 36 | 8442 | Short | Diffuse | 699 | 351 | -0.64 | -0.96 | 1.15 | 9 | | |
| | 4 | 6 | 46 | 7395 | Short | Diffuse | 705 | 352 | 0.32 | -1.12 | 1.16 | 10 | | |
| | 4 | 35 | 0 | 8251 | Short | Sharp | 710 | 342 | 1.12 | 0.48 | 1.22 | 11 | | |
| | 3 | 43 | 23 | 6698 | Short | Sharp | 710 | 349 | 1.12 | -0.64 | 1.29 | 12 | | |
| | 4 | 14 | 26 | 7629 | Long | Sharp | 699 | 338 | -0.64 | 1.12 | 1.29 | 13 | | |
| | 4 | 4 | 33 | 7331 | Short | Diffuse | 705 | 336 | 0.32 | 1.44 | 1.48 | 14 | | |
| | 4 | 5 | 70 | 7372 | Short | Sharp | 705 | 336 | 0.32 | 1.44 | 1.48 | 15 | | |
| | 4 | 29 | 10 | 8074 | Short | Diffuse | 710 | 338 | 1.12 | 1.12 | 1.58 | 16 | | |
| | 3 | 51 | 20 | 6937 | Long | Diffuse | 713 | 345 | 1.6 | 0 | 1.60 | 17 | | |
| | 3 | 51 | 66 | 6951 | Short | Sharp | 713 | 345 | 1.6 | 0 | 1.60 | 18 | | |
| | 4 | 15 | 26 | 7656 | Long | Diffuse | 693 | 344 | -1.6 | 0.16 | 1.61 | 19 | | |
| | 4 | 19 | 36 | 7782 | Short | Diffuse | 710 | 337 | 1.12 | 1.28 | 1.70 | 20 | | |
| | 4 | 41 | 16 | 8436 | Short | Diffuse | 712 | 351 | 1.44 | -0.96 | 1.73 | 21 | | |
| | 3 | 45 | 23 | 6758 | Short | Sharp | 711 | 354 | 1.28 | -1.44 | 1.93 | 22 | | |

| | | | | | | | | | | | |
|---|---|---|---|---|---|---|---|---|---|---|---|
| 3 | 45 | 46 | 6765 | Long | Sharp | 694 | 337 | -1.44 | 1.28 | 1.93 | 23 |
| 4 | 22 | 33 | 7871 | Short | Diffuse | 694 | 353 | -1.44 | -1.28 | 1.93 | 24 |
| 4 | 10 | 26 | 7509 | Long | Sharp | 714 | 340 | 1.76 | 0.8 | 1.93 | 25 |
| 4 | 29 | 56 | 8088 | Short | Diffuse | 707 | 333 | 0.64 | 1.92 | 2.02 | 26 |
| 3 | 53 | 43 | 7004 | Short | Diffuse | 712 | 336 | 1.44 | 1.44 | 2.04 | 27 |
| 4 | 19 | 46 | 7785 | Short | Diffuse | 713 | 337 | 1.6 | 1.28 | 2.05 | 28 |
| 4 | 29 | 10 | 8074 | Long | Sharp | 716 | 345 | 2.08 | 0 | 2.08 | 29 |
| 4 | 17 | 23 | 7718 | Long | Sharp | 714 | 337 | 1.76 | 1.28 | 2.18 | 30 |
| 3 | 24 | 86 | 6147 | Long | Sharp | 715 | 338 | 1.92 | 1.12 | 2.22 | 31 |
| 4 | 0 | 56 | 7218 | Short | Diffuse | 701 | 359 | -0.32 | -2.24 | 2.26 | 32 |
| 4 | 32 | 50 | 8176 | Long | Sharp | 717 | 343 | 2.24 | 0.32 | 2.26 | 33 |
| 4 | 16 | 30 | 7690 | Long | Sharp | 708 | 331 | 0.8 | 2.24 | 2.38 | 34 |
| 3 | 43 | 93 | 6719 | Short | Diffuse | 715 | 336 | 1.92 | 1.44 | 2.40 | 35 |
| 3 | 47 | 46 | 6825 | Long | Sharp | 712 | 332 | 1.44 | 2.08 | 2.53 | 36 |
| 4 | 21 | 86 | 7857 | Long | Sharp | 716 | 336 | 2.08 | 1.44 | 2.53 | 37 |
| 4 | 24 | 30 | 7930 | Short | Sharp | 698 | 330 | -0.8 | 2.4 | 2.53 | 38 |
| 3 | 59 | 90 | 7198 | Short | Diffuse | 717 | 337 | 2.24 | 1.28 | 2.58 | 39 |
| 4 | 25 | 73 | 7973 | Short | Sharp | 706 | 329 | 0.48 | 2.56 | 2.60 | 40 |
| 4 | 15 | 26 | 7666 | Long | Sharp | 710 | 330 | 1.12 | 2.4 | 2.65 | 41 |
| 4 | 13 | 93 | 7619 | Long | Sharp | 709 | 329 | 0.96 | 2.56 | 2.73 | 42 |
| 4 | 28 | 50 | 8056 | Short | Sharp | 716 | 357 | 2.08 | -1.92 | 2.83 | 43 |
| 4 | 25 | 43 | 7964 | Short | Diffuse | 720 | 340 | 2.72 | 0.8 | 2.84 | 44 |
| 4 | 24 | 43 | 7934 | Short | Diffuse | 702 | 327 | -0.16 | 2.88 | 2.88 | 45 |
| 4 | 36 | 30 | 8290 | Short | Sharp | 716 | 358 | 2.08 | -2.08 | 2.94 | 46 |
| 3 | 41 | 46 | 6645 | Long | Sharp | 715 | 331 | 1.92 | 2.24 | 2.95 | 47 |
| 3 | 59 | 60 | 7189 | Long | Sharp | 703 | 326 | 0 | 3.04 | 3.04 | 48 |
| 4 | 15 | 56 | 7677 | Long | Diffuse | 722 | 347 | 3.04 | -0.32 | 3.06 | 49 |
| 3 | 57 | 86 | 7137 | Long | Sharp | 688 | 358 | -2.4 | -2.08 | 3.18 | 50 |
| 3 | 57 | 50 | 7126 | Short | Sharp | 721 | 354 | 2.88 | -1.44 | 3.22 | 51 |
| 3 | 29 | 60 | 6289 | Short | Diffuse | 723 | 349 | 3.2 | -0.64 | 3.26 | 52 |
| 3 | 44 | 33 | 6731 | Short | Sharp | 688 | 331 | -2.4 | 2.24 | 3.28 | 53 |
| 3 | 52 | 36 | 6972 | Short | Sharp | 721 | 335 | 2.88 | 1.6 | 3.29 | 54 |
| 4 | 3 | 10 | 7294 | Short | Sharp | 723 | 339 | 3.2 | 0.96 | 3.34 | 55 |
| 4 | 42 | 53 | 8477 | Long | Sharp | 724 | 341 | 3.36 | 0.64 | 3.42 | 56 |
| 4 | 3 | 73 | 7313 | Short | Sharp | 725 | 346 | 3.52 | -0.16 | 3.52 | 57 |
| 3 | 37 | 80 | 6535 | Short | Sharp | 706 | 323 | 0.48 | 3.52 | 3.55 | 58 |

| | | | | | | | | | | | | |
|---|---|---|---|---|---|---|---|---|---|---|---|---|
| | 3 | 55 | 56 | 7068 | Short | Diffuse | 726 | 346 | 3.68 | -0.16 | 3.68 | 59 | |
| | 4 | 3 | 53 | 7307 | Short | Diffuse | 726 | 350 | 3.68 | -0.8 | 3.77 | 60 | |
| | 3 | 38 | 26 | 6549 | Long | Sharp | 723 | 332 | 3.2 | 2.08 | 3.82 | 61 | |
| | 3 | 40 | 23 | 6608 | Short | Diffuse | 680 | 352 | -3.68 | -1.12 | 3.85 | 62 | |
| | 3 | 50 | 46 | 6915 | Short | Sharp | 682 | 332 | -3.36 | 2.08 | 3.95 | 63 | |
| | 3 | 36 | 0 | 6481 | Short | Diffuse | 680 | 355 | -3.68 | -1.6 | 4.01 | 64 | |
| | 4 | 22 | 60 | 7879 | Short | Diffuse | 678 | 349 | -4 | -0.64 | 4.05 | 65 | |
| | 4 | 28 | 43 | 8054 | Short | Sharp | 681 | 358 | -3.52 | -2.08 | 4.09 | 66 | |
| | 3 | 52 | 70 | 6982 | Short | Diffuse | 694 | 369 | -1.44 | -3.84 | 4.10 | 67 | |
| | 3 | 50 | 80 | 6925 | Long | Sharp | 704 | 319 | 0.16 | 4.16 | 4.16 | 68 | |
| | 3 | 53 | 26 | 6999 | Short | Sharp | 685 | 325 | -2.88 | 3.2 | 4.31 | 69 | |
| | 4 | 15 | 73 | 7658 | Long | Sharp | 686 | 366 | -2.72 | -3.36 | 4.32 | 70 | |
| | 3 | 49 | 23 | 6878 | Short | Diffuse | 730 | 351 | 4.32 | -0.96 | 4.43 | 71 | |
| | 4 | 7 | 6 | 7413 | Long | Sharp | 687 | 322 | -2.56 | 3.68 | 4.48 | 72 | |
| | 3 | 57 | 16 | 7116 | Short | Diffuse | 728 | 358 | 4 | -2.08 | 4.51 | 73 | |
| | 2 | 49 | 3 | 5072 | Long | Sharp | 682 | 326 | -3.36 | 3.04 | 4.53 | 74 | |
| | 3 | 32 | 0 | 6361 | Long | Sharp | 731 | 340 | 4.48 | 0.8 | 4.55 | 75 | |
| | 4 | 42 | 66 | 8481 | Short | Diffuse | 684 | 323 | -3.04 | 3.52 | 4.65 | 76 | |
| | 3 | 47 | 96 | 6840 | Long | Sharp | 704 | 315 | 0.16 | 4.8 | 4.80 | 77 | |
| | 4 | 12 | 60 | 7579 | Long | Sharp | 692 | 317 | -1.76 | 4.48 | 4.81 | 78 | |
| | 4 | 14 | 23 | 7628 | Long | Sharp | 675 | 334 | -4.48 | 1.76 | 4.81 | 79 | |
| | 3 | 37 | 60 | 6529 | Long | Sharp | 688 | 318 | -2.4 | 4.32 | 4.94 | 80 | |
| | 3 | 57 | 43 | 7124 | Short | Diffuse | 729 | 362 | 4.16 | -2.72 | 4.97 | 81 | |
| | 3 | 37 | 30 | 6520 | Long | Sharp | 732 | 357 | 4.64 | -1.92 | 5.02 | 82 | |
| | 3 | 55 | 3 | 7052 | Long | Sharp | 675 | 361 | -4.48 | -2.56 | 5.16 | 83 | |
| | 4 | 9 | 0 | 7471 | Long | Sharp | 732 | 360 | 4.64 | -2.4 | 5.22 | 84 | |
| | 3 | 40 | 23 | 6608 | Long | Sharp | 734 | 356 | 4.96 | -1.76 | 5.26 | 85 | |
| | 4 | 9 | 33 | 7481 | Long | Sharp | 676 | 364 | -4.32 | -3.04 | 5.28 | 86 | |
| | 3 | 40 | 53 | 6617 | Long | Sharp | 676 | 325 | -4.32 | 3.2 | 5.38 | 87 | |
| | 3 | 22 | 53 | 6077 | Short | Diffuse | 735 | 356 | 5.12 | -1.76 | 5.41 | 88 | |
| | 3 | 34 | 10 | 6424 | Short | Sharp | 728 | 322 | 4 | 3.68 | 5.44 | 89 | |
| | 3 | 39 | 30 | 6580 | Long | Sharp | 735 | 357 | 5.12 | -1.92 | 5.47 | 90 | |
| | 3 | 34 | 30 | 6430 | Long | Sharp | 686 | 315 | -2.72 | 4.8 | 5.52 | 91 | |
| | 3 | 17 | 56 | 5928 | Short | Diffuse | 726 | 371 | 3.68 | -4.16 | 5.55 | 92 | |
| | 3 | 30 | 23 | 6308 | Short | Diffuse | 703 | 310 | 0 | 5.6 | 5.60 | 93 | |
| | 3 | 34 | 33 | 6431 | Short | Diffuse | 705 | 310 | 0.32 | 5.6 | 5.61 | 94 | |

| | | | | | | | | | | | |
|---|---|---|---|---|---|---|---|---|---|---|---|
| 3 | 39 | 70 | 6592 | Long | Sharp | 726 | 318 | 3.68 | 4.32 | 5.67 | 95 |
| 3 | 39 | 23 | 6578 | Long | Sharp | 738 | 351 | 5.6 | -0.96 | 5.68 | 96 |
| 3 | 41 | 70 | 6652 | Long | Sharp | 671 | 361 | -5.12 | -2.56 | 5.72 | 97 |
| 3 | 16 | 33 | 5891 | Short | Diffuse | 729 | 370 | 4.16 | -4 | 5.77 | 98 |
| 4 | 11 | 0 | 7531 | Short | Diffuse | 738 | 356 | 5.6 | -1.76 | 5.87 | 99 |
| 3 | 34 | 23 | 6428 | Short | Diffuse | 685 | 377 | -2.88 | -5.12 | 5.87 | 100 |
| 4 | 2 | 0 | 7261 | Long | Sharp | 675 | 370 | -4.48 | -4 | 6.01 | 101 |
| 3 | 16 | 30 | 5890 | Long | Sharp | 712 | 382 | 1.44 | -5.92 | 6.09 | 102 |
| 4 | 15 | 93 | 7679 | Long | Diffuse | 665 | 351 | -6.08 | -0.96 | 6.16 | 103 |
| 3 | 21 | 3 | 6032 | Short | Diffuse | 717 | 381 | 2.24 | -5.76 | 6.18 | 104 |
| 3 | 28 | 16 | 6246 | Long | Sharp | 702 | 306 | -0.16 | 6.24 | 6.24 | 105 |
| 3 | 17 | 56 | 5928 | Short | Sharp | 727 | 376 | 3.84 | -4.96 | 6.27 | 106 |
| 3 | 25 | 53 | 6167 | Long | Diffuse | 699 | 384 | -0.64 | -6.24 | 6.27 | 107 |
| 3 | 50 | 3 | 6902 | Long | Sharp | 742 | 341 | 6.24 | 0.64 | 6.27 | 108 |
| 4 | 6 | 56 | 7398 | Short | Diffuse | 741 | 355 | 6.08 | -1.6 | 6.29 | 109 |
| 3 | 28 | 66 | 6261 | Short | Sharp | 734 | 320 | 4.96 | 4 | 6.37 | 110 |
| 4 | 17 | 50 | 7726 | Short | Diffuse | 682 | 311 | -3.36 | 5.44 | 6.39 | 111 |
| 3 | 44 | 13 | 6725 | Long | Sharp | 663 | 348 | -6.4 | -0.48 | 6.42 | 112 |
| 3 | 44 | 16 | 6726 | Long | Sharp | 662 | 345 | -6.56 | 0 | 6.56 | 113 |
| 4 | 11 | 6 | 7533 | Long | Sharp | 662 | 350 | -6.56 | -0.8 | 6.61 | 114 |
| 4 | 17 | 50 | 7726 | Short | Diffuse | 706 | 303 | 0.48 | 6.72 | 6.74 | 115 |
| 3 | 47 | 80 | 6835 | Long | Sharp | 746 | 346 | 6.88 | -0.16 | 6.88 | 116 |
| 4 | 2 | 56 | 7278 | Long | Sharp | 661 | 356 | -6.72 | -1.76 | 6.95 | 117 |
| 3 | 26 | 53 | 6197 | Short | Diffuse | 741 | 368 | 6.08 | -3.68 | 7.11 | 118 |
| 3 | 29 | 6 | 6273 | Long | Sharp | 746 | 357 | 6.88 | -1.92 | 7.14 | 119 |
| 3 | 26 | 26 | 6189 | Long | Sharp | 725 | 306 | 3.52 | 6.24 | 7.16 | 120 |
| 3 | 44 | 90 | 6748 | Long | Sharp | 663 | 324 | -6.4 | 3.36 | 7.23 | 121 |
| 3 | 32 | 83 | 6386 | Short | Diffuse | 686 | 303 | -2.72 | 6.72 | 7.25 | 122 |
| 3 | 46 | 36 | 6792 | Long | Sharp | 748 | 337 | 7.2 | 1.28 | 7.31 | 123 |
| 3 | 30 | 56 | 6318 | Short | Diffuse | 680 | 385 | -3.68 | -6.4 | 7.38 | 124 |
| 3 | 42 | 36 | 6672 | Short | Diffuse | 745 | 365 | 6.72 | -3.2 | 7.44 | 125 |
| 4 | 14 | 50 | 7636 | Long | Sharp | 661 | 325 | -6.72 | 3.2 | 7.44 | 126 |
| 4 | 7 | 13 | 7415 | Long | Diffuse | 749 | 338 | 7.36 | 1.12 | 7.44 | 127 |
| 4 | 31 | 66 | 8151 | Short | Sharp | 662 | 367 | -6.56 | -3.52 | 7.44 | 128 |
| 4 | 2 | 86 | 7287 | Short | Diffuse | 746 | 363 | 6.88 | -2.88 | 7.46 | 129 |
| 3 | 35 | 20 | 6457 | Long | Sharp | 743 | 319 | 6.4 | 4.16 | 7.63 | 130 |
| 4 | 34 | 33 | 8231 | Long | Diffuse | 656 | 335 | -7.52 | 1.6 | 7.69 | 131 |
| 3 | 19 | 26 | 5979 | Short | Diffuse | 674 | 384 | -4.64 | -6.24 | 7.78 | 132 |
| 3 | 44 | 46 | 6735 | Long | Sharp | 694 | 297 | -1.44 | 7.68 | 7.81 | 133 |
| 4 | 33 | 6 | 8193 | Long | Sharp | 661 | 320 | -6.72 | 4 | 7.82 | 134 |
| 3 | 53 | 6 | 6993 | Short | Diffuse | 752 | 344 | 7.84 | 0.16 | 7.84 | 135 |
| 3 | 58 | 63 | 7160 | Long | Diffuse | 749 | 328 | 7.36 | 2.72 | 7.85 | 136 |
| 3 | 27 | 53 | 6227 | Short | Diffuse | 673 | 384 | -4.8 | -6.24 | 7.87 | 137 |
| 3 | 27 | 53 | 6227 | Long | Sharp | 658 | 324 | -7.2 | 3.36 | 7.95 | 138 |
| 3 | 36 | 90 | 6508 | Long | Sharp | 657 | 364 | -7.36 | -3.04 | 7.96 | 139 |
| 4 | 23 | 26 | 7899 | Long | Diffuse | 734 | 306 | 4.96 | 6.24 | 7.97 | 140 |
| 3 | 30 | 50 | 6316 | Long | Sharp | 748 | 367 | 7.2 | -3.52 | 8.01 | 141 |
| 4 | 8 | 0 | 7441 | Long | Sharp | 714 | 296 | 1.76 | 7.84 | 8.04 | 142 |
| 3 | 24 | 86 | 6147 | Long | Sharp | 718 | 297 | 2.4 | 7.68 | 8.05 | 143 |
| 3 | 36 | 43 | 6494 | Short | Diffuse | 748 | 368 | 7.2 | -3.68 | 8.09 | 144 |
| 4 | 6 | 30 | 7390 | Long | Sharp | 753 | 337 | 8 | 1.28 | 8.10 | 145 |
| 3 | 40 | 93 | 6629 | Long | Diffuse | 746 | 318 | 6.88 | 4.32 | 8.12 | 146 |
| 4 | 1 | 40 | 7243 | Long | Sharp | 666 | 310 | -5.92 | 5.6 | 8.15 | 147 |
| 3 | 37 | 73 | 6533 | Long | Sharp | 661 | 374 | -6.72 | -4.64 | 8.17 | 148 |
| 4 | 22 | 80 | 7885 | Short | Diffuse | 658 | 320 | -7.2 | 4 | 8.24 | 149 |
| 3 | 28 | 63 | 6260 | Long | Sharp | 655 | 366 | -7.68 | -3.36 | 8.38 | 150 |
| 4 | 6 | 26 | 7389 | Long | Sharp | 755 | 336 | 8.32 | 1.44 | 8.44 | 151 |
| 4 | 22 | 60 | 7879 | Short | Diffuse | 681 | 393 | -3.52 | -7.68 | 8.45 | 152 |
| 4 | 10 | 3 | 7502 | Long | Sharp | 658 | 317 | -7.2 | 4.48 | 8.48 | 153 |
| 4 | 12 | 6 | 7563 | Long | Sharp | 742 | 382 | 6.24 | -5.92 | 8.60 | 154 |
| 4 | 23 | 26 | 7899 | Short | Diffuse | 666 | 384 | -5.92 | -6.24 | 8.60 | 155 |
| 3 | 48 | 3 | 6842 | Short | Sharp | 757 | 342 | 8.64 | 0.48 | 8.65 | 156 |
| 3 | 47 | 86 | 6837 | Long | Sharp | 662 | 309 | -6.56 | 5.76 | 8.73 | 157 |
| 3 | 29 | 23 | 6278 | Short | Diffuse | 656 | 317 | -7.52 | 4.48 | 8.75 | 158 |
| 3 | 35 | 96 | 6480 | Long | Sharp | 655 | 372 | -7.68 | -4.32 | 8.81 | 159 |
| 3 | 17 | 46 | 5925 | Short | Diffuse | 735 | 390 | 5.12 | -7.2 | 8.83 | 160 |
| 3 | 15 | 20 | 5857 | Long | Sharp | 731 | 393 | 4.48 | -7.68 | 8.89 | 161 |
| 3 | 18 | 93 | 5969 | Short | Sharp | 729 | 295 | 4.16 | 8 | 9.02 | 162 |
| 3 | 18 | 43 | 5954 | Long | Sharp | 759 | 352 | 8.96 | -1.12 | 9.03 | 163 |
| 3 | 27 | 66 | 6231 | Long | Diffuse | 739 | 389 | 5.76 | -7.04 | 9.10 | 164 |
| 4 | 17 | 93 | 7739 | Long | Sharp | 654 | 316 | -7.84 | 4.64 | 9.11 | 165 |
| 3 | 31 | 10 | 6334 | Short | Diffuse | 655 | 314 | -7.68 | 4.96 | 9.14 | 166 |
| 3 | 23 | 73 | 6113 | Long | Sharp | 653 | 375 | -8 | -4.8 | 9.33 | 167 |
| 4 | 23 | 73 | 7913 | Long | Diffuse | 653 | 315 | -8 | 4.8 | 9.33 | 168 |
| 3 | 19 | 93 | 5999 | Long | Sharp | 720 | 289 | 2.72 | 8.96 | 9.36 | 169 |
| 4 | 27 | 76 | 8034 | Short | Sharp | 685 | 289 | -2.88 | 8.96 | 9.41 | 170 |
| 3 | 38 | 30 | 6550 | Long | Sharp | 746 | 304 | 6.88 | 6.56 | 9.51 | 171 |
| 4 | 36 | 26 | 8289 | Short | Diffuse | 658 | 306 | -7.2 | 6.24 | 9.53 | 172 |
| 3 | 27 | 90 | 6238 | Short | Sharp | 758 | 322 | 8.8 | 3.68 | 9.54 | 173 |
| 3 | 34 | 63 | 6440 | Short | Sharp | 657 | 306 | -7.36 | 6.24 | 9.65 | 174 |
| 3 | 20 | 43 | 6014 | Long | Sharp | 686 | 287 | -2.72 | 9.28 | 9.67 | 175 |
| 4 | 35 | 23 | 8258 | Short | Diffuse | 644 | 359 | -9.44 | -2.24 | 9.70 | 176 |
| 3 | 33 | 96 | 6420 | Long | Sharp | 723 | 286 | 3.2 | 9.44 | 9.97 | 177 |
| 3 | 36 | 56 | 6498 | Short | Diffuse | 762 | 368 | 9.44 | -3.68 | 10.13 | 178 |
| 3 | 31 | 10 | 6334 | Short | Diffuse | 669 | 291 | -5.44 | 8.64 | 10.21 | 179 |
| 3 | 36 | 36 | 6492 | Short | Diffuse | 698 | 410 | -0.8 | -10.4 | 10.43 | 180 |
| 3 | 21 | 96 | 6060 | Short | Diffuse | 720 | 282 | 2.72 | 10.08 | 10.44 | 181 |
| 3 | 30 | 23 | 6308 | Long | Sharp | 737 | 289 | 5.44 | 8.96 | 10.48 | 182 |
| 3 | 28 | 0 | 6241 | Long | Sharp | 769 | 346 | 10.56 | -0.16 | 10.56 | 183 |
| 3 | 22 | 86 | 6087 | Long | Sharp | 715 | 280 | 1.92 | 10.4 | 10.58 | 184 |
| 3 | 48 | 40 | 6853 | Short | Diffuse | 748 | 296 | 7.2 | 7.84 | 10.64 | 185 |
| 3 | 17 | 90 | 5938 | Long | Sharp | 700 | 278 | -0.48 | 10.72 | 10.73 | 186 |
| 3 | 29 | 6 | 6273 | Long | Sharp | 767 | 371 | 10.24 | -4.16 | 11.05 | 187 |
| 3 | 32 | 96 | 6390 | Long | Sharp | 736 | 281 | 5.28 | 10.24 | 11.52 | 188 |
| 4 | 20 | 30 | 7810 | Long | Sharp | 733 | 278 | 4.8 | 10.72 | 11.75 | 189 |
| 3 | 32 | 20 | 6367 | Short | Sharp | 641 | 305 | -9.92 | 6.4 | 11.81 | 190 |
| 3 | 27 | 23 | 6218 | Long | Sharp | 687 | 418 | -2.56 | -11.68 | 11.96 | 191 |
| 4 | 11 | 96 | 7560 | Long | Sharp | 638 | 306 | -10.4 | 6.24 | 12.13 | 192 |
| 3 | 34 | 73 | 6443 | Long | Sharp | 640 | 302 | -10.08 | 6.88 | 12.20 | 193 |
| 4 | 41 | 73 | 8453 | Short | Diffuse | 627 | 361 | -12.16 | -2.56 | 12.43 | 194 |
| 3 | 15 | 56 | 5868 | Short | Diffuse | 757 | 285 | 8.64 | 9.6 | 12.92 | 195 |
| 3 | 47 | 23 | 6818 | Short | Diffuse | 781 | 367 | 12.48 | -3.52 | 12.97 | 196 |
| 3 | 17 | 20 | 5917 | Long | Diffuse | 758 | 285 | 8.8 | 9.6 | 13.02 | 197 |
| 3 | 35 | 20 | 6457 | Long | Sharp | 658 | 277 | -7.2 | 10.88 | 13.05 | 198 |
| 3 | 43 | 73 | 6713 | Long | Sharp | 754 | 281 | 8.16 | 10.24 | 13.09 | 199 |
| 3 | 15 | 70 | 5872 | Short | Diffuse | 762 | 286 | 9.44 | 9.44 | 13.35 | 200 |
| 3 | 19 | 60 | 5989 | Short | Diffuse | 769 | 397 | 10.56 | -8.32 | 13.44 | 201 |
| 3 | 18 | 26 | 5949 | Long | Sharp | 769 | 292 | 10.56 | 8.48 | 13.54 | 202 |

| | | | | | | | | | | | | |
|---|---|---|---|---|---|---|---|---|---|---|---|---|
| | 3 | 17 | 26 | 5919 | Short | Diffuse | 653 | 276 | -8 | 11.04 | 13.63 | 203 |
| | 3 | 15 | 20 | 5857 | Short | Diffuse | 735 | 424 | 5.12 | -12.64 | 13.64 | 204 |
| | 3 | 16 | 43 | 5894 | Long | Diffuse | 788 | 334 | 13.6 | 1.76 | 13.71 | 205 |
| | 3 | 36 | 23 | 6488 | Long | Sharp | 739 | 266 | 5.76 | 12.64 | 13.89 | 206 |
| | 4 | 20 | 20 | 7807 | Long | Sharp | 750 | 267 | 7.52 | 12.48 | 14.57 | 207 |
| | 3 | 17 | 86 | 5937 | Short | Sharp | 789 | 376 | 13.76 | -4.96 | 14.63 | 208 |
| | 4 | 14 | 63 | 7640 | Short | Sharp | 644 | 272 | -9.44 | 11.68 | 15.02 | 209 |
| | 3 | 50 | 30 | 6910 | Short | Diffuse | 610 | 331 | -14.88 | 2.24 | 15.05 | 210 |
| | 4 | 15 | 76 | 7672 | Short | Diffuse | 791 | 310 | 14.08 | 5.6 | 15.15 | 211 |
| | 3 | 23 | 70 | 6112 | Long | Diffuse | 625 | 291 | -12.48 | 8.64 | 15.18 | 212 |
| | 3 | 25 | 30 | 6160 | Short | Diffuse | 742 | 257 | 6.24 | 14.08 | 15.40 | 213 |
| | 3 | 23 | 23 | 6098 | Long | Sharp | 743 | 256 | 6.4 | 14.24 | 15.61 | 214 |
| | 4 | 27 | 53 | 8027 | Long | Diffuse | 626 | 285 | -12.32 | 9.6 | 15.62 | 215 |
| | 3 | 50 | 16 | 6906 | Short | Diffuse | 605 | 329 | -15.68 | 2.56 | 15.89 | 216 |
| | 3 | 16 | 23 | 5888 | Short | Sharp | 619 | 404 | -13.44 | -9.44 | 16.42 | 217 |
| | 3 | 16 | 53 | 5897 | Short | Diffuse | 608 | 304 | -15.2 | 6.56 | 16.56 | 218 |
| | 4 | 33 | 6 | 8193 | Short | Diffuse | 606 | 305 | -15.52 | 6.4 | 16.79 | 219 |
| | 3 | 21 | 23 | 6038 | Long | Diffuse | 800 | 294 | 15.52 | 8.16 | 17.53 | 220 |
| | 3 | 18 | 66 | 5961 | Long | Sharp | 818 | 312 | 18.4 | 5.28 | 19.14 | 221 |
| | 3 | 24 | 43 | 6134 | Long | Sharp | 595 | 282 | -17.28 | 10.08 | 20.01 | 222 |
| | 3 | 15 | 93 | 5879 | Long | Diffuse | 827 | 380 | 19.84 | -5.6 | 20.62 | 223 |
| | 4 | 18 | 53 | 7757 | Short | Diffuse | 626 | 236 | -12.32 | 17.44 | 21.35 | 224 |
| | 3 | 50 | 20 | 6907 | Short | Diffuse | 569 | 315 | -21.44 | 4.8 | 21.97 | 225 |
| | 4 | 26 | 40 | 7993 | Long | Diffuse | 567 | 372 | -21.76 | -4.32 | 22.18 | 226 |
| | 3 | 27 | 23 | 6218 | Long | Sharp | 663 | 525 | -6.4 | -28.8 | 29.50 | 227 |
| | 3 | 23 | 70 | 6112 | Short | Diffuse | 515 | 335 | -30.08 | 1.6 | 30.12 | 228 |
| **CORRESPONDING IDEAL CUMULATIVE TRACK DISTRIBUTION - NO CUTOFF** | | | | | | | | | | | | |
| **COPIED FROM PRECEDING TAB** | | | | | | | | | | | | |
| | | | | | | | | | | | 0.25 | 14.00 |
| | | | | | | | | | | | 0.50 | 25.71 |
| | | | | | | | | | | | 0.75 | 35.83 |
| | | | | | | | | | | | 1.00 | 44.85 |
| | | | | | | | | | | | 1.25 | 53.06 |
| | | | | | | | | | | | 1.50 | 60.61 |
| | | | | | | | | | | | 1.75 | 67.62 |
| | | | | | | | | | | | 2.00 | 74.16 |
| | | | | | | | | | | | 2.25 | 80.30 |
| | | | | | | | | | | | 2.50 | 86.08 |
| | | | | | | | | | | | 2.75 | 91.53 |
| | | | | | | | | | | | 3.00 | 96.68 |
| | | | | | | | | | | | 3.25 | 101.57 |
| | | | | | | | | | | | 3.50 | 106.21 |
| | | | | | | | | | | | 3.75 | 110.63 |
| | | | | | | | | | | | 4.00 | 114.85 |
| | | | | | | | | | | | 4.25 | 118.88 |
| | | | | | | | | | | | 4.50 | 122.73 |
| | | | | | | | | | | | 4.75 | 126.42 |
| | | | | | | | | | | | 5.00 | 129.96 |
| | | | | | | | | | | | 5.25 | 133.36 |
| | | | | | | | | | | | 5.50 | 136.63 |
| | | | | | | | | | | | 5.75 | 139.79 |
| | | | | | | | | | | | 6.00 | 142.82 |
| | | | | | | | | | | | 6.25 | 145.76 |

TAB #4: CUTOFF MODELS, DIVIDED BY PI

| | | | | | |
|---|---|---|---|---|---|
| a = | 4 mm | | | h= | 0.5 |
| b = | 3.5 mm | | | Active region scale | 1.05 |
| a-b= | 0.5 mm | | | | |
| K = | 22.5 mm | | | | |
| K/b= | 6.43 | | | | |
| sqrt(K^2-b^2)= | 22.23 mm | | | | |
| K/(a-b)= | 45.00 | | | | |
| sqrt(K^2-(a-b)^2)= | 22.49 mm | | | | |
| simple scale= | 9.18 | | | | |
| subtractive scale= | 10.93 | | | | |

| R | R/b | Simple cutoff b part (C1) | R/(a-b) | Simple cutoff a-b part (C1) | Simple scaled (C1) |
|---|---|---|---|---|---|
| 0.25 | 0.07 | 0.77 | 0.50 | 0.67 | 13.15 |
| 0.50 | 0.14 | 1.50 | 1.00 | 1.13 | 24.16 |
| 0.75 | 0.21 | 2.20 | 1.50 | 1.47 | 33.68 |
| 1.00 | 0.29 | 2.86 | 2.00 | 1.73 | 42.16 |
| 1.25 | 0.36 | 3.49 | 2.50 | 1.94 | 49.87 |
| 1.50 | 0.43 | 4.09 | 3.00 | 2.12 | 56.97 |
| 1.75 | 0.50 | 4.66 | 3.50 | 2.27 | 63.56 |
| 2.00 | 0.57 | 5.20 | 4.00 | 2.40 | 69.71 |
| 2.25 | 0.64 | 5.71 | 4.50 | 2.51 | 75.48 |
| 2.50 | 0.71 | 6.20 | 5.00 | 2.62 | 80.91 |
| 2.75 | 0.79 | 6.66 | 5.50 | 2.71 | 86.03 |
| 3.00 | 0.86 | 7.10 | 6.00 | 2.80 | 90.87 |
| 3.25 | 0.93 | 7.52 | 6.50 | 2.88 | 95.47 |
| 3.50 | 1.00 | 7.92 | 7.00 | 2.95 | 99.83 |
| 3.75 | 1.07 | 8.31 | 7.50 | 3.02 | 103.99 |
| 4.00 | 1.14 | 8.68 | 8.00 | 3.08 | 107.95 |
| 4.25 | 1.21 | 9.03 | 8.50 | 3.14 | 111.74 |
| 4.50 | 1.29 | 9.36 | 9.00 | 3.20 | 115.36 |
| 4.75 | 1.36 | 9.69 | 9.50 | 3.25 | 118.82 |
| 5.00 | 1.43 | 10.00 | 10.00 | 3.30 | 122.15 |
| 5.25 | 1.50 | 10.30 | 10.50 | 3.35 | 125.35 |
| 5.50 | 1.57 | 10.59 | 11.00 | 3.40 | 128.43 |
| 5.75 | 1.64 | 10.87 | 11.50 | 3.44 | 131.39 |
| 6.00 | 1.71 | 11.13 | 12.00 | 3.49 | 134.25 |
| 6.25 | 1.79 | 11.39 | 12.50 | 3.53 | 137.00 |
| 6.50 | 1.86 | 11.65 | 13.00 | 3.57 | 139.67 |
| 6.75 | 1.93 | 11.89 | 13.50 | 3.60 | 142.25 |
| 7.00 | 2.00 | 12.12 | 14.00 | 3.64 | 144.74 |
| 7.25 | 2.07 | 12.35 | 14.50 | 3.67 | 147.16 |
| 7.50 | 2.14 | 12.57 | 15.00 | 3.71 | 149.50 |
| 7.75 | 2.21 | 12.79 | 15.50 | 3.74 | 151.78 |
| 8.00 | 2.29 | 13.00 | 16.00 | 3.77 | 153.99 |
| 8.25 | 2.36 | 13.20 | 16.50 | 3.80 | 156.14 |
| 8.50 | 2.43 | 13.40 | 17.00 | 3.83 | 158.23 |
| 8.75 | 2.50 | 13.59 | 17.50 | 3.86 | 160.27 |
| 9.00 | 2.57 | 13.78 | 18.00 | 3.89 | 162.25 |
| 9.25 | 2.64 | 13.96 | 18.50 | 3.92 | 164.18 |
| 9.50 | 2.71 | 14.14 | 19.00 | 3.94 | 166.07 |
| 9.75 | 2.79 | 14.32 | 19.50 | 3.97 | 167.91 |
| 10.00 | 2.86 | 14.49 | 20.00 | 4.00 | 169.70 |
| 10.25 | 2.93 | 14.65 | 20.50 | 4.02 | 171.46 |
| 10.50 | 3.00 | 14.82 | 21.00 | 4.04 | 173.17 |
| 10.75 | 3.07 | 14.97 | 21.50 | 4.07 | 174.85 |
| 11.00 | 3.14 | 15.13 | 22.00 | 4.09 | 176.49 |
| 11.25 | 3.21 | 15.28 | 22.50 | 4.11 | 178.10 |
| 11.50 | 3.29 | 15.43 | 23.00 | 4.14 | 179.67 |
| 11.75 | 3.36 | 15.58 | 23.50 | 4.16 | 181.21 |
| 12.00 | 3.43 | 15.72 | 24.00 | 4.18 | 182.72 |
| 12.25 | 3.50 | 15.86 | 24.50 | 4.20 | 184.20 |
| 12.50 | 3.57 | 16.00 | 25.00 | 4.22 | 185.65 |
| 12.75 | 3.64 | 16.14 | 25.50 | 4.24 | 187.07 |
| 13.00 | 3.71 | 16.27 | 26.00 | 4.26 | 188.47 |
| 13.25 | 3.79 | 16.40 | 26.50 | 4.28 | 189.84 |
| 13.50 | 3.86 | 16.53 | 27.00 | 4.30 | 191.19 |
| 13.75 | 3.93 | 16.65 | 27.50 | 4.31 | 192.51 |
| 14.00 | 4.00 | 16.78 | 28.00 | 4.33 | 193.81 |
| 14.25 | 4.07 | 16.90 | 28.50 | 4.35 | 195.09 |
| 14.50 | 4.14 | 17.02 | 29.00 | 4.37 | 196.34 |
| 14.75 | 4.21 | 17.13 | 29.50 | 4.38 | 197.58 |
| 15.00 | 4.29 | 17.25 | 30.00 | 4.40 | 198.79 |
| 15.25 | 4.36 | 17.36 | 30.50 | 4.42 | 199.99 |
| 15.50 | 4.43 | 17.48 | 31.00 | 4.43 | 201.17 |
| 15.75 | 4.50 | 17.59 | 31.50 | 4.45 | 202.32 |
| 16.00 | 4.57 | 17.69 | 32.00 | 4.47 | 203.46 |
| 16.25 | 4.64 | 17.80 | 32.50 | 4.48 | 204.59 |
| 16.50 | 4.71 | 17.91 | 33.00 | 4.50 | 205.69 |
| 16.75 | 4.79 | 18.01 | 33.50 | 4.51 | 206.79 |
| 17.00 | 4.86 | 18.11 | 34.00 | 4.53 | 207.86 |
| 17.25 | 4.93 | 18.21 | 34.50 | 4.54 | 208.92 |
| 17.50 | 5.00 | 18.31 | 35.00 | 4.56 | 209.96 |
| 17.75 | 5.07 | 18.41 | 35.50 | 4.57 | 210.99 |
| 18.00 | 5.14 | 18.51 | 36.00 | 4.58 | 212.01 |
| 18.25 | 5.21 | 18.60 | 36.50 | 4.60 | 213.01 |
| 18.50 | 5.29 | 18.70 | 37.00 | 4.61 | 214.00 |
| 18.75 | 5.36 | 18.79 | 37.50 | 4.62 | 214.98 |
| 19.00 | 5.43 | 18.88 | 38.00 | 4.64 | 215.94 |
| 19.25 | 5.50 | 18.97 | 38.50 | 4.65 | 216.89 |
| 19.50 | 5.57 | 19.06 | 39.00 | 4.66 | 217.83 |
| 19.75 | 5.64 | 19.15 | 39.50 | 4.68 | 218.76 |
| 20.00 | 5.71 | 19.24 | 40.00 | 4.69 | 219.67 |
| 20.25 | 5.79 | 19.32 | 40.50 | 4.70 | 220.58 |
| 20.50 | 5.86 | 19.41 | 41.00 | 4.71 | 221.47 |
| 20.75 | 5.93 | 19.49 | 41.50 | 4.73 | 222.36 |
| 21.00 | 6.00 | 19.57 | 42.00 | 4.74 | 223.23 |
| 21.25 | 6.07 | 19.66 | 42.50 | 4.75 | 224.09 |
| 21.50 | 6.14 | 19.74 | 43.00 | 4.76 | 224.94 |
| 21.75 | 6.21 | 19.82 | 43.50 | 4.77 | 225.79 |
| 22.00 | 6.29 | 19.90 | 44.00 | 4.78 | 226.62 |
| 22.25 | 6.36 | 19.98 | 44.50 | 4.80 | 227.44 |
| 22.50 | 6.43 | 20.03 | 45.00 | 4.81 | 228.00 |
| 22.75 | 6.50 | 20.03 | 45.50 | 4.81 | 228.00 |
| 23.00 | 6.57 | 20.03 | 46.00 | 4.81 | 228.00 |
| 23.25 | 6.64 | 20.03 | 46.50 | 4.81 | 228.00 |
| 23.50 | 6.71 | 20.03 | 47.00 | 4.81 | 228.00 |
| 23.75 | 6.79 | 20.03 | 47.50 | 4.81 | 228.00 |
| 24.00 | 6.86 | 20.03 | 48.00 | 4.81 | 228.00 |
| 24.25 | 6.93 | 20.03 | 48.50 | 4.81 | 228.00 |
| 24.50 | 7.00 | 20.03 | 49.00 | 4.81 | 228.00 |
| 24.75 | 7.07 | 20.03 | 49.50 | 4.81 | 228.00 |
| 25.00 | 7.14 | 20.03 | 50.00 | 4.81 | 228.00 |
| 25.25 | 7.21 | 20.03 | 50.50 | 4.81 | 228.00 |
| 25.50 | 7.29 | 20.03 | 51.00 | 4.81 | 228.00 |
| 25.75 | 7.36 | 20.03 | 51.50 | 4.81 | 228.00 |
| 26.00 | 7.43 | 20.03 | 52.00 | 4.81 | 228.00 |
| 26.25 | 7.50 | 20.03 | 52.50 | 4.81 | 228.00 |
| 26.50 | 7.57 | 20.03 | 53.00 | 4.81 | 228.00 |
| 26.75 | 7.64 | 20.03 | 53.50 | 4.81 | 228.00 |
| 27.00 | 7.71 | 20.03 | 54.00 | 4.81 | 228.00 |
| 27.25 | 7.79 | 20.03 | 54.50 | 4.81 | 228.00 |
| 27.50 | 7.86 | 20.03 | 55.00 | 4.81 | 228.00 |
| 27.75 | 7.93 | 20.03 | 55.50 | 4.81 | 228.00 |
| 28.00 | 8.00 | 20.03 | 56.00 | 4.81 | 228.00 |
| 28.25 | 8.07 | 20.03 | 56.50 | 4.81 | 228.00 |
| 28.50 | 8.14 | 20.03 | 57.00 | 4.81 | 228.00 |
| 28.75 | 8.21 | 20.03 | 57.50 | 4.81 | 228.00 |
| 29.00 | 8.29 | 20.03 | 58.00 | 4.81 | 228.00 |
| 29.25 | 8.36 | 20.03 | 58.50 | 4.81 | 228.00 |
| 29.50 | 8.43 | 20.03 | 59.00 | 4.81 | 228.00 |
| 29.75 | 8.50 | 20.03 | 59.50 | 4.81 | 228.00 |
| 30.00 | 8.57 | 20.03 | 60.00 | 4.81 | 228.00 |
| 30.25 | 8.64 | 20.03 | 60.50 | 4.81 | 228.00 |
| 30.50 | 8.71 | 20.03 | 61.00 | 4.81 | 228.00 |
| 30.75 | 8.79 | 20.03 | 61.50 | 4.81 | 228.00 |
| 31.00 | 8.86 | 20.03 | 62.00 | 4.81 | 228.00 |
| 31.25 | 8.93 | 20.03 | 62.50 | 4.81 | 228.00 |
| 31.50 | 9.00 | 20.03 | 63.00 | 4.81 | 228.00 |
| 31.75 | 9.07 | 20.03 | 63.50 | 4.81 | 228.00 |
| 32.00 | 9.14 | 20.03 | 64.00 | 4.81 | 228.00 |
| 32.25 | 9.21 | 20.03 | 64.50 | 4.81 | 228.00 |
| 32.50 | 9.29 | 20.03 | 65.00 | 4.81 | 228.00 |
| 32.75 | 9.36 | 20.03 | 65.50 | 4.81 | 228.00 |
| 33.00 | 9.43 | 20.03 | 66.00 | 4.81 | 228.00 |
| 33.25 | 9.50 | 20.03 | 66.50 | 4.81 | 228.00 |
| 33.50 | 9.57 | 20.03 | 67.00 | 4.81 | 228.00 |
| 33.75 | 9.64 | 20.03 | 67.50 | 4.81 | 228.00 |
| 34.00 | 9.71 | 20.03 | 68.00 | 4.81 | 228.00 |
| 34.25 | 9.79 | 20.03 | 68.50 | 4.81 | 228.00 |
| 34.50 | 9.86 | 20.03 | 69.00 | 4.81 | 228.00 |
| 34.75 | 9.93 | 20.03 | 69.50 | 4.81 | 228.00 |
| 35.00 | 10.00 | 20.03 | 70.00 | 4.81 | 228.00 |
| 35.25 | 10.07 | 20.03 | 70.50 | 4.81 | 228.00 |
| 35.50 | 10.14 | 20.03 | 71.00 | 4.81 | 228.00 |
| 35.75 | 10.21 | 20.03 | 71.50 | 4.81 | 228.00 |
| 36.00 | 10.29 | 20.03 | 72.00 | 4.81 | 228.00 |
| 36.25 | 10.36 | 20.03 | 72.50 | 4.81 | 228.00 |
| 36.50 | 10.43 | 20.03 | 73.00 | 4.81 | 228.00 |
| 36.75 | 10.50 | 20.03 | 73.50 | 4.81 | 228.00 |
| 37.00 | 10.57 | 20.03 | 74.00 | 4.81 | 228.00 |
| 37.25 | 10.64 | 20.03 | 74.50 | 4.81 | 228.00 |
| 37.50 | 10.71 | 20.03 | 75.00 | 4.81 | 228.00 |
| 37.75 | 10.79 | 20.03 | 75.50 | 4.81 | 228.00 |
| 38.00 | 10.86 | 20.03 | 76.00 | 4.81 | 228.00 |
| 38.25 | 10.93 | 20.03 | 76.50 | 4.81 | 228.00 |
| 38.50 | 11.00 | 20.03 | 77.00 | 4.81 | 228.00 |
| 38.75 | 11.07 | 20.03 | 77.50 | 4.81 | 228.00 |
| 39.00 | 11.14 | 20.03 | 78.00 | 4.81 | 228.00 |
| 39.25 | 11.21 | 20.03 | 78.50 | 4.81 | 228.00 |
| 39.50 | 11.29 | 20.03 | 79.00 | 4.81 | 228.00 |
| 39.75 | 11.36 | 20.03 | 79.50 | 4.81 | 228.00 |
| 40.00 | 11.43 | 20.03 | 80.00 | 4.81 | 228.00 |
| 40.25 | 11.50 | 20.03 | 80.50 | 4.81 | 228.00 |
| 40.50 | 11.57 | 20.03 | 81.00 | 4.81 | 228.00 |
| 40.75 | 11.64 | 20.03 | 81.50 | 4.81 | 228.00 |
| 41.00 | 11.71 | 20.03 | 82.00 | 4.81 | 228.00 |
| 41.25 | 11.79 | 20.03 | 82.50 | 4.81 | 228.00 |
| 41.50 | 11.86 | 20.03 | 83.00 | 4.81 | 228.00 |
| 41.75 | 11.93 | 20.03 | 83.50 | 4.81 | 228.00 |
| 42.00 | 12.00 | 20.03 | 84.00 | 4.81 | 228.00 |
| 42.25 | 12.07 | 20.03 | 84.50 | 4.81 | 228.00 |
| 42.50 | 12.14 | 20.03 | 85.00 | 4.81 | 228.00 |
| 42.75 | 12.21 | 20.03 | 85.50 | 4.81 | 228.00 |
| 43.00 | 12.29 | 20.03 | 86.00 | 4.81 | 228.00 |
| 43.25 | 12.36 | 20.03 | 86.50 | 4.81 | 228.00 |
| 43.50 | 12.43 | 20.03 | 87.00 | 4.81 | 228.00 |
| 43.75 | 12.50 | 20.03 | 87.50 | 4.81 | 228.00 |
| 44.00 | 12.57 | 20.03 | 88.00 | 4.81 | 228.00 |
| 44.25 | 12.64 | 20.03 | 88.50 | 4.81 | 228.00 |
| 44.50 | 12.71 | 20.03 | 89.00 | 4.81 | 228.00 |
| 44.75 | 12.79 | 20.03 | 89.50 | 4.81 | 228.00 |
| 45.00 | 12.86 | 20.03 | 90.00 | 4.81 | 228.00 |
| 45.25 | 12.93 | 20.03 | 90.50 | 4.81 | 228.00 |
| 45.50 | 13.00 | 20.03 | 91.00 | 4.81 | 228.00 |
| 45.75 | 13.07 | 20.03 | 91.50 | 4.81 | 228.00 |
| 46.00 | 13.14 | 20.03 | 92.00 | 4.81 | 228.00 |
| 46.25 | 13.21 | 20.03 | 92.50 | 4.81 | 228.00 |
| 46.50 | 13.29 | 20.03 | 93.00 | 4.81 | 228.00 |
| 46.75 | 13.36 | 20.03 | 93.50 | 4.81 | 228.00 |
| 47.00 | 13.43 | 20.03 | 94.00 | 4.81 | 228.00 |
| 47.25 | 13.50 | 20.03 | 94.50 | 4.81 | 228.00 |
| 47.50 | 13.57 | 20.03 | 95.00 | 4.81 | 228.00 |
| 47.75 | 13.64 | 20.03 | 95.50 | 4.81 | 228.00 |
| 48.00 | 13.71 | 20.03 | 96.00 | 4.81 | 228.00 |
| 48.25 | 13.79 | 20.03 | 96.50 | 4.81 | 228.00 |
| 48.50 | 13.86 | 20.03 | 97.00 | 4.81 | 228.00 |
| 48.75 | 13.93 | 20.03 | 97.50 | 4.81 | 228.00 |
| 49.00 | 14.00 | 20.03 | 98.00 | 4.81 | 228.00 |
| 49.25 | 14.07 | 20.03 | 98.50 | 4.81 | 228.00 |
| 49.50 | 14.14 | 20.03 | 99.00 | 4.81 | 228.00 |
| 49.75 | 14.21 | 20.03 | 99.50 | 4.81 | 228.00 |
| 50.00 | 14.29 | 20.03 | 100.00 | 4.81 | 228.00 |

mm     height of source above hypothetical active region

| Subtraction b part (C1-C2) | Subtraction a-b part (C1-C2) | Subtractive scaled (C2) | Active region solid angle |
|---|---|---|---|
| 0.00 | 0.00 | 15.65 | 25.27 |
| 0.00 | 0.00 | 28.74 | 70.12 |
| 0.00 | 0.00 | 40.04 | 106.60 |
| 0.01 | 0.00 | 50.10 | 132.34 |
| 0.01 | 0.00 | 59.23 | 150.49 |
| 0.02 | 0.00 | 67.62 | 163.70 |
| 0.02 | 0.00 | 75.39 | 173.63 |
| 0.03 | 0.00 | 82.64 | 181.34 |
| 0.04 | 0.01 | 89.41 | 187.47 |
| 0.04 | 0.01 | 95.77 | 192.45 |
| 0.05 | 0.01 | 101.75 | 196.57 |
| 0.06 | 0.01 | 107.40 | 200.04 |
| 0.07 | 0.01 | 112.73 | 203.00 |
| 0.08 | 0.01 | 117.78 | 205.54 |
| 0.10 | 0.01 | 122.57 | 207.76 |
| 0.11 | 0.02 | 127.12 | 209.71 |
| 0.12 | 0.02 | 131.45 | 211.43 |
| 0.14 | 0.02 | 135.57 | 212.96 |
| 0.16 | 0.02 | 139.50 | 214.34 |
| 0.17 | 0.02 | 143.25 | 215.58 |
| 0.19 | 0.03 | 146.84 | 216.70 |
| 0.21 | 0.03 | 150.27 | 217.73 |
| 0.23 | 0.03 | 153.55 | 218.66 |
| 0.25 | 0.04 | 156.70 | 219.52 |
| 0.27 | 0.04 | 159.71 | 220.31 |
| 0.29 | 0.04 | 162.61 | 221.04 |
| 0.32 | 0.05 | 165.39 | 221.72 |
| 0.34 | 0.05 | 168.07 | 222.34 |
| 0.36 | 0.05 | 170.64 | 222.93 |
| 0.39 | 0.06 | 173.11 | 223.48 |
| 0.42 | 0.06 | 175.49 | 223.99 |
| 0.44 | 0.06 | 177.78 | 224.47 |
| 0.47 | 0.07 | 179.99 | 224.92 |
| 0.50 | 0.07 | 182.12 | 225.34 |
| 0.53 | 0.08 | 184.17 | 225.74 |
| 0.56 | 0.08 | 186.15 | 226.12 |
| 0.59 | 0.08 | 188.05 | 226.48 |
| 0.62 | 0.09 | 189.89 | 226.82 |
| 0.66 | 0.09 | 191.67 | 227.14 |
| 0.69 | 0.10 | 193.38 | 227.44 |
| 0.73 | 0.10 | 195.03 | 227.74 |
| 0.76 | 0.11 | 196.62 | 228.01 |
| 0.80 | 0.11 | 198.16 | 228.28 |
| 0.84 | 0.12 | 199.64 | 228.53 |
| 0.88 | 0.13 | 201.08 | 228.77 |
| 0.91 | 0.13 | 202.46 | 229.00 |
| 0.95 | 0.14 | 203.79 | 229.22 |
| 1.00 | 0.14 | 205.07 | 229.43 |
| 1.04 | 0.15 | 206.31 | 229.64 |
| 1.08 | 0.15 | 207.50 | 229.83 |
| 1.12 | 0.16 | 208.65 | 230.02 |
| 1.17 | 0.17 | 209.76 | 230.20 |
| 1.21 | 0.17 | 210.82 | 230.37 |
| 1.26 | 0.18 | 211.85 | 230.54 |
| 1.31 | 0.19 | 212.83 | 230.70 |
| 1.36 | 0.19 | 213.78 | 230.86 |
| 1.40 | 0.20 | 214.69 | 231.01 |
| 1.45 | 0.21 | 215.57 | 231.15 |
| 1.50 | 0.21 | 216.41 | 231.29 |
| 1.56 | 0.22 | 217.21 | 231.42 |
| 1.61 | 0.23 | 217.98 | 231.56 |
| 1.66 | 0.24 | 218.72 | 231.68 |
| 1.72 | 0.25 | 219.42 | 231.80 |
| 1.77 | 0.25 | 220.10 | 231.92 |
| 1.83 | 0.26 | 220.74 | 232.04 |
| 1.88 | 0.27 | 221.35 | 232.15 |
| 1.94 | 0.28 | 221.93 | 232.26 |
| 2.00 | 0.29 | 222.48 | 232.36 |
| 2.06 | 0.29 | 223.00 | 232.46 |
| 2.12 | 0.30 | 223.49 | 232.56 |
| 2.18 | 0.31 | 223.96 | 232.66 |
| 2.24 | 0.32 | 224.40 | 232.75 |
| 2.30 | 0.33 | 224.81 | 232.84 |
| 2.37 | 0.34 | 225.19 | 232.93 |
| 2.43 | 0.35 | 225.55 | 233.02 |
| 2.50 | 0.36 | 225.88 | 233.10 |
| 2.56 | 0.37 | 226.19 | 233.18 |
| 2.63 | 0.38 | 226.47 | 233.26 |
| 2.70 | 0.39 | 226.72 | 233.34 |
| 2.77 | 0.40 | 226.96 | 233.42 |
| 2.84 | 0.41 | 227.16 | 233.49 |
| 2.91 | 0.42 | 227.35 | 233.56 |
| 2.98 | 0.43 | 227.51 | 233.63 |
| 3.05 | 0.44 | 227.65 | 233.70 |
| 3.12 | 0.45 | 227.76 | 233.77 |
| 3.20 | 0.46 | 227.85 | 233.83 |
| 3.27 | 0.47 | 227.92 | 233.90 |
| 3.35 | 0.48 | 227.97 | 233.96 |
| 3.42 | 0.49 | 227.99 | 234.02 |
| 3.47 | 0.50 | 228.00 | 234.08 |
| 3.47 | 0.50 | 228.00 | 234.14 |
| 3.47 | 0.50 | 228.00 | 234.20 |
| 3.47 | 0.50 | 228.00 | 234.25 |
| 3.47 | 0.50 | 228.00 | 234.31 |
| 3.47 | 0.50 | 228.00 | 234.36 |
| 3.47 | 0.50 | 228.00 | 234.41 |
| 3.47 | 0.50 | 228.00 | 234.46 |
| 3.47 | 0.50 | 228.00 | 234.52 |
| 3.47 | 0.50 | 228.00 | 234.56 |
| 3.47 | 0.50 | 228.00 | 234.61 |
| 3.47 | 0.50 | 228.00 | 234.66 |
| 3.47 | 0.50 | 228.00 | 234.71 |
| 3.47 | 0.50 | 228.00 | 234.75 |
| 3.47 | 0.50 | 228.00 | 234.80 |
| 3.47 | 0.50 | 228.00 | 234.84 |
| 3.47 | 0.50 | 228.00 | 234.88 |
| 3.47 | 0.50 | 228.00 | 234.93 |
| 3.47 | 0.50 | 228.00 | 234.97 |
| 3.47 | 0.50 | 228.00 | 235.01 |
| 3.47 | 0.50 | 228.00 | 235.05 |
| 3.47 | 0.50 | 228.00 | 235.09 |
| 3.47 | 0.50 | 228.00 | 235.13 |
| 3.47 | 0.50 | 228.00 | 235.16 |
| 3.47 | 0.50 | 228.00 | 235.20 |
| 3.47 | 0.50 | 228.00 | 235.24 |
| 3.47 | 0.50 | 228.00 | 235.27 |
| 3.47 | 0.50 | 228.00 | 235.31 |
| 3.47 | 0.50 | 228.00 | 235.34 |
| 3.47 | 0.50 | 228.00 | 235.38 |
| 3.47 | 0.50 | 228.00 | 235.41 |
| 3.47 | 0.50 | 228.00 | 235.44 |
| 3.47 | 0.50 | 228.00 | 235.48 |
| 3.47 | 0.50 | 228.00 | 235.51 |
| 3.47 | 0.50 | 228.00 | 235.54 |
| 3.47 | 0.50 | 228.00 | 235.57 |
| 3.47 | 0.50 | 228.00 | 235.60 |
| 3.47 | 0.50 | 228.00 | 235.63 |
| 3.47 | 0.50 | 228.00 | 235.66 |
| 3.47 | 0.50 | 228.00 | 235.69 |
| 3.47 | 0.50 | 228.00 | 235.72 |
| 3.47 | 0.50 | 228.00 | 235.75 |
| 3.47 | 0.50 | 228.00 | 235.77 |
| 3.47 | 0.50 | 228.00 | 235.80 |
| 3.47 | 0.50 | 228.00 | 235.83 |
| 3.47 | 0.50 | 228.00 | 235.85 |
| 3.47 | 0.50 | 228.00 | 235.88 |
| 3.47 | 0.50 | 228.00 | 235.91 |
| 3.47 | 0.50 | 228.00 | 235.93 |
| 3.47 | 0.50 | 228.00 | 235.96 |
| 3.47 | 0.50 | 228.00 | 235.98 |
| 3.47 | 0.50 | 228.00 | 236.00 |
| 3.47 | 0.50 | 228.00 | 236.03 |
| 3.47 | 0.50 | 228.00 | 236.05 |
| 3.47 | 0.50 | 228.00 | 236.08 |
| 3.47 | 0.50 | 228.00 | 236.10 |
| 3.47 | 0.50 | 228.00 | 236.12 |
| 3.47 | 0.50 | 228.00 | 236.14 |
| 3.47 | 0.50 | 228.00 | 236.17 |
| 3.47 | 0.50 | 228.00 | 236.19 |
| 3.47 | 0.50 | 228.00 | 236.21 |
| 3.47 | 0.50 | 228.00 | 236.23 |
| 3.47 | 0.50 | 228.00 | 236.25 |
| 3.47 | 0.50 | 228.00 | 236.27 |
| 3.47 | 0.50 | 228.00 | 236.29 |
| 3.47 | 0.50 | 228.00 | 236.31 |
| 3.47 | 0.50 | 228.00 | 236.33 |
| 3.47 | 0.50 | 228.00 | 236.35 |
| 3.47 | 0.50 | 228.00 | 236.37 |
| 3.47 | 0.50 | 228.00 | 236.39 |
| 3.47 | 0.50 | 228.00 | 236.41 |
| 3.47 | 0.50 | 228.00 | 236.43 |
| 3.47 | 0.50 | 228.00 | 236.44 |
| 3.47 | 0.50 | 228.00 | 236.46 |
| 3.47 | 0.50 | 228.00 | 236.48 |
| 3.47 | 0.50 | 228.00 | 236.50 |
| 3.47 | 0.50 | 228.00 | 236.52 |
| 3.47 | 0.50 | 228.00 | 236.53 |
| 3.47 | 0.50 | 228.00 | 236.55 |
| 3.47 | 0.50 | 228.00 | 236.57 |
| 3.47 | 0.50 | 228.00 | 236.58 |
| 3.47 | 0.50 | 228.00 | 236.60 |
| 3.47 | 0.50 | 228.00 | 236.62 |
| 3.47 | 0.50 | 228.00 | 236.63 |
| 3.47 | 0.50 | 228.00 | 236.65 |
| 3.47 | 0.50 | 228.00 | 236.66 |
| 3.47 | 0.50 | 228.00 | 236.68 |
| 3.47 | 0.50 | 228.00 | 236.70 |
| 3.47 | 0.50 | 228.00 | 236.71 |
| 3.47 | 0.50 | 228.00 | 236.73 |
| 3.47 | 0.50 | 228.00 | 236.74 |
| 3.47 | 0.50 | 228.00 | 236.75 |
| 3.47 | 0.50 | 228.00 | 236.77 |
| 3.47 | 0.50 | 228.00 | 236.78 |
| 3.47 | 0.50 | 228.00 | 236.80 |
| 3.47 | 0.50 | 228.00 | 236.81 |
| 3.47 | 0.50 | 228.00 | 236.83 |
| 3.47 | 0.50 | 228.00 | 236.84 |
| 3.47 | 0.50 | 228.00 | 236.85 |
| 3.47 | 0.50 | 228.00 | 236.87 |
| 3.47 | 0.50 | 228.00 | 236.88 |
| 3.47 | 0.50 | 228.00 | 236.89 |
| 3.47 | 0.50 | 228.00 | 236.91 |
| 3.47 | 0.50 | 228.00 | 236.92 |
| 3.47 | 0.50 | 228.00 | 236.93 |
| 3.47 | 0.50 | 228.00 | 236.94 |
| 3.47 | 0.50 | 228.00 | 236.96 |
| 3.47 | 0.50 | 228.00 | 236.97 |
| 3.47 | 0.50 | 228.00 | 236.98 |
| 3.47 | 0.50 | 228.00 | 236.99 |
| 3.47 | 0.50 | 228.00 | 237.01 |

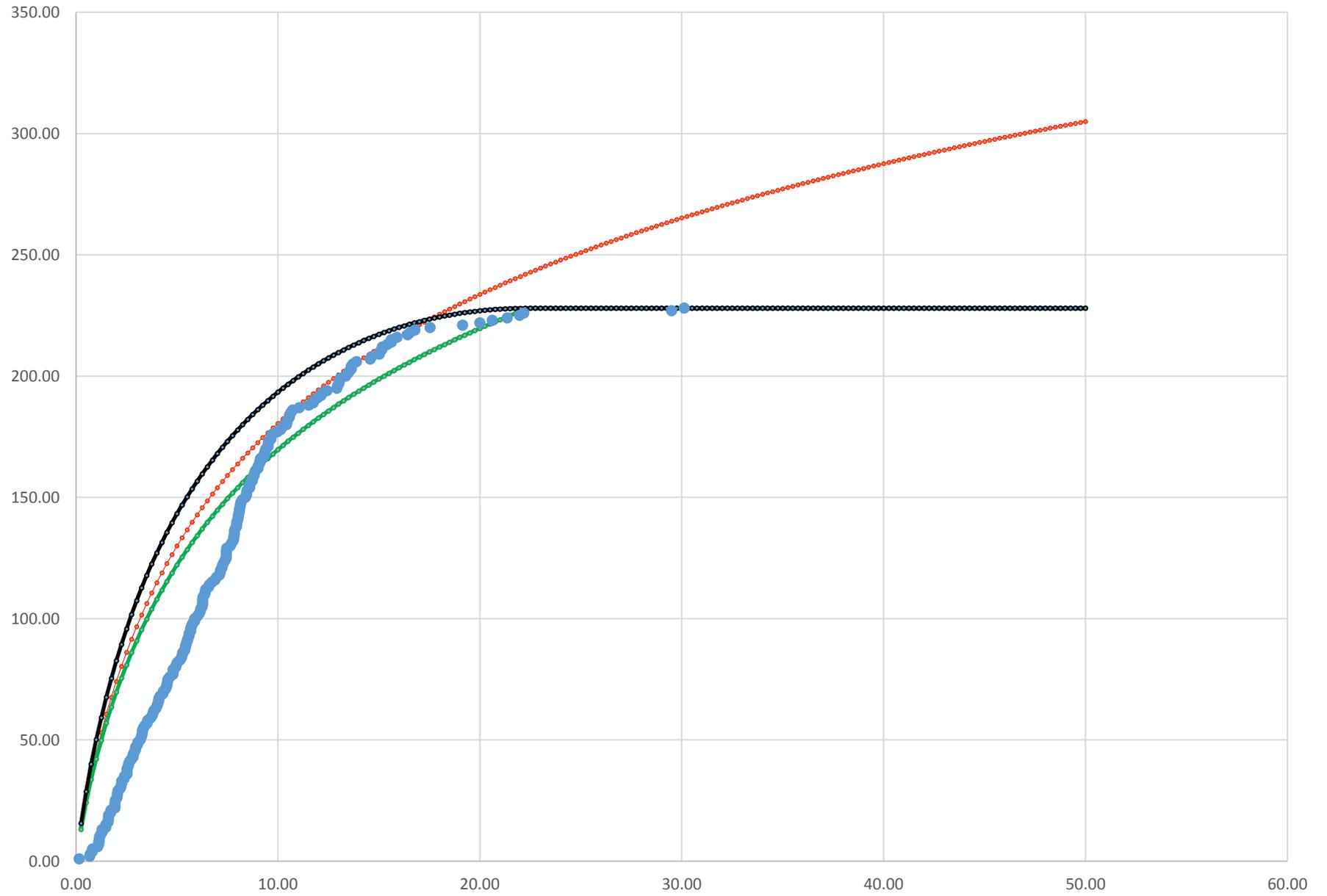